\documentclass[11pt,a4paper]{article}
\pdfoutput=1
\usepackage{jheppub}

\usepackage[T1]{fontenc} 
\usepackage{amsmath,amsfonts,amssymb,physics,eqnarray,bm}
\usepackage{graphicx,float,caption,subcaption,xcolor,enumitem}
\usepackage{dcolumn}

\graphicspath{{img/}}

\title{Generic ETH: Eigenstate thermalization beyond the microcanonical}

\author[a]{Elena Caceres,}
\author[b,c]{Stefan Eccles,}
\author[d]{Jason Pollack,}
\author[e]{and Sarah Racz}

\affiliation[a]{Theory Group, Department of Physics, The University of Texas at Austin, Austin TX 78712, USA}
\affiliation[b]{Institute for Gravitation and the Cosmos, Pennsylvania State University, University Park, PA 16801, USA}
\affiliation[c]{Okinawa Institute of Science and Technology Graduate University, Onna, Okinawa 904 0495, Japan}
\affiliation[d]{Institute for Quantum \& Information Sciences and Department of Electrical Enginneering \& Computer Science, Syracuse University, Syracuse, NY 13244, USA}
\affiliation[e]{Department of Physics, Sarah Lawrence College, Bronxville, NY 10708, USA}

\emailAdd{elenac@utexas.edu}
\emailAdd{sve5267@psu.edu}
\emailAdd{japollac@syr.edu}
\emailAdd{sracz@sarahlawrence.edu}

\abstract{
The Eigenstate Thermalization Hypothesis (ETH) has played a key role in recent advances in the high energy and condensed matter communities. It explains how an isolated quantum system in a far-from-equilibrium initial state can evolve to a state that is indistinguishable from thermal equilibrium, with observables relaxing to almost 
time-independent results that can be described using traditional statistical mechanics ensembles.
In this work we probe the limits of ETH, pushing it outside its prototypical applications in several directions.  We design a qutrit lattice system with conserved quasilocal charge, in which we verify a form of generalized eigenstate thermalization.  We also observe signatures of thermalization in states well outside microcanonical windows of both charge and energy, which we dub `generic ETH.'}

\begin{document}
\maketitle

\tableofcontents



\newcommand{\jp}[1]{\textcolor{violet}{jp: #1}}
\newcommand{\sv}[1]{\textcolor{red}{se: #1}}
\newcommand{\ec}[1]{\textcolor{blue}{ec: #1}}
\newcommand{\sr}[1]{\textcolor{green}{sr: #1}}

\section{Introduction}
\label{sec:intro}

A standard course in statistical mechanics and thermodynamics conveys that in many scenarios, a few macroscopic properties of a system such as the temperature, pressure, and volume of an ideal gas suffice to characterize its behavior completely.  In such scenarios, the precise microscopic details of the state do not matter except in so far as they determine the system’s energy and the values of other conserved parameters.   Observables are then predicted using thermal ensembles, time-independent distributions of states that maximize entropy while remaining consistent with constraints on conserved quantities.  Even if an initial state is prepared out-of-equilibrium, measurable quantities often evolve toward those thermal values and remain close to them for most times thereafter.  Indeed, this is what is meant by “thermalization”.  The fact that such behavior is common, but not universal, begs the question:  Is there a simple criterion that clarifies when a given system, state, or operator will thermalize? 

In the case of a closed quantum system, the Eigenstate Thermalization Hypothesis (ETH)\cite{Srednicki:1994mfb, Deutsch:1991msp,Jensen:1984gu} provides the prevailing framework for addressing the question of which operators and states will thermalize in a given system.  Most directly, the ETH is an ansatz for the form of operator matrix elements in the energy eigenbasis, along with the suggestion that it is such operators that thermalize, at least in states that have narrow energy support.  The ETH has been well-supported both numerically and analytically in various settings \cite{rigol2008thermalization,mondaini2016eigenstate}, and applied to domains ranging from conformal field theory \cite{Lashkari:2016vgj,Dymarsky:2019etq} to holography \cite{Pollack:2020gfa,Saad:2019pqd,Altland:2021rqn}. Nevertheless, open questions still remain regarding its domain of applicability.  An informal, but well-motivated expectation is that it is the local and otherwise ``simple'' operators in chaotic systems which take the ETH form.  By contrast, integrable systems (here meaning systems with an extensive number of conserved charges) generally do not thermalize and correspondingly have simple observables which are not well-characterized by ETH.  In fact, this is not a strict dichotomy.  Many states in integrable systems have been shown to relax to a Generalized Gibbs Ensemble \cite{Rigol_Dunjko_Yurovsky_PhysRevLett.98.050405,Cramer_Dawson_Eisert_Osborne_PhysRevLett.100.030602} which take account of conserved parameters additional to the energy.  In such cases, a generalized eigenstate thermalization hypothesis \cite{cassidy_2011,He_Santos_Wright_Rigol_2013,Rigol:2016itf} offers an explanation for this behavior again in terms of operator matrix elements.  Once again, the necessary and sufficient conditions for the occurrence of generalized ETH, and the extent to which it suffices to account for all instances of thermalization or lack thereof, is the subject of ongoing research in varied contexts \cite{Dymarsky:2019etq,PhysRevLett.131.060401,Pappalardi:2023nsj}.  Various additional generalizations of standard ETH have also been considered, for instance to the case of noncommuting charges \cite{Murthy:2022dao}, and in another direction to account for higher moment correlations in operator coefficients \cite{Foini_Kurchan_2019}. In this work, we do not attempt a complete review of these topics, but we do intend to provide a gentle introduction for the uninitiated, replete with examples in lattice systems.

Our novel contributions include further probes of the link between ETH, chaos, and thermalization in lattice systems, including the onset of generalized eigenstate thermalization in a system with conserved charge.  We design a qutrit spin chain system specifically for this purpose.  The system parameters allow a tunable degree of chaos and charge diffusion, as well as independent control over the local charge density, energy density, and entanglement entropy of initial states to an extent not possible in some simpler models.  We verify the expectation that, in the presence of a conserved charge, ETH applies and entails thermalization within individual charge sectors, provided these sectors are sufficiently chaotic.  We additionally observe thermalizing behavior for some states that are prepared well outside a microcanonical window (across both energy and charge sectors), which illustrates thermalization beyond the prototypical regime of ETH. We dub  this behavior `generic eigenstate thermalization'.  We provide an explanation for why we believe this is to be expected, but we leave further investigation of this phenomenon to future work.

The paper is organized as follows.  
Section \ref{sec: Background, thermalization and ETH} provides some theoretical background.  We begin by establishing our working definitions of some key terms.  We introduce ETH as a mechanism for quantum thermalization, and discuss its connection to quantum chaos.  We also explain our use of the term generic ETH.
In section \ref{sec: qubit spinchain models} we introduce a simple one-dimensional qubit spin chain model and use it to illustrate the simultaneous onset of chaos, thermalization, and the ETH form of simple operators in these systems.  This section is primarily illustrative and pedagogical, but the model introduced here provides a basis for the model introduced in the next section.  Readers only interested in our results in systems with conserved charge may skip straight to section \ref{sec: Thermalization with conserved charge},
in which we extend the qubit model of the preceding section to a qutrit model in a manner that includes a quasilocal conserved charge. There is a corresponding notion of local charge density which is not conserved but undergoes diffusive local transport. By design, we find that within individual charge sectors, this Hamiltonian is chaotic.  ETH applies with increasing accuracy as the Hilbert space dimension of these sectors increases, and to the extent that it applies, thermalization occurs for states drawn from a small energy window within charge sectors.  We then study the behavior of more generic states with a spread across charge sectors and again confirm thermalizing behavior.  Thermalization of such states is consistent with the broad mechanism of generalized ETH, but outside the prototypical states to which ETH is usually applied, and so we refer to this behavior as generic ETH.  In section \ref{sec: conclusions} we conclude with a brief discussion and directions for future work.

\section{Background: thermalization and ETH}
\label{sec: Background, thermalization and ETH}
In this section we briefly introduce the Eigenstate Thermalization Hypothesis as a framework for understanding thermalization in quantum systems. Before doing so we define some key terms, since various authors and subdisciplines may differ in their usage.  By no means do we intend a complete review of the topics mentioned here, but rather a minimal orientation.   A focused and succinct introduction to ETH is given by \cite{Deutsch:2018ulr}.  A much larger, but still pedagogical overview is available in \cite{DAlessio:2015qtq}, which we draw on heavily also to inspire some of our later numerical illustrations.  Another overview of related topics is \cite{Mori:2017qhg}, which includes extended discussion of timescales involved in the relaxation process to thermal states.   A theoretically rigorous discussion of many foundational aspects of quantum thermalization is found in \cite{Gogolin:2015gts}, which includes rigorous criteria for various notions of thermalization and a summary of results available for each.  Although not phrased primarily in the language of ETH, the review \cite{Borgonovi:2016nrx} provides a complementary perspective on the link between chaos and thermalization, with particular emphasis on examples from realistic models of atomic and nuclear systems.

\subsection*{Equilibration vs. Thermalization}
Roughly speaking, we say that a measurable quantity \textit{equilibrates} if it begins far from and then evolves toward some particular value, designated the equilibrium value, and remains near this value for most times thereafter.  Generically, this can be any time-dependent property of a system (using the terminology of \cite{Gogolin:2015gts}), such as the value of a classical phase space function, the expectation value of a Hermitian operator in a quantum system, or the standard deviation of that same quantum operator, ...etc.  With the understanding that we will be focused on measurable quantities in a closed quantum system (including measurements made on identically prepared systems) we include various ``state properties'' \cite{Keeler:2023shl} along with the usual operator observables. An important example of a state property in quantum systems is the von Neumann entropy, $S(\rho):=-\Tr\rho\log\rho$, of a subsystem density matrix $\rho$, which will serve as one of our chief diagnostics of equilibration.  When evaluated on a subsystem of a pure state, it quantifies the entanglement between the subsystem and complement system, so we hereafter refer to it as the entanglement entropy.  

We say that such properties or observables \textit{thermalize} if they equilibrate to values matching the prediction of some thermal ensemble.  Thus the latter condition is stronger than the former. An observable might equilibrate but not thermalize, but not vice versa, and if an observable has equilibrated, in order to say whether or not it has thermalized requires a choice of thermal ensemble for comparison.

We can also speak of equilibration (thermalization) of states themselves, rather than particular observables, if some pre-chosen set of characteristic observables are found to equilibrate (thermalize) in these states.  Likewise we can speak of the equilibration (thermalization) of a system on the whole, e.g. a specific Hamiltonian, with respect to some characteristic set of (or measure over) states and observable properties.  With these considerations in mind, it is apparent that a huge variety of alternative notions of equilibration and thermalization may be considered as relevant to different contexts.  See \cite{Gogolin:2015gts} for an excellent overview of such notions and for various technical results available for the different notions.  

\subsection*{What's the appropriate ensemble?}
Investigating whether and to what extent thermalization occurs requires the identification of a target thermal configuration for comparison.  Contrast this with equilibration: if equilibration occurs at all, the relevant configuration can be identified by time-averaging the objects of interest, either over all time or a specified interval.  Time-dependent deviations from these averages (i.e. the fluctuations around equilibrium) are then quantified and judged to be small enough and rare enough by some standard, or not, but no extra conditions are placed on the equilibrium values themselves.\footnote{Sometimes the thermalization properties of a system refer specifically and strictly to its  behavior in a thermodynamic limit.  In this work we are explicitly considering finite, closed systems, and we only discuss the thermodynamic limit in a few scenarios.}  For \textit{thermal equilibrium}, these values must match the predictions of some thermal ensemble, the quintessential examples of which are the microcanonical, canonical, and grand-canonical ensembles. As a matter of terminology, we will refer to the canonical case as the \textit{Gibbs ensemble}, and also consider \textit{generalized Gibbs ensembles} when additional nontrivial conserved charges are present. Sometimes this term is applied to integrable systems when when an extensive number of conserved charges are present, though we do not require this.  See \cite{Rigol_Dunjko_Yurovsky_PhysRevLett.98.050405,Cramer_Dawson_Eisert_Osborne_PhysRevLett.100.030602} for early examples using such ensembles, \cite{rigol_2016} and section 5.2 of \cite{Gogolin:2015gts} for some pertinent review, and \cite{PhysRevA.104.L031303} for some interesting related work.  All of these ensembles are static configurations (for time-independent Hamiltonians) that maximize Gibbs entropy, or von Neumann entropy in the quantum case, subject to some constraints. This fact, taken together with an appeal to the second law of thermodynamics, is often considered justification enough for their relevance \cite{PhysRev.106.620,PhysRev.108.171}. A complete description of thermalization must also include dynamical explication of when, whether, and why such configurations appear to be approached, but regardless these ensembles are here taken to \textit{define} thermality.  

For open systems, the relevant ensemble depends on which macroscopic quantities are used to parameterize the system, or equivalently on what type of exchange is allowed with some idealized, external bath.  For states on a finite closed system, the relevance of any particular ensemble is less clear, though a pertinent and standard derivation shows that if a microcanonical distribution is assumed on a large system, it often leads to a reduced Gibbs state on its subsystems.  In quantum systems, an even stronger result is available \cite{Goldstein:2005aib} which essentially leverages this result for pure states, stating that almost any state drawn from a microcanonical window also reduces to a Gibbs state on its small subsystems.  Indeed, this result is closely tied to ETH.  In fact, the primary implication of ETH is often summarized by saying that for states with \textit{narrow energy support}, certain operator expectation values are well approximated by predictions of a microcanonical ensemble.  In this work we will illustrate this standard result, but also emphasize that where ETH occurs, it is more predictive than just that.  Many operators can be shown to thermalize in states that do not have narrow energy support, a behavior which we dub `generic ETH'.  The most predictive static ensemble associated with a given state is in principle determined not just by the expectation values of conserved charges but also their higher moments. Choosing an ensemble that depends on arbitrarily high moments of the conserved charges ultimately culminates in employing the diagonal ensemble \cite{PhysRevA.78.013626} when only energy is conserved, and more generally, the infinite time average of the state's density matrix. This is a static ensemble that trivially predicts any equilibrium quantities for the state, but to refer to this as a ``thermal'' prediction in general would clearly undercut the spirit of the term regarding late-time insensitivity to details of the initial state and other standard connotations of thermality.  Regardless, for local observables in chaotic systems, expectation values are often remarkably insensitive to these higher moments. See section \ref{subsec: generic ETH in theory} for further discussion of this point.  For many methods of state preparation, it is sufficient to choose between the a (generalized) Gibbs state or a (generalized) microcanonical state. There are standard results about equivalence of ensembles in a thermodynamic limit (e.g. \cite{Lima1971, Lima1972}), so that usually the physics predicted by these ensembles is equivalent for subsystems of large systems.  But for finite systems their predictions may differ, and we explore a few such cases in \ref{subsec: generic ETH in qubit system} and \ref{subsec: generic ETH in Qutrit system}.  We will therefore always take care to specify which ensemble we are comparing to for ``thermal'' predictions.

\subsection{Quantum thermalization}
\label{subsec: Quantum thermalization}

The question of why a closed quantum system ever seems to thermalize can be posed starkly in terms of a generic operator expectation value.  Consider a Hilbert space $\mathcal{H}$ of finite dimension, which we denote $|\mathcal{H}|$. Suppose the system Hamiltonian $\hat{H}$ has a completely non-degenerate spectrum (as will be the case if it is drawn from common random matrix ensembles) specified by $H\ket{E_i}=E_i\ket{E_i}$, and by default suppose these energy labels are sorted such that $E_i<E_j$ if $i<j$.   We write a generic initial state in the energy  eigenbasis as $\ket{\Psi(t=0)}=\sum_i c_i \ket{E_i}$ and denote the energy expectation value in this state $E_{\Psi} := \bra{\Psi}\hat{H}\ket{\Psi}$.   We write the matrix elements of an operator $\hat{\mathcal{O}}$ in the energy basis as $\mathcal{O}_{ij}:=\bra{E_i}\hat{\mathcal{O}}\ket{E_j}$.  The expectation value of a general operator is then given by
\begin{equation}\begin{split}\label{eq: operator expectation}
    \mathcal{O}_\Psi(t)
    &:=\bra{\Psi(t)}\hat{\mathcal{O}}\ket{\Psi(t)}
    =\sum_i |c_i|^2 \mathcal{O}_{ii} +\sum_{i,j\ne i}c_i^\star c_j e^{-i(E_j-E_i)t}\mathcal{O}_{ij}.\\
\end{split}\end{equation}
A microcanonical thermal expectation value is, of course, obtained by tracing an operator with the microcanonical density matrix at energy $E$, which we denote $\hat{\rho}^{\text{(mc)}}_E$.\footnote{
\label{label: microcanonical}
Technically the microcanonical ensemble in a finite system requires the specification of a small energy window. We could write
\[
\hat{\rho}_{[E_{\text{min}},E_\text{max}]} := \hat{\Pi}_{[E_{\text{min}}, E_\text{max}]}/\Tr\left(\hat{\Pi}_{[E_{\text{min}}, E_\text{max}]}\right)
\]
where $\hat{\Pi}_{[E_{\text{min}}, E_\text{max}]}$ is the projector onto the sub-Hilbert space of states given by $\text{span}\{\ket{E_i} \big| E_i \in [E_{\text{min}}, E_\text{max}])\}$.  With the understanding that the window is taken to be small, and that with a high density of states the dependence on the width is very weak for physical operators, we drop this two-parameter notation.  In a sense we're getting ahead of ourselves with such assumptions, which are themselves further justified in the ETH framework.  It is important for some later expressions, however, that the $E$ in $\hat{\rho}^{(\text{mc})}_E$ denotes the average energy in the window as opposed to the central energy.}
We can then ask, under what circumstances does an operator expectation value approach or approximate the thermal expectation value?  
\begin{equation}\label{eq: time-dependent op thermalization}
    \mathcal{O}_\Psi(t)
    \overset{?}{\approx}\Tr\left(\hat{\rho}^{(\text{mc})}_{E_{\Psi}}\hat{\mathcal{O}}\right)
\end{equation}
At first glance, it is puzzling that it should ever be the case; the expectation value $\mathcal{O}_\Psi(t)$ depends on all the details of the initial state, encoded in all the $c_i$'s in equation \eqref{eq: operator expectation}, whereas the microcanonical value knows none of these details except through the energy $E_\Psi$.  If the system equilibrates at all, we can at least extract the equilibrium value by taking the infinite time-average of \eqref{eq: operator expectation}, leading to
\begin{equation}\begin{split}\label{eq: time averaged operator}
    \mathcal{O}^*_\Psi&:=\underset{T\rightarrow\infty}{\lim}T^{-1}\int_{t=0}^{t=T}\mathcal{O}_\Psi(t') dt'
    =\sum_i |c_i|^2 \mathcal{O}_{ii}.\\
\end{split}\end{equation}
This allows us to ask a simpler, time-averaged version of the question \eqref{eq: time-dependent op thermalization}.  What is required to ensure that just the time-averaged expectation value matches the thermal value?
\begin{equation}\begin{split}\label{eq: time averaged thermalization}
    \mathcal{O}^*_\Psi\overset{?}{\approx}\Tr\left(\hat{\rho}^{(\text{mc})}_{E_\Psi}\hat{\mathcal{O}}\right).
\end{split}\end{equation}
Even the time-averaged quantity $\mathcal{O}^*_\Psi$ retains a dependence on all the magnitudes $|c_i|$, so in any system we can certainly write down states and operators where this approximation is badly violated.  Why would this behavior ever be generic?  To answer this we now describe the Eigenstate Thermalization Hypothesis (ETH) \cite{Srednicki:1994mfb, Deutsch:1991msp,Jensen:1984gu}. 

\subsection{Eigenstate Thermalization Hypothesis}
\label{subsec: eigenstate thermalization hypothesis}
In its simplest form, the ETH is the recognition that if the diagonal matrix elements of $\hat{\mathcal{O}}$ are well-approximated by a smooth function of the energy, which we denote simply by $\mathcal{O}:\mathbb{R} \rightarrow \mathbb{R}$, and if the state is supported in a narrow energy window, then the time average $\mathcal{O}^*_\Psi=\sum_i |c_i|^2 \mathcal{O}_{ii}$ is well approximated by replacing $\mathcal{O}_{ii}$ with the constant value $\mathcal{O}(E_\Psi)$ and pulling it out of the sum.  The $|c_i|^2$'s then sum to one by normalization, and the expectation value is approximately just $\mathcal{O}(E_\Psi)$.  The same approximate value is obtained if the ETH assumption is employed in a microcanonical expectation value at the same energy, resulting in the sought-after thermal equality:
\begin{equation}\label{eq: time averaged result}
    \mathcal{O}^*_\Psi \approx \mathcal{O}(E_\Psi) \approx \Tr\left(\hat{\rho}^{(\text{mc})}_{E_\Psi} \hat{\mathcal{O}}\right).
\end{equation}
This condition on the diagonal matrix elements entails that for all states with narrow energy support, the time average will approximate the microcanonical value, without invoking `typical state' statistics or other special conditions on the nonzero coefficients $c_i$.  In fact, even in \textit{individual} eigenstates, observables of this form appear thermal, leading to the name ``eigenstate thermalization.''

In the modern formulation, a more complete statement of ETH includes the requirement that the off-diagonal elements of $\hat{O}$ are exponentially suppressed in the size of the system. This is crucial in anticipating real-time thermalization, as opposed to just the time-averaged result demonstrated by equation \eqref{eq: time averaged result}. Following the notation of \cite{Srednicki:1999bhx,DAlessio:2015qtq}, we write the complete form of the ETH ansatz for operator matrix elements as 
\begin{equation}\label{eq: complete ETH}
        \mathcal{O}^{(ETH)}_{ij}=\mathcal{O}(\bar{E})\delta_{ij} + e^{-S(\bar{E})/2}f_{\mathcal{O}}(\bar{E},\omega)R_{ij},
\end{equation}
where $\bar{E}$ is the average energy $\frac{1}{2}(E_i+E_j)$ and  $\omega:=E_i -E_j$ the energy difference.  This form is expected to hold for many local and physically relevant operators in chaotic systems, at least out to near the edges of the spectrum.  Importantly, all off-diagonal elements are exponentially suppressed, with $S(\bar{E})$ representing the thermal entropy at energy $\bar{E}$.  This entropy can be thought of as representing the log of the number of active states in a thermal (Gibbs) state at energy expectation $\bar{E}$, so that the overall suppression factor scales inversely with (the square root of) the density of states at $\bar{E}$.  More generally, this suppression factor could be written as $S(\bar{E},\omega)$, but it is common to restrict $\omega$-dependence to the ``envelope function'' $f_\mathcal{O}(\bar{E},\omega)$.  This function is largely unspecified and context dependent.  It is expected to fall off with increasing $|\omega|$ and becomes flat for $|\omega|< \hbar/\tau^*$, where $\tau^*$ is a system timescale beyond which random matrix theory results apply\footnote{In systems with conserved charges, this is often a time associated with diffusion across the full system, called the Thouless time \cite{JTEdwards_1972}.  See \cite{Friedman:2019gyi} for analysis of this timescale in a Floquet system with conserved charge, and \cite{Roy:2020tgl, Roy:2022zig,Kumar:2023jvb} for analysis in periodically driven many body systems with and without a $U(1)$ symmetry for fermionic, bosonic, and mixed species, respectively. }.  The detailed form of $f_{\mathcal{O}}(\bar{E},\omega)$, particularly its falloff with $|\omega|$, has been related to the decay of nonequal-time correlation functions and linear response relaxation times after perturbations \cite{PhysRevLett.111.050403}.  In \cite{Sorokhaibam:2022tgq} it is found that consistency with the second law of thermodynamics constrains the generic form of $f_{\mathcal{O}}(\bar{E},\omega)$ to monotonically increase with $S(\bar{E})$, in the sense that this leads to nondecreasing entropy when $\mathcal{O}$ is a perturbing operator.

Both the matrix $R_{ij}$ and $f_{\mathcal{O}}(\bar{E},\omega)$ are required to satisfy different conditions, depending on whether the Hamiltonian obeys time-reversal symmetry or not.  In the former case, a basis can be chosen such that the eigenstates of the Hamiltonian and the matrix elements of hermitian operators are real, and this basis is used by default.  $R_{ij}$ then represents a real random matrix with independent gaussian elements of mean $\overline{R_{ij}}=0$ and variance $\overline{R^2_{ij}}=1$, except the diagonal elements which instead have $\overline{R^2_{ij}}=2$. Here the overbar denotes an ensemble average; of course, for a specific operator and Hamiltonian, the $R_{ij}$ are not truly drawn from a random ensemble, but the elements appear uncorrelated and satisfy the stated statistics.  In the case of time-reversal symmetry, the condition on the real envelope function is $f_{\mathcal{O}}(\bar{E},\omega)=f_{\mathcal{O}}(\bar{E},-\omega)$.
If there is no time-reversal symmetry, the envelope function and the $R_{ij}$ may be complex, and the above conditions are replaced with $\overline{R_{ij}}=0$, $\overline{|R_{ij}|^2}=1$, and $f^*_{\mathcal{O}}(\bar{E},\omega)=f_{\mathcal{O}}(\bar{E},-\omega)$. 

Regardless of the precise form of the off-diagonal elements, their smallness is crucial.  One can show that, in the absence of level-spacing degeneracies\footnote{Note that we here refer to degeneracies of level \textit{spacing} rather than in the energy levels themselves.  This is a sort of genericity condition that excludes Hamiltonians with noninteracting factors, for instance.}, the average squared deviation of $\mathcal{O}_\Psi(t)$ from $\mathcal{O}_{\Psi}^*$ is bounded as
\begin{equation}\label{eq: small fluctuations}
\underset{t\rightarrow\infty}{\lim}\frac{1}{T}\int_{t=0}^{t=T} \left(\mathcal{O}(t)-\mathcal{O}^*\right)^2
=\sum_{i,j\ne i}|c_i|^2|c_j|^2 |\mathcal{O}_{ij}|^2
\le \underset{i\ne j}{\max}|\mathcal{O}_{ij}|^2.
\end{equation}
Thus when the off-diagonal elements are exponentially suppressed, $\mathcal{O}_\Psi(t)$ is close to $\mathcal{O}_\Psi^*$ \textit{at most times}. At specific times, the expectation value can still be far from equilibrium, but this requires a conspiratorial collaboration between the $c^*_i c_j$ and the $\mathcal{O}_{ij}$ such that their products sum coherently to overcome the exponential suppression of individual off-diagonal matrix elements $\mathcal{O}_{ij}$.  This can certainly occur, but it describes a fine-tuned scenario of a sort that will generically be destroyed by time evolution (i.e.\ through dephasing), corresponding to the intuition that an out-of-equilibrium state is ``atypical'' and equilibration is just a move toward typicality.  

It should be noted that although we have excluded exact degeneracies in the preceding discussion for simplicity, certain analogous statements still apply in the presence of degeneracies.  For instance, in the presence of an abelian algebra of conserved charges, the above discussion can be applied separately within each mutual eigensector of definite charge.  Alternatively, for any given state, one may choose a basis that aligns the entire support within each degenerate energy subspace to a single basis vector (such that all $c_i$ are zero except one per sector). Working in a reduced, nondegenerate basis by dropping the unsupported states, one sees that \eqref{eq: small fluctuations} still holds exactly.  
Regardless of whether exact degeneracies are present, the rate at which thermalization actually occurs depends on the details of the off-diagonal elements and the characteristics of the spacings of the energy levels.  See \cite{Volya_2020} for analysis of relaxation dynamics of observables in a variety of systems including a chaotic random matrix models and more realistic nuclear shell models.  A complete discussion of these factors is beyond the scope of this work, but some intuition regarding both is gained by considering a standard connection to random matrix theory, which we turn to in the next subsection.

Before proceeding we here pause to give a very brief and inevitably incomplete historical summary, highlighting a few seminal works that initiated research into ETH. The seeds of this approach can be traced back to Von Neumann's early work on quantum ergodicity \cite{RN9}, which considers systems satisfying criteria closely related to the modern formulation of ETH \cite{PhysRevLett.108.110601}.  
Many decades later, in 1985, Jensen and Shankar \cite{Jensen:1984gu} provided an early numerical demonstration of the emergence of thermal behavior in the eigenstates of a non-integrable spin chain, which is often cited as a precurser study to the modern ETH.  
In 1991, Deutsch \cite{Deutsch:1991msp} suggested a model of chaotic many-body systems as arising from a noninteracting Hamiltonian overlaid with random matrix interactions.  By averaging over a suitable class of interactions, he showed that eigenstate expectation values inherit a smooth energy dependence corresponding to microcanonical values and leading to thermalization.  
In a 1994 work \cite{Srednicki:1994mfb}, Srednicki invoked the Berry conjecture \cite{Berry_1977} \footnote{This states that in classically chaotic systems of many particles, high energy eigenfunctions look like Gaussian random superpositions of plane waves.  A related, but distinct, Berry-Tabor conjecture will be described in section \ref{subsec: connection to RMT and chaos}.} in hard-shell gas models to show that the eigenstates have single particle momentum distributions matching standard thermal equilibrium distributions.  In later work \cite{Srednicki:1995pt,Srednicki:1999bhx} he established what is now the standard form of the ETH ansatz that we employ in this work. 
A more or less contemporary line of research established similar phenomena in complex nuclear systems \cite{ZELEVINSKY199685}, including thermal behavior of eigenstates as seen through single-particle occupations and chaotic level-spacing statistics (which we also review below).  In the review \cite{Borgonovi:2016nrx}, the case is made that a recognition of thermalization via the properties of individual eigenstates can also be traced back to Landau and Lifshitz (see \cite{Borgonovi:2016nrx} for discussion and citations therein), leaving the primary task for modern research as being the identification of practical conditions for this occurrence. Although not phrased entirely in the language of ETH, this line of research emphasizes the chaotization of eigenstates via a split between mean field and residual Hamiltonian which is very similar in spirit to the one advocated by Deutch \cite{Deutsch:1991msp,Deutsch:2018ulr} as described above.

\subsection{Connection to random matrix theory and quantum chaos}
\label{subsec: connection to RMT and chaos}

One way to motivate the ETH form \eqref{eq: complete ETH} is via random matrix theory (RMT), under the recognition that if the eigenstates of the Hamiltonian look like random vectors\footnote{Here we mean that the state vector coefficients follow statistics as if they were independently Gaussian distributed variables with mean zero and variance as required by normalization.  See section 2 of \cite{DAlessio:2015qtq}.} in a basis that diagonalizes the operator in question, then in the energy basis the same operator will take the form
 \begin{equation}
		\mathcal{\hat{O}}_{ij}= \overline{O}\delta_{ij} + R_{ij}\sqrt{\frac{\overline{O^2}}{\mathcal{D}_T}}.
	\end{equation}
Here $\overline{O}$ and $\overline{O^2}$ are the mean of the operator $\hat{\mathcal{O}}$'s spectrum and squared spectrum, respectively. $R_{ij}$ are random complex matrix elements with mean 0 and variance 1 (except in the presence of time-reversal symmetry, in which case these are real elements and the diagonal elements have variance of 2), and $\mathcal{D}_T$ is the total Hilbert space dimension.  The ETH ansatz is essentially a refinement on this form, where the random matrix theory result applies within a microcanonical window.  

To further understand the connection between random matrix theory and thermalization, we here briefly summarize a few basic results from this field.  This also allows us to motivate the definition of quantum chaos employed in this work.  See \cite{Forrester:2003um} for a succinct historical overview of applications of random matrix theory in physics, along with relevant citations.  See \cite{Kota2014book}, for instance, for an introduction to the properties of common matrix ensembles, which we will now only briefly introduce.

Early physics applications of RMT due to Wigner \cite{wigner1951,wigner1955,wigner1957}, noted that the energy spectra of large atomic nuclei, though too complex to be modelled exactly, could be well-characterized statistically through comparison to random Hamiltonians, meaning Hamiltonians drawn from ensembles that do not enforce any particular structure outside a few simple symmetries.  The most commonly employed ensembles are the so-called classical gaussian ensembles, consisting of the Gaussian Unitary Ensemble (GUE), Gaussian Orthogonal Ensemble (GOE), and Gaussian Symplectic Ensemble (GSE).  Each of these consider ensembles of Hermitian matrices, here thought of as Hamiltonians, with components that are independent Gaussian variables of mean zero and variance one (except the diagonal elements, which are always real and have variance 2).  The three types represent a classification of systems' behavior under the time-reversal operation $T$ \cite{dyson2004I,dyson2004II,dyson2004III,dysonThreefoldWay}. The GUE case represents Hamiltonians that do not enforce time-reversal or any other particular symmetry, and it is a complex ensemble invariant under arbitrary unitary transformations.  The GOE applies to systems which do enforce time reversal symmetry and in which $T^2=1$. This ensemble is real and invariant under orthogonal transformations. The Gaussian Symplectic Ensemble corresponds to systems with a time-reversal invariance but where the $T^2=-1$.  These matrices are self-dual quaternion, with an invariance under symplectic transformations.  

A primary quantitative task of random matrix theory is to deduce the statistical properties of the spectra of Hamiltonians drawn from a given ensemble. One such characterization concerns the distribution of gaps between neighboring eigenvalues, or the ``level-spacing'' distribution.  The distributions associated with the classical Gaussian ensembles are known as the Wigner-Dyson distributions, and they are well approximated by the Wigner Surmise:
\begin{equation}\label{eq: wigner surmise}
    P_\beta(s):= A_\beta s^\beta \exp\left[-B_\beta s^2\right].
\end{equation}
Here $s$ is a normalized energy parameter indicating the size of the gaps, scaled such that the full distribution has mean level-spacing of one.  $P(s)ds$ represents the proportion of level-spacings in the range $[s,s+ds]$.  The $\beta$ parameter in equation \eqref{eq: wigner surmise} is an index that takes values $1, 2,$ and $4$ in the case of GOE, GUE, and GSE, respectively.  As it appears in $A_\beta$, and $B_\beta$, $\beta$ merely indexes the different constants that occur in the normalized distributions.  But the $\beta$ parameter is quantitatively important in that it distinguishes the decay rate of $P(s)$ as $s\rightarrow 0$ among the three cases.  The fact that for each value of $\beta$, $P(s)$ tends to zero at $s=0$ is known as level-repulsion, and it is a hallmark feature of chaotic systems.  It is to be contrasted with another frequently-occuring distribution which is associated with integral systems, the so-called Poisson distribution, which describes the case that the energy levels themselves are completely uncorrelated:
\begin{equation}\label{eq: poisson distribution}
    P_{\text{Poisson}}(s):= e^{-s}.
\end{equation}
The association of Poissonian statistics with integrability, and Wigner-Dyson statistics with chaos, came first from investigations that noticed these associations in systems with clear classical counterparts.  The Berry-Tabor conjecture \cite{BerryTabor} asserts that systems with an integrable classical counterpart exhibit Poissonian level-spacing statistics. The analogous BGS conjecture of Bohigas, Giannoni, and Schmit \cite{Bohigas:1983er} posits that systems with a chaotic classical counterpart exhibit level-spacing statistics associated with random matrix theory, i.e.\ the Wigner-Dyson distributions.  Although pathological or non-generic exceptions may occur, the overwhelmingly success of these conjectures has led to the use of Wigner-Dyson level-spacing statistics as a \textit{definition} of quantum chaos, even for quantum systems far from or without any clear classical limit.\footnote{This is certainly not the only definition of quantum chaos in the literature, but we employ it in this work. A very pedagogical summary of this material can be found in \cite{Gubin2012}, along with demonstrations in spin 1/2 spin chain systems.}

Aside from the empirical connection noted in the Berry-Tabor and BGS conjectures, the use of level-spacing statistics to define quantum chaos seems to bear little resemblance to classical definitions, which are usually based on the rapid divergence of nearby phase space trajectories.  Qualitative parallels can be drawn, however. Level-repulsion is an avoidance of degeneracies, which can be thought of as an avoidance of conserved (or approximately conserved) quantities. This is commensurate with the classical dichotomy between chaotic and integrable systems.  The same level-repulsion helps ensure that chaotic systems undergo rapid and uniform dephasing, as is crucial for thermalization.  More rigorously, recent work \cite{Vikram:2022xiq} has connected spectral statistics to the dynamical criteria of quantum cyclic ergodicity and aperiodicity (notions which generalize classical ergodicity on a discretized phase space to arbitrary quantum systems), finding that systems with Wigner-Dyson statistics satisfy both criteria.  Of course, the occurence of a Wigner-Dyson distribution in nearest-neighbor eigenvalue spacings is far from a complete characterization of a system's spectrum; it occurs in systems with much more structure than the Gaussian random ensembles, which represent Hamiltonians that are in some sense ``as chaotic as possible''.  Further characterizations of the spectrum can be found in quantities such as the spectral form factor \cite{Mehta1967book, Brezin:1997rze}, which is essentially the Fourier transform of the two point correlation function of the eigenvalue distribution, and has been studied in a wide variety of systems (for example, see \cite{Cotler:2016fpe,Gharibyan:2018jrp,Friedman:2019gyi}). The extent to which any given characterisation of chaos, such as the onset of Wigner-Dyson level-spacings employed in this work, guarantees other behaviors traditionally associated with chaotic systems, and appearance of the ETH form of simple operators, is a question that we will probe in the remainder of this work.

\subsection{Generic ETH}
\label{subsec: generic ETH in theory}
Let us return to the ETH prediction for an operator's equilibrium value, given by the infinite-time average of the expectation value, $\mathcal{O}^*$.  Employing the ETH ansatz, this can be written
\begin{equation}\begin{split}\label{eq: eth prediction}
    \mathcal{O}^*_{\Psi}&:=\sum_i |c_i|^2 \mathcal{O}_{ii}
    =\sum_i |c_i|^2\left(\mathcal{O}(E_i) +  R_{ii}e^{-S(E_i)/2}f_{\mathcal{O}}(E_i,0)\right)\\
\end{split}\end{equation}
Assuming no special correlation between the $c_i$ and $R_{ii}$ terms, the second term is a sum of exponentially small terms of mean zero, so it represents a small correction.  The first term in \eqref{eq: eth prediction} can be rewritten by expanding the smooth function $\mathcal{O}$ around $E_\Psi:=\bra{\Psi}\hat{H}\ket{\Psi}$, leading to
\begin{equation}\begin{split}\label{eq: moment expansion}
    \mathcal{O}^*_{\Psi}\approx \mathcal{O}(E_\Psi) + \sum_{n=2}^\infty\frac{1}{n!}\delta^{(n)}E_\Psi \frac{\dd^n\mathcal{O}(E_\Psi)}{\dd E^n}.
\end{split}\end{equation}
Here $\delta^{(n)}E_\Psi:=\sum_i |c_i|^2(E_i-E_\Psi)^n$ denotes the $n$th central moment of the energy distribution in state $\Psi$.  The first central moment vanishes by definition, having expanded around the expectation value $E_\Psi$.  This expression entails that if we choose another state with the same energy expectation value, we can express the difference between this operator's equilibrium values in the two states as
\begin{equation}\begin{split}\label{eq: difference in moments}
    \mathcal{O}^*_{\Psi_1}-\mathcal{O}^*_{\Psi_2}\approx \sum_{n=2}^\infty\frac{1}{n!}\left(\delta^{(n)}E_{\Psi_1}-\delta^{(n)}E_{\Psi_2}\right) \frac{\dd^n\mathcal{O}(E_\Psi)}{\dd E^n}.
\end{split}\end{equation}
Likewise, if we replace $\Psi_2$ with a mixed state, the difference will again be expressed in terms of the differences of the central moments of the energy distributions. The prototypical application of ETH is to choose $\Psi_1$ to be supported only in a microcanonical window and replace $\Psi_2$ with the microcanonical ensemble itself at the same energy.  Then the right hand side of \eqref{eq: difference in moments} is small because all the moments are individually small.  More generally though, the terms in \eqref{eq: difference in moments}  can be small for any of several reasons.  Even if the moments are not small their differences might be: for an arbitrary $\Psi$, this entails that increasingly accurate static ensembles for ETH observables can be attained by matching ever higher moments of the initial state's energy distribution. Such distributions may be sought in accordance with Jaynes' principle \cite{PhysRev.106.620,PhysRev.108.171}. For instance, for some positive integer $N<|\mathcal{H}|$ we may define $\rho_N$ to be the normalized mixed state that maximizes the von Neumann entropy $S(\rho_N):=-\Tr(\rho_N\log(\rho_N))$ subject to constraints $\Tr(\rho_N (\hat{H}-E_\Psi)^n) = \delta^{(n)}E_\Psi$ for $n=0$ through $N$.  The $N=1$ case gives the usual Gibbs state, while for higher $N$ this is a nontrivial optimization problem, including questions of existence and uniqueness \cite{Mead1984}. 
As a limiting case, for $N=|\mathcal{H}|$ one must obtain the diagonal ensemble which matches \textit{all} energy moments of the state.\footnote{Recall that the diagonal ensemble \cite{PhysRevA.78.013626} simply retains \textit{all} the initial state information on the diagonal of $\rho_\Psi=\ket{\Psi}\bra{\Psi}$ in the energy basis.}  In systems where the microcanonical and canonical ensembles become equivalent in the thermodynamic limit, one anticipates that states which thermalize are those which likewise have subextensive fluctuations.  In finite systems, however, Gibbs state fluctuations can be appreciable, and states with similarly large energy fluctuations are best described as thermalizing to the predictions of a Gibbs state. Indeed we will see examples of this in the small lattice systems studied in this work.  When the equilibrium predictions of these two ensembles are approximately equal despite the latter having substantial fluctuations, it points to a third reason that \eqref{eq: difference in moments} can be small, which is simply that the second and higher derivatives of the function $\mathcal{O}(E)$ are themselves small over the scale of the fluctuations.  Such observables have a much broader insensitivity to the details of the initial state than occurs just in a microcanonical window.   That this is does in fact occur for many local variables $\hat{\mathcal{O}}$ is also natural: since $\mathcal{O}(E)$ can be thought of as the microcanonical expectation value of $\hat{\mathcal{O}}$ as a function of energy, this is unsurprisingly often a smooth and slowly varying function for physically relevant observables.

We emphasize the multiplicity of scenarios under which equation \eqref{eq: difference in moments} can be made small to emphasize several directions in which the ETH can be pushed beyond its the prototypical application to microcanonical states, by implying a predictive utility for non-standard equilibrium distributions, and implying the thermalization of many observables in a broader class of initial states.  We refer to these behaviors collectively as `generic ETH,' and illustrate a few such instances in the lattice systems in the next few sections.  A similar term, `generalized eigenstate thermalization' \cite{cassidy_2011, He_Santos_Wright_Rigol_2013, Rigol:2016itf} is used to describe thermalization in the presence of conserved charge, positing that an analogous mechanism to standard ETH can occur in such systems.  This encompasses the statement that within individual charge sectors, standard ETH may apply, and further posits that states with (generalized) microcanonical support in both an energy and charge window will probe a smooth, now multivariate function on a generalized ``diagonal'' of matrix elements.   A multivariate analogue to the expansion in equation \eqref{eq: moment expansion} then implies analogous statements of generic ETH which may be brought to bear also in systems where conserved charges are present.  Thus, both generalized ETH and generic ETH can be in play simultaneously, as we will see in section \ref{subsec: The qutrit Hamiltonian}.

Let us briefly discuss the effect of the fluctuations of diagonal terms of the ETH ansatz. When the ETH fails, for instance if nonvanishing fluctuations remain in thermodynamic limit for diagonal elements of local operators, this has been directly linked to the presence of (quasi)local integrals of motion in the system \cite{PhysRevLett.124.040603}. On the other hand, if the ETH ansatz applies, the affect of these terms on the equilibrium prediction in a finite system is captured in the second term in equation \eqref{eq: eth prediction}.   If the $c_i$ and $R_{ii}$ are uncorrelated, the sum vanishes on average, while its variance is described by $\overline{\sum_i |c_i|^4 |R_{ii}|^2 |f(E_i,0)|^2 e^{-S(E_i)}}$. The overbar denotes an average over instantiations of $c_i$, thus to define it rigorously would require specifying some ensemble of states to consider.  Of course, the entropic suppression serves to keep the sum small.  Likewise, if the $c_i$ express a random vector across an energy window with $N_\Psi$ states, with an average squared magnitude of $1/N_\Psi$, then increasing $N_\Psi$ can also serve to reduce these fluctuations as $1/N_\Psi$.  This happens in any fixed microcanonical window as the density of states increases with system size, but also by broadening the support outside a microcanonical window at fixed system size.  If the support moves closer to the edge of the spectrum (where the density of states drops), the entropic suppression is less effective, but the effect of including more states may still be suppressive.  This will be evident in the thermalization of broadly-supported (non-microcanonical) states shown in later examples, e.g. in figure \ref{fig: charge pauli x}.

In the remainder of the paper, we turn to illustrations of all the concepts described in this section in lattice systems.  First, in section \ref{sec: qubit spinchain models} we will introduce a 1-d qubit spin chain system with tunable parameters that allow us to see the transition between Poissonian and Wigner-Dyson level spacing statistics.   With this system we illustrate the connection between ETH form of simple operators and the thermalization in a wide class of states.  In section \ref{sec: Thermalization with conserved charge}, we then turn to the case of thermalization in the presence of conserved charge, designing a qutrit Hamiltonian that allows us to illustrate both generalized ETH and expand on what we have described in this section as generic ETH.

\section{Qubit spinchain models}
\label{sec: qubit spinchain models}

Having discussed the underpinnings of ETH and its connection to chaos, we now present a study of 1-d qubit spinchain models to demonstrate the connection between chaos, ETH, and thermalization. This section provides pedagogical illustrations of the concepts of the preceding section\footnote{Plots in this section are created using Mathematica \cite{Mathematica}.}, and serves as a reference for model we will describe in the next section.  Section \ref{subsec: generic ETH in qubit system} also illustrates our use of the term `generic ETH' for the qubit system.

\subsection{Qubit Hamiltonian}
\label{subsec: qubit hamiltonian}
We begin by considering a 1-d Ising model of $L$ qubits in the presence of transverse and parallel magnetic fields.  This Hamiltonian has been frequently used in the study of chaos and thermalization.\footnote{The particular choice of coefficients that optimizes for chaotic behavior goes back to \cite{Banuls:2011vuw}, but the same Hamiltonian has been employed in numerous studies of chaotic scrambling and information spreading, for example \cite{Shenker:2013pqa,  Hosur:2015ylk, Couch:2019zni, Eccles:2021zum}.} 
\begin{equation}\begin{split}\label{eq: qubit hamiltonian}
		H^{\text{qubit}} &= \left(J\sum_{r=1}^{L-1}\sigma_{z}^{(r)} \sigma_{z}^{(r+1)}
		+h_x\sum_{r=1}^{L}\sigma_{x}^{(r)}  +h_z\sum_{r=1}^{L}\sigma_{z}^{(r)}\right)L^{-1}.\\
\end{split}\end{equation}
Here $\sigma^{(r)}_{i}$  for $i \in \{x,y,z\}$ denotes a local Pauli operator at the $r$'th site of an $L$ qubit spin chain. We choose to include an overall factor of $1/L$ in the Hamiltonian because it will keeps the total spectral range roughly of order 1 as we change the system size, which will be helpful for some later comparisons.  We will always set the two-body interaction coefficient $J$ to 1 (arbitrary units).  The coefficients $h_x$ and $h_z$ specify the strength and orientation of the external magnetic field.  For these we will usually employ $h_x=1.05$ and $h_z=0.5$, which is known to bring the Hamiltonian far from integrability \cite{Banuls:2011vuw}.  In cases where we employ other coefficients below, it will be indicated explicitly.

\subsection{Parity operator $P$: isolating symmetry sectors}
\label{subsec: parity operator P}
\begin{figure}[H]
	\centering
	\begin{subfigure}{.3\textwidth}
		\centering
		\includegraphics[width=\linewidth]{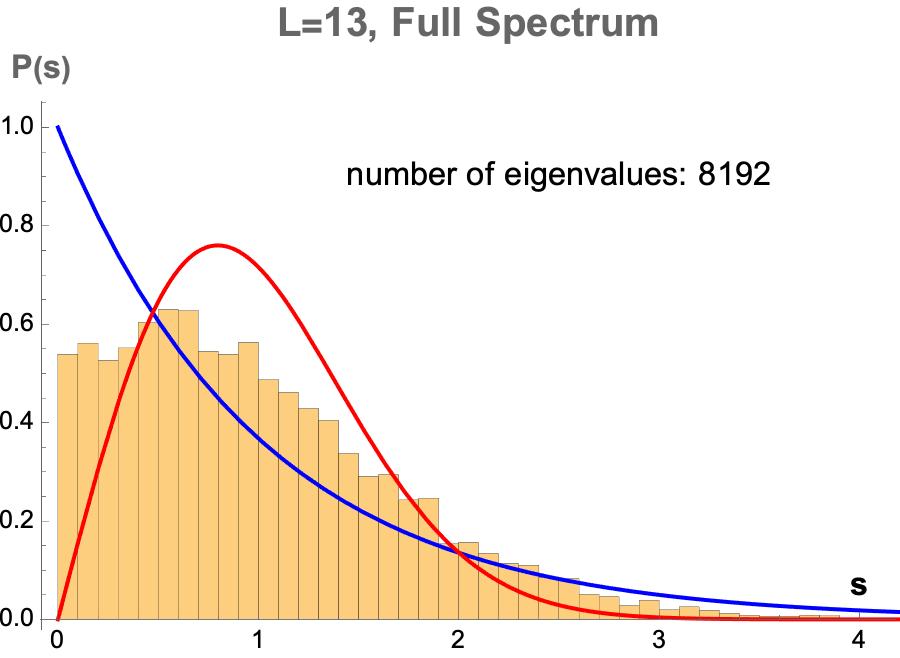}
		\caption{Full}
		\label{fig:parity both}
	\end{subfigure}%
	\begin{subfigure}{.3\textwidth}
		\centering
		\includegraphics[width=\linewidth]{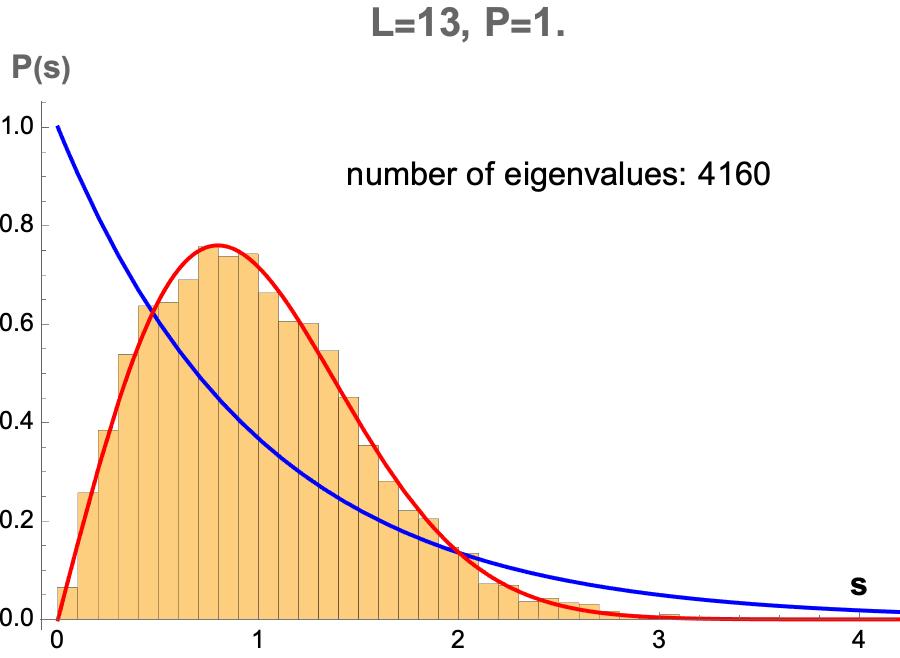}
		\caption{$P=1$}
		\label{fig:parity plus}
	\end{subfigure}%
	\begin{subfigure}{.3\textwidth}
		\centering
		\includegraphics[width=\linewidth]{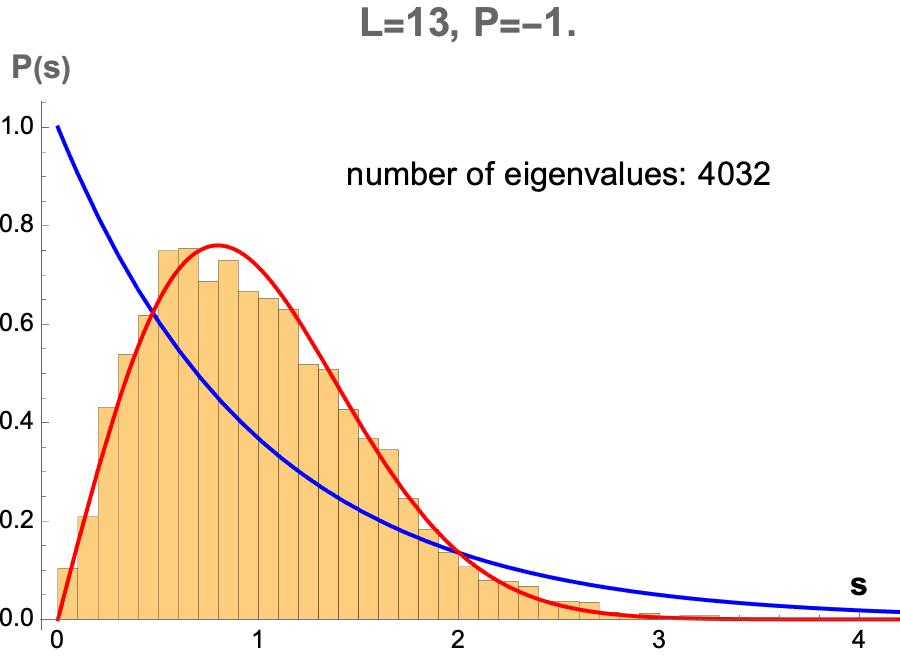}
		\caption{$P=-1$}
		\label{fig:parity minus}
	\end{subfigure}
	\caption{The appearance of a Wigner-Dyson distribution (red line) as opposed to a Poisson distribution (blue line) in level spacing histograms is our chief diagnostic of chaos.  It is only achieved within individual charge sectors, as illustrated here with the conserved parity operator $P$.  Hamiltonian \eqref{eq: qubit hamiltonian} has a symmetry under the inversion of qubit sites enacted by $P$: $r\leftrightarrow L- (r-1)$,  $r\in {1,...,L}$.  Failure to work in a subspace of definite parity = $\pm1$ results in an intermediate distribution, as in figure \ref{fig:parity both}.}
	\label{fig: parity sectors}
\end{figure}

Since level-spacing distributions play a prominent role in this work, we here emphasize a few related points.  Firstly, in order to consider these distributions as a diagnostic of chaos, we must first isolate the spectra within individual symmetry sectors.\footnote{Also note that all level-spacing distributions reported in this work are obtained after first processing the spectra via an unfolding procedure that sets the mean level spacing is 1 and ensures a uniform density of states.  We follow the same unfolding procedure described in \cite{Santos:2010iji}.  Some arbitrariness is introduced by the choice of unfolding procedure which may affect some spectral properties, but level-spacing distributions are largely insensitive to these choices.}  To illustrate this fact, note that for \textit{any} choice of coefficients, the Hamiltonian \eqref{eq: qubit hamiltonian} has at least one obvious symmetry:  the parity operation, $P$, which swaps the order of qubits by sending site $i$ to $L+1-i$. \footnote{If the spin chain were instead placed on a ring by including the coupling term $J \sigma_z^{(1)}\sigma_z^{(L)}$ there would be additional shift symmetries (discrete translation invariance) generated by the smallest shift of sending all sites $i$ to $i+1$ mod $L$.}    The $P$ operator has eigenvalues $\pm 1$, and correspondingly divides the Hamiltonian eigenstates into those which are even under this inversion, and those which are odd.  Figure \ref{fig: parity sectors} illustrates three level spacing distributions associated with the Hamiltonian \eqref{eq: qubit hamiltonian} with the standard coefficients:  the full spectrum, the $P=1$ sector, and the $P=-1$ sectors.  Only within individual sectors do the level-spacing distributions match the Wigner-Dyson form.  Failing to segregate these sectors results in a distribution that is intermediate between Poisson and Wigner-Dyson.  More generally, if a Hamiltonian is decomposable as a direct sum of multiple chaotic sectors, overlapping these tends to wash out the Wigner-Dyson distribution, such that many such overlapping sectors result in an approximate Poisson distribution.   This implies that, given a general spectrum with a Poisson distribution, it may be possible to decompose it into many chaotic subsectors.  We are not aware of any general algorithm for doing so, though see \cite{fremling2022} for some related analytical results.

One of the goals of this work is to explore the interplay between the presence of conserved charges and the thermalization process in chaotic systems. It is worth noting at this point that the presence of the conserved parity operator $P$ in this qubit Hamiltonian is insufficient for our purposes for several reasons.  Firstly, this operator represents a global charge, with no corresponding quasi-local operator representing the ``amount'' of parity present in local subsystems.  This is not a problem per se, except that we wish to investigate the latter scenario.  Secondly, the sizes of the corresponding state spaces for $P=1$ and $P=-1$ are approximately equal (both scaling as $2^{L-1}$ for large $L$), while we are interested in examining the ability of differing state space volumes to affect the thermalization across subsectors.  Lastly, it is not possible to build completely unentangled states supported only in the negative parity sector.  This is an interesting asymmetry between the sectors, but it means that we cannot compare the thermalization behavior of the two sectors in terms of the growth and saturation of subsystem entanglement entropy, which is one of our most useful diagnostics.  We will construct an alternative Hamiltonian in section \ref{subsec: The qutrit Hamiltonian} that includes a conserved charge that contrasts these features of the parity operator.  In particular, it will include a notion of local charge (though we will explicitly break local charge conservation and conserve only the total). There will also be considerable imbalance in the state space volumes of various charge subsectors, and the subsystem entanglement entropy will be tunable to a large extent independently of this charge.

\subsection{From integrability to chaos (and thermalization)}
\label{subsec: from integrability to chaos}

\begin{figure}
	\centering
	\begin{subfigure}{.3\textwidth}
		\centering
		\includegraphics[width=\linewidth]{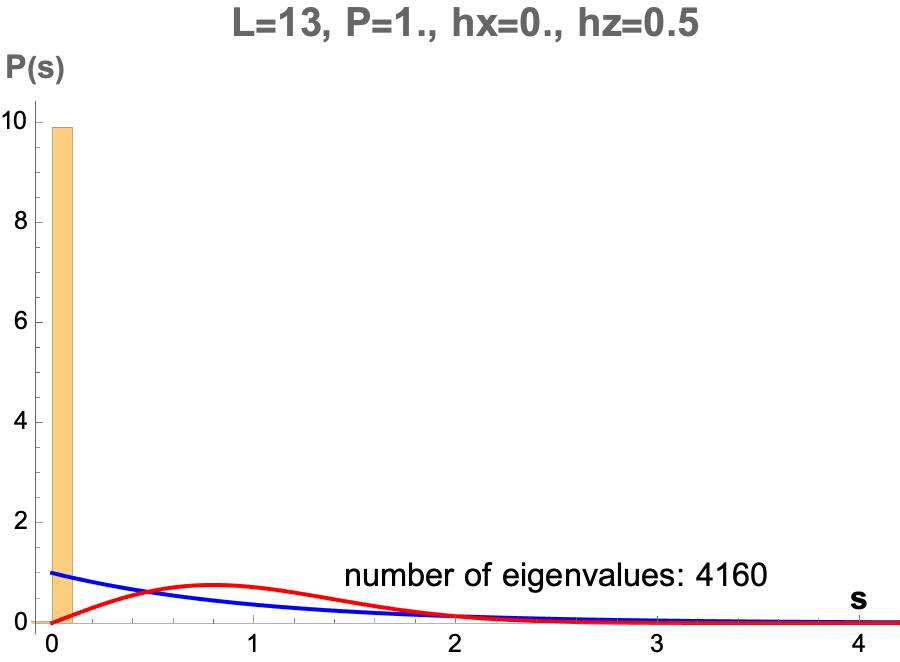}
		\label{fig:thermalization failure 1 hist}
	\end{subfigure}%
	\begin{subfigure}{.3\textwidth}
		\centering
		\includegraphics[width=\linewidth]{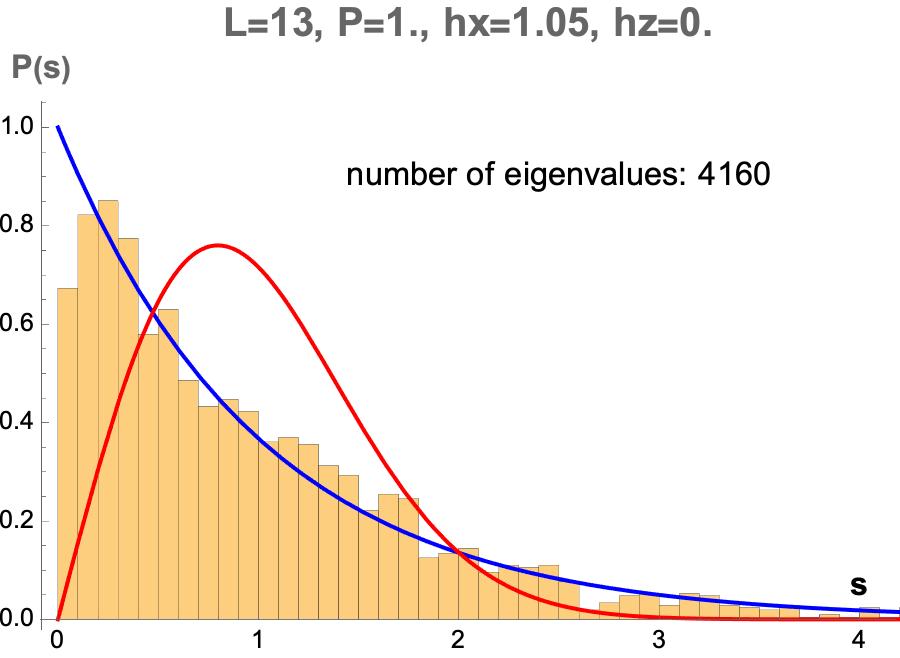}
		\label{fig:thermalization failure 2 hist}
	\end{subfigure}%
	\begin{subfigure}{.3\textwidth}
		\centering
		\includegraphics[width=\linewidth]{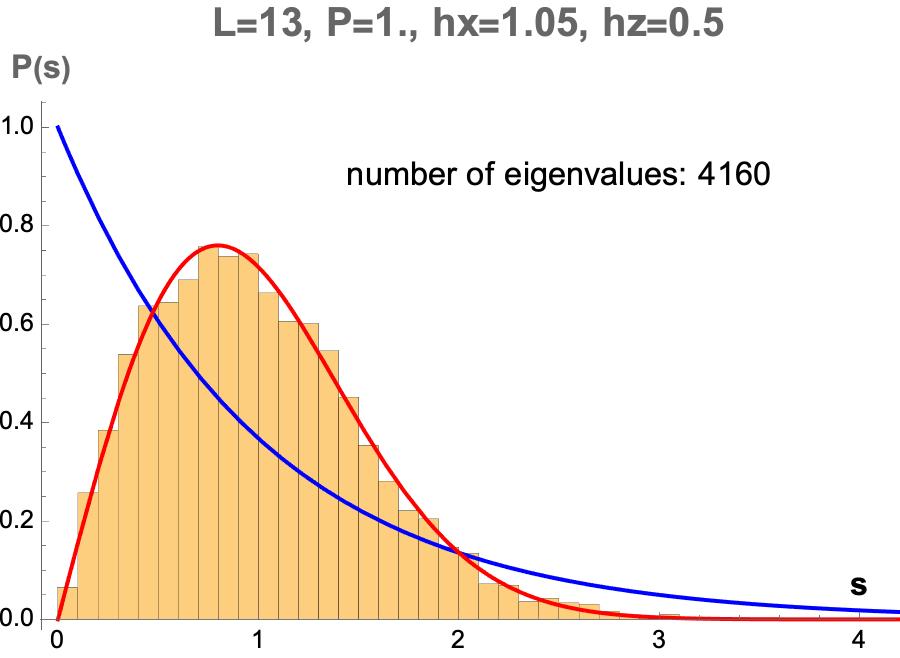}
		\label{fig:thermalization success hist}
	\end{subfigure}
	\begin{subfigure}{.3\textwidth}
		\centering
		\includegraphics[width=\linewidth]{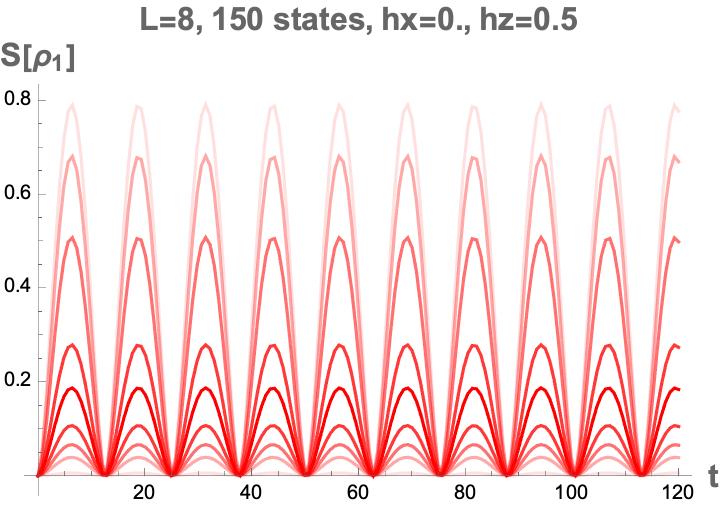}
		\caption{$h_x = 0, h_z = 0.5$}
		\label{fig:thermalization failure 1}
	\end{subfigure}%
	\begin{subfigure}{.3\textwidth}
		\centering
		\includegraphics[width=\linewidth]{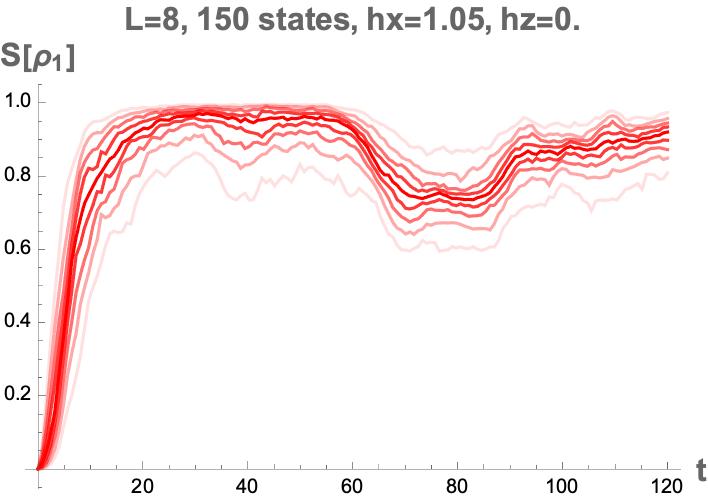}
		\caption{$h_x = 1.05,  h_z = 0$}
		\label{fig:thermalization failure 2}
	\end{subfigure}%
	\begin{subfigure}{.3\textwidth}
		\centering
		\includegraphics[width=\linewidth]{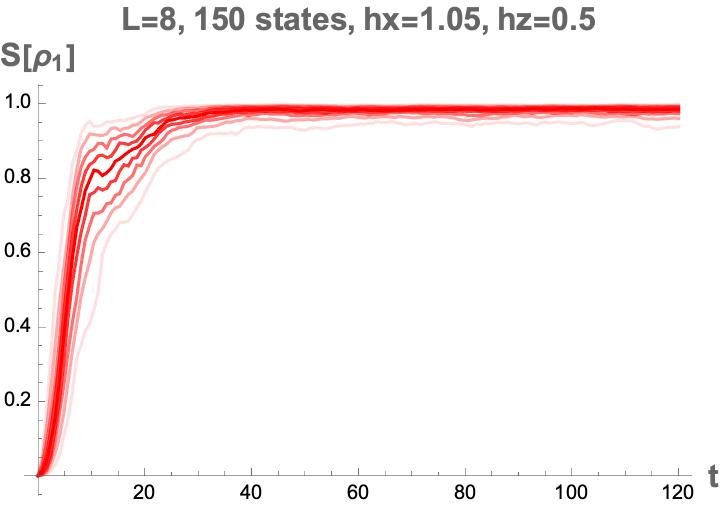}
		\caption{$h_x = 1.05, h_z = 0.5$}
		\label{fig:thermalization success}
	\end{subfigure}
	\caption{Three different system coefficients are compared, one per column, illustrating the association between chaotic spectral statistics (top row) and thermalization as seen the behavior of single qubit entanglement entropy (bottom row).  The leftmost column uses the standard coefficients except that $h_x$ has been set to zero, the central column instead sets $h_z$ to zero, and the rightmost column uses the standard coefficients ($J=1, h_x=1.05, h_z=0.5$).  Each top row figure shows the nearest neighbor level-spacing distributions in the $P=1$ sector.  To create the plots on the bottom row, 150 random unentangled states were initialized and time-evolved through exact diagonalization of the corresponding Hamiltonian. Von Neumann entropy of the first qubit was computed as a function of time. The red lines correspond to quantiles ($10\%$  through $90\%$ in increments of $10\%$) of the entanglement entropy distributions at each time step.}
	\label{fig: thermalization onset}
\end{figure}

Under different choices of coefficients, the Hamiltonian \eqref{eq: qubit hamiltonian} has wide ranging phenomenology.   Figure \ref{fig: thermalization onset} considers three different choices of coefficients, juxtaposing the associated level-spacing distributions and a time-dependent illustration of thermalization based on entanglement growth of the first qubit reduced state.  We now discuss the contents of this figure in detail.

Each column of figure \ref{fig: thermalization onset} corresponds to a different choices of Hamiltonian coefficients.  
For each choice, the top figure shows a normalized level-spacing histogram for the positive-parity sector of the system. Two ideal probability distributions for integrable and chaotic Hamiltonians are overlaid for comparison. The blue line is a Poisson distribution $p(s)=e^{-s}$, which encodes a simple pattern of exponentially decreasing likelihood for larger energy displacements between neighboring eigenvalues, which corresponds to integrable systems.  The red line we shall refer to as the Wigner-Dyson distribution $p(s)=\frac{\pi}{2}s e^{-\frac{\pi}{4}s^2}$  (in this case it is that of the Gaussian orthogonal ensemble, see section \ref{subsec: connection to RMT and chaos}).  A good fit to this distribution is a hallmark feature of a chaotic system.  The bottom row of figure \ref{fig: thermalization onset} shows the dynamics of the entanglement entropy of the first qubit under the three different Hamiltonians, starting from randomly-generated unentangled initial states.\footnote{\label{footnote: random unentangled}Each state is prepared as an unentangled tensor product of single qubit states $\ket{\psi}_i=\cos(\theta_i/2)\ket{0}+\sin(\theta_i/2)e^{i\phi_i}\ket{1}$ for $i\in 1,...,L$ with each $\theta_i \in [0,\pi], \phi_i \in [0,2\pi]$ randomly selected from a flat distribution on their respective ranges.} For each plot, 150 unentangled initial states were time-evolved through exact diagonalization and the entanglement entopies of the first qubit binned at each time step.   Each red line then indicates a different 10\% quantile (e.g. the central, darkest line represents the value at which 50\% of the qubits have lower entanglement entropy).

Consider first the leftmost column, where $h_x=0$.  This Hamiltonian has a highly degenerate spectrum, indicating many conserved quantities, both local and nonlocal (for instance, not just the total z-spin $\sum_{r=1}^{L}\sigma_z^{(r)}$ but also each individual spin $\sigma_z^{r}$ is conserved).  The ``distribution'' of level spacings is thus hardly a distribution at all and thermalization badly fails, as is evident in the entanglement entropy dynamics which undergo continual oscillations and do not equilibrate at all.   In the central column, the system instead has $h_z$ set to zero.  The same quantities are no longer conserved, and the spectrum is now closer to a Poisson distribution than Wigner-Dyson (though somewhat intermediate). The entanglement dynamics indicates a much narrower spread, quickly proceeding to a near-maximal value, but still undergoing large fluctuations (a pattern which continues if the time axis is extended).    The final column uses both coefficients nonzero.  The level-spacing statistics exhibit a clean Wigner-Dyson distribution, and the entanglement entropy promptly saturates its thermal value (in this case near maximal, see next section) with small fluctuations.  This illustrates the prototypical association between chaos and thermalization which we will further explore in the following subsections.

\subsection{Thermalization in more detail}
\label{subsec: thermalization in more detail}

In the previous section we illustrated the link between chaos and thermalization, as seen through Wignor-Dyson level-spacings and entanglement saturation, respectively.   The fact that this latter behavior is ``thermal'' could be almost taken for granted, but as emphasized in section \ref{sec: Background, thermalization and ETH}, there are many possible criteria of thermalization, and each should involve a comparison of observables (or any time-dependent properties) to the predictions of a target thermal distribution.  In this section we continue to use the entanglement entropy of the first qubit as a diagnostic of thermalization, but now we consider a wider array of initial states and a more careful comparison to thermal predictions over a corresponding range of thermal parameters.\footnote{In fact \cite{Banuls:2011vuw} studied the same Hamiltonian and found a rich phenomenology, with different degrees of thermalization achieved for specific initial states.  Our results do not contradict these, but we use both different classes of initial states and different diagnostics of thermalization.}

For the system considered here with the standard/chaotic coefficients, the Hamiltonian itself is the only quasi-local conserved charge, so we expect the energy (or temperature) to be the only parameter in the relevant thermal ensemble.  In order to test thermalization across a range of energies/temperatures, we employ random unentangled states with variable energy expectation values.   By randomly preparing unentangled states (see footnote \ref{footnote: random unentangled}), the energy of the resulting states tends to be close to zero, the center of the spectrum.  We can, however, choose to keep only those states that fall near some other target energy $E:=\bra{\Psi}\hat{H}\ket{\Psi}$.  A corresponding Gibbs ensemble
\begin{equation}\label{eq: Gibbs}
	\hat{\rho}^{(\text{Gibbs})}_{\beta(E)}:= e^{-\beta \hat{H}}/\tr(e^{-\beta \hat{H}}),
\end{equation}
where $\beta(E)$ indicates that the inverse temperature $\beta$ has been chosen such that $\langle \hat{H} \rangle_\beta=E$
then provides the relevant thermal prediction for such a state.  The choice to use a Gibbs ensemble will be discussed more fully in section \ref{subsec: generic ETH in qubit system}.

\begin{figure}[H]
	\centering
	\begin{subfigure}{.5\textwidth}
		\centering
		\includegraphics[width=\linewidth]{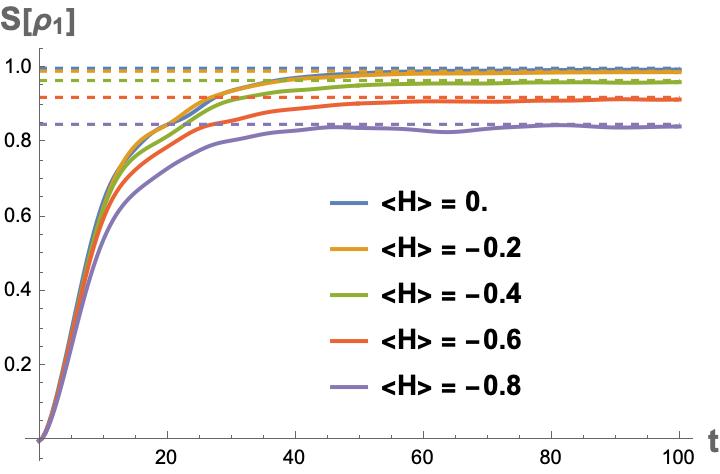}
		\caption{Entropy averages, $\langle H\rangle<0$.}
		\label{fig:entropy growth H<0}
	\end{subfigure}%
	\begin{subfigure}{.5\textwidth}
		\centering
		\includegraphics[width=\linewidth]{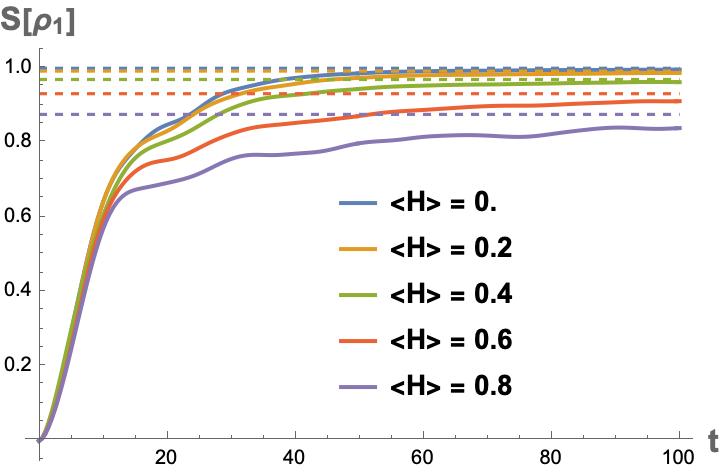}
		\caption{Entropy averages, $\langle H\rangle>0$.}
		\label{fig:entropy growth H>0}
	\end{subfigure}\\%
	\begin{subfigure}{.5\textwidth}
		\centering
		\includegraphics[width=\linewidth]{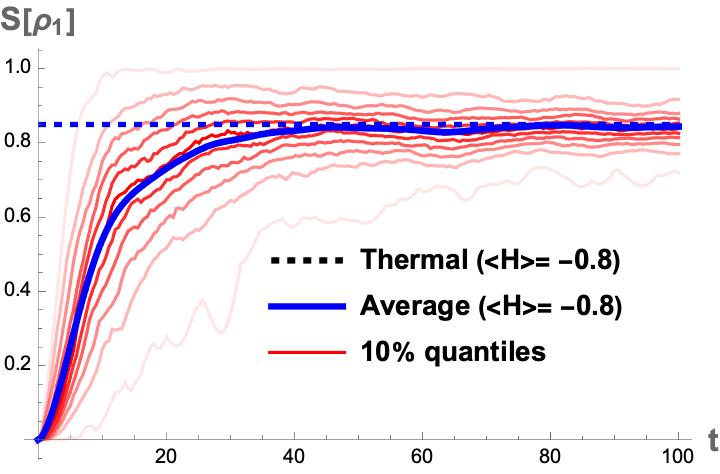}
		\caption{Entropy spread shown in quantiles.}
		\label{fig:entropy growth quantiles}
	\end{subfigure}%
		\begin{subfigure}{.5\textwidth}
		\centering
		\includegraphics[width=\linewidth]{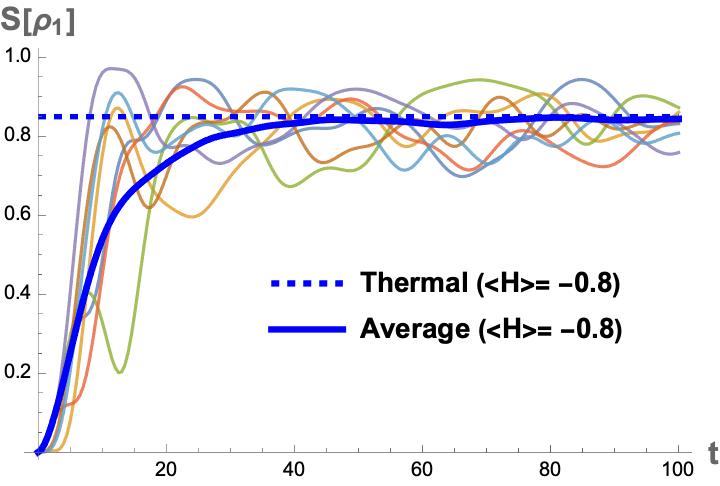}
		\caption{Samples of individual entropy curves.}
		\label{fig:entropy growth sample curves}
	\end{subfigure}%
	\caption{Top Panels:  Thermalization is illustrated in a system of size $L=10$ using the entanglement growth on qubit 1 from initially-unentangled states. Each solid line represents the average (at each time step) over 300 randomly generated unentangled states satisfying a constraint on $\langle \hat{H} \rangle$.  The dashed lines show the corresponding Gibbs state thermal prediction.  The left/right panels consider energies below/above zero, respectively.
	Bottom Panels:  Using the case of $\langle \hat{H} \rangle = -0.8$, The spread around the average value is illustrated.  The left panel shows 10\% quantiles from 300 entropy curves as a function of time, while the right panel shows a sample of individual entropy curves.}
	\label{fig: qubit entropy thermalization curves}
\end{figure}

In figure \ref{fig: qubit entropy thermalization curves}, we display the entanglement dynamics for initially-unentangled states at various energies in comparison to the corresponding thermal values.  To avoid overlap, the top left panel shows sample curves with $\langle H \rangle < 0$ while the top right panel shows curves with $\langle H\rangle > 0$.  Each solid line represents the average entanglement entropy of 300 initially-unentangled states at the target energy ($\pm0.001$).  Dashed lines of matching color indicate the corresponding thermal predictions.  The match between saturation values and thermal predictions is evident, thought note that these lines represent an average over many states.  The fluctuations in individual entropy curves are still appreciable, particularly for energies further from the center of the spectrum.  This is illustrated in the bottom two panels, which focus on the case of $\langle \hat{H}\rangle=-0.8$.  In the bottom left panel, red curves give 10\% quantiles in the entropy distribution from the 300 states in the same manner as figure \ref{fig: thermalization onset}.  The bottom right panel overlays a random selection of 6 individual state entropy curves.  These panels illustrate that fluctuations are still appreciable for this system size ($L=10$), and the smooth behavior emerges only in the average.  These fluctuations also have the effect of pulling the average down slightly below the thermal expectation.  

\begin{figure}[H]
	\centering
	\begin{subfigure}{.45\textwidth}
		\centering
		\includegraphics[width=\linewidth]{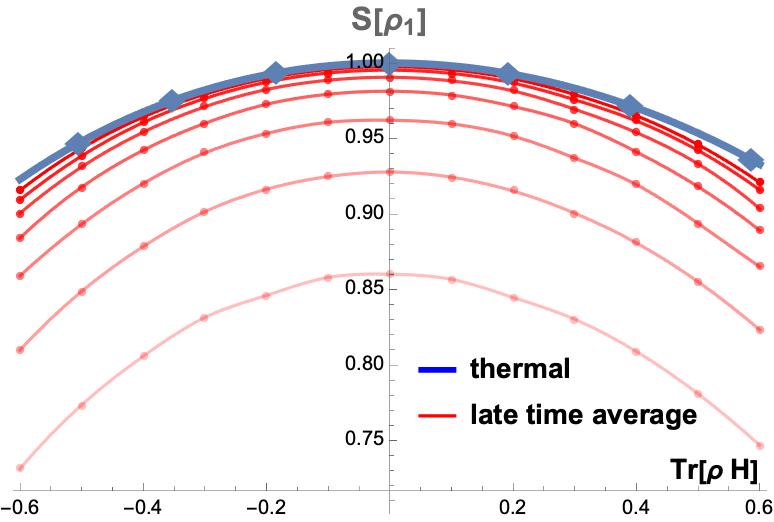}
		\label{fig:late time entropy saturation values}
	\end{subfigure}%
	\begin{subfigure}{.55\textwidth}
		\centering
		\includegraphics[width=\linewidth]{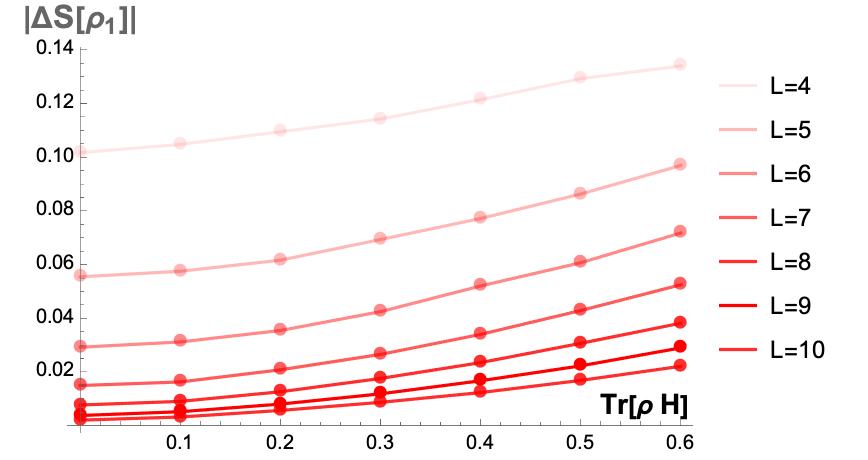}
		\label{fig:matrix element ratios}
	\end{subfigure}
	\caption{Illustrations of the improving thermalization with increasing system size $L$. In the left panel, the blue curve is a thermal expectation value for the first qubit entanglement entropy in a system of size $L=12$, which we take as representative of larger systems.  In both plots, red datapoints of increasing opacity represent data for increasing system size, from $L=4$ to $L=10$, interpolated for clarity.  In the left (right) panel, each point corresponds to the average (standard deviation) of the first qubit entanglement entropy, sampled over a large late time window ($\Delta t \sim 1000$). The resulting average (standard deviation) was then averaged over 300 initially-unentangled states at the corresponding energy expectation value.}
 \label{fig: improving thermalization with L}
\end{figure}

Thermalization is expected to improve, both in the sense of better average fit to the thermal values and smaller fluctuations, as the system size increases.  This is illustrated in figure \ref{fig: improving thermalization with L}.   In the left panel, we compare the thermal predictions from a system of size $L=12$ to the late time average obtained from initially-unentangled states of increasing system size (in this case, the averaging is over both a set of 300 states and also a late-time window $\Delta t \sim 1500$).  The thermal prediction (blue curve) differs negligibly if we use a system size $L=10$ to $L=12$, which we take to indicate that the $L \gtrsim 10$ curves already well-represent the thermal entropy expectation values for larger systems.  The red data points show late time averages for unentangled states across a range of energies, for system sizes $L=4$ through $L=10$.  With increasing size the late time averages approach the $L=12$ thermal prediction, though notable deviations persist far from the center of the spectrum.  The right panel of figure \ref{fig: improving thermalization with L} shows the average size of temporal fluctuations, which are seen to decrease with system size, but in all cases are larger for energies farther from the spectral average of zero.  Together, figures \ref{fig: qubit entropy thermalization curves} and \ref{fig: improving thermalization with L} illustrate thermalization of this system, at least as seen through subsystem entanglement entropies, which only improves as the system size is increased.  

\subsection{ETH in the qubit spin chain}
\label{subsec: ETH in qubit spin chain}

So far we have illustrated the link between chaos and thermalization, but we have not yet explicitly connected thermalization to the mechanism of ETH explained in section \ref{subsec: eigenstate thermalization hypothesis}.  In this section we check that in the chaotic regime, this qubit spin chain does adhere to the predictions of ETH.  Recall that the ETH posits that operators which thermalize take on a particular form when expressed in the basis of energy eigenstates.  We repeat this form here for convenience (see section \ref{subsec: eigenstate thermalization hypothesis} for discussion):
\begin{equation}\label{eq: complete ETH repeated}
	\mathcal{O}_{ij}=\mathcal{O}(\bar{E})\delta_{ij} + e^{-S(\bar{E})/2}f_{\mathcal{O}}(\bar{E},\omega)R_{ij}.
\end{equation}
In the usual case, operators which take this form include the local operators, meaning those with support on few qubit factors.  We will focus on two aspects of such operators:

\begin{itemize}[noitemsep,topsep=0pt]
    \item The eigenstates themselves are approximately thermal.
	\item Off-diagonal matrix elements are exponentially suppressed.
\end{itemize}

\begin{figure}[H]
	\centering
	\begin{subfigure}{.59\textwidth}
		\centering
		\includegraphics[width=\linewidth]{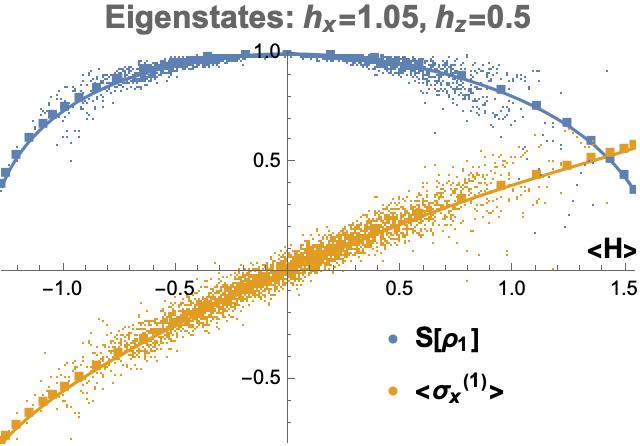}
		\caption{Thermal properties of eigenstates, $L=12$.}
		\label{fig: thermal eigenstates}
	\end{subfigure}%
	\begin{subfigure}{.41\textwidth}
		\centering
		\includegraphics[width=\linewidth]{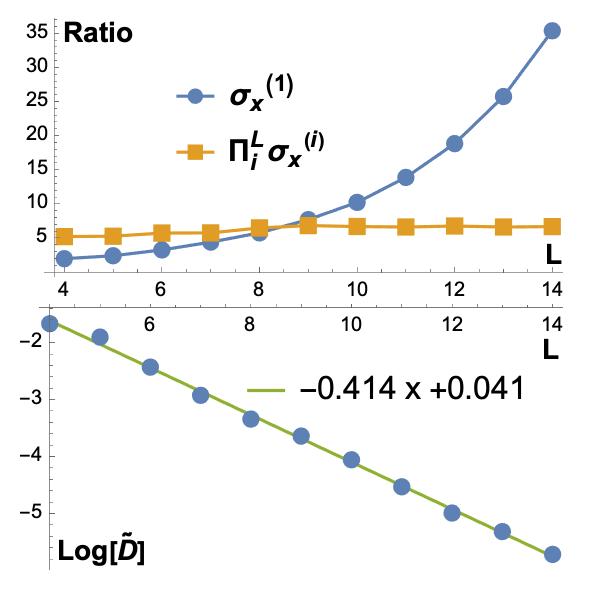}
		\caption{Suppression of off-diagonals.}
		\label{fig: diag to off-diag}
	\end{subfigure}
	\caption{The left figure (a) shows a comparison between thermal values (solid lines) and individual eigenstate values (points) for both the entanglement entropy and a Pauli operator expectation at the first site.
	The top panel of (b) shows the diagonal to off-diagonal element ratio
    in the energy basis for a local operator and a nonlocal operator as system size increases.
    The bottom panel of (b) shows the decay of the average magnitude on the anti-diagonal, averaging over 50 single site operators of fixed spectrum. See main text for discussion.}
	\label{fig: qubit ETH checks}
\end{figure}

We test these expectations in figure \ref{fig: qubit ETH checks}. The left panel \ref{fig: thermal eigenstates} focuses on the first point.  In every eigenstate, the expectation value of the local operator $\sigma_x^{(1)}$ as well as the entropy on the first qubit are plotted in comparison to the thermal predictions at the same energy (from the Gibbs state with corresponding energy expectation value).  The conformity between the eigenstates and thermal predictions is apparent, though it becomes weaker towards the edges of the spectrum.
The right panels \ref{fig: diag to off-diag} both serve to illustrate the suppression of off-diagonal matrix elements with increasing system size.  The top panel considers the ratio of the average diagonal element magnitude to that of off-diagonal elements for two operators:
$\text{Ratio}:=\overline{|\text{diag}|}\big/\overline{|\text{off-diag}|}$.
For the local operator, $\sigma_x^{(1)}$, this ratio is seen to increase exponentially with system size. By contrast the operator $\prod_i^L \sigma_{x}^{(i)}$, which has the same spectrum but is supported along the entire spin chain, shows a ratio approximately independent of system size.  To compare more explicitly with the rate of exponential suppression appearing as $\exp{-S(\bar{E})/2}$ in the ETH ansatz, we do a second test in the bottom panel of the figure.  Instead of averaging over all off-diagonals, we average specifically over the counter-diagonal, meaning we compute
\begin{equation}\label{eq: counter diagonal}
    \tilde{D}:=\frac{1}{2^L}\sum_i \big|\bra{E_i}\hat{O}\ket{E_{i'}}\big|
    \text{ where } i':=2^L+1-i.
\end{equation}
The terms in this average all have $\bar{E}:=\frac{1}{2}(E_i+E_{i'})\approx 0$, thus the suppression associated with each term should be given by the same, maximal thermodynamic entropy $S(\bar{E}=0) = \log(2^L)$.  On a log scale this should appear as a decay rate of roughly $-\log(2)/2 \approx -0.35$.  We took the average of this quantity from fifty random single-site operators of fixed spectrum in systems of size $L=4$ through $L=14$.  Results are displayed on a log scale in the bottom panel of figure \ref{fig: diag to off-diag}, along with a best fit line.  The exponential suppression is again evident, with a log scale coefficient $\approx -0.4$, which is close but in fact sightly faster than the ideal ETH suppression coming from $e^{-S(\bar{E})/2}$ alone. Some difference is not surprising, however, since there is additional dependence coming from $f_{\mathcal{O}}(\bar{E},\omega)$ in \eqref{eq: complete ETH repeated} along the counter-diagonal which we have not attempted to model at all.

\subsection{`Generic ETH' in the qubit system}
\label{subsec: generic ETH in qubit system}

\begin{figure}[H]
	\centering
	\begin{subfigure}{.5\textwidth}
		\centering
		\includegraphics[width=\linewidth]{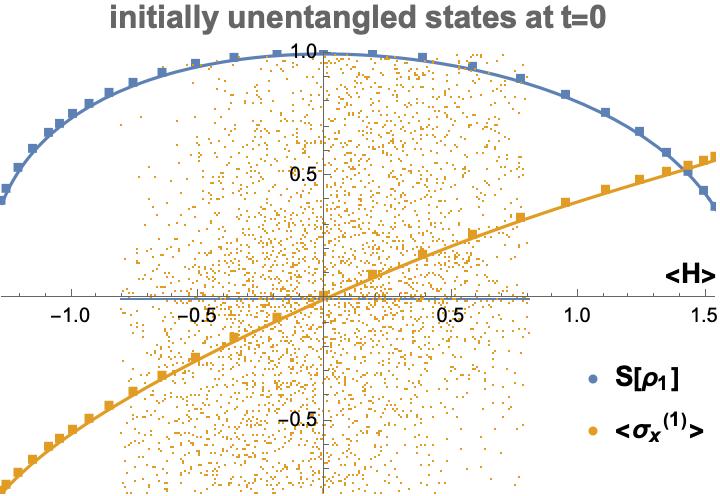}
		\caption{$t=0$}
		\label{fig: unentangled t=0}
	\end{subfigure}%
	\begin{subfigure}{.5\textwidth}
		\centering
		\includegraphics[width=\linewidth]{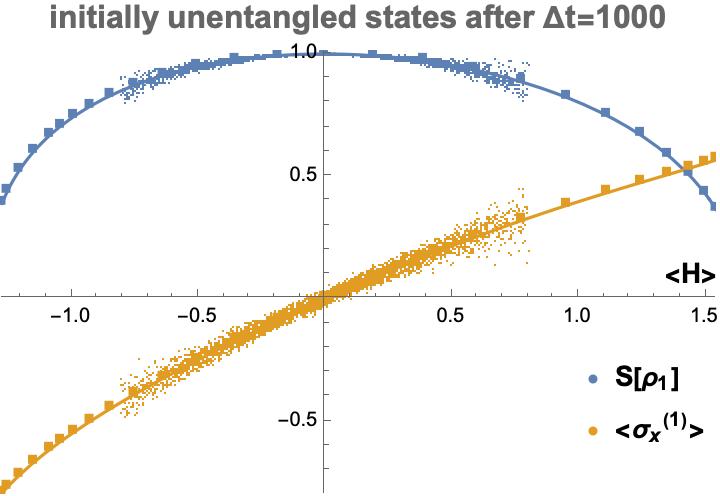}
		\caption{$t=1000$}
		\label{fig: unentangled t=1000}
	\end{subfigure}
	\caption{A comparison between thermal values (solid lines) and values computed in random unentangled states (small points) across the energy range $-0.8 < \langle H \rangle < 0.8$ for the quantities $S[\rho_1]$ and $\langle \sigma_x^{(1)}\rangle$.  The left panel shows initial values (note the entropy values are all on the energy axis, as these are unentangled states) while the right panel shows values obtained after evolving the same states for $\Delta t = 1000$. See main text for discussion.}
	\label{fig: thermalization of initially unentangled states}
\end{figure}

A final visualization of thermalization in this system is given in figure \ref{fig: thermalization of initially unentangled states}.  Instead of considering the eigenstates themselves, for each point in the spectrum between $\langle H\rangle=-.8$ and $\langle H \rangle =0.8$, a random unentangled state is generated at the same energy ($\pm 0.002$).\footnote{The reason we generate one state per spectral point is merely to match the density of states appearing in figure \ref{fig: qubit ETH checks} for the eigenstates themselves.  The reason we limit to the range $-0.8 < \langle H \rangle < 0.8$ is that it becomes prohibitively time consuming to generate random unentangled states at energy further from the center of the spectrum.}  The initial values for $\langle \sigma_x^{(1)}\rangle$ and $S[\rho_1]$ are displayed in the left panel.  The same states are then evolved for a large time interval ($\Delta t =1000$) and the same quantities computed in the resulting states.  Thermalization is evident in the grouping of these values around the thermal predictions.  

\begin{figure}[H]
	\centering
	\begin{subfigure}{.5\textwidth}
		\centering
		\includegraphics[width=\linewidth]{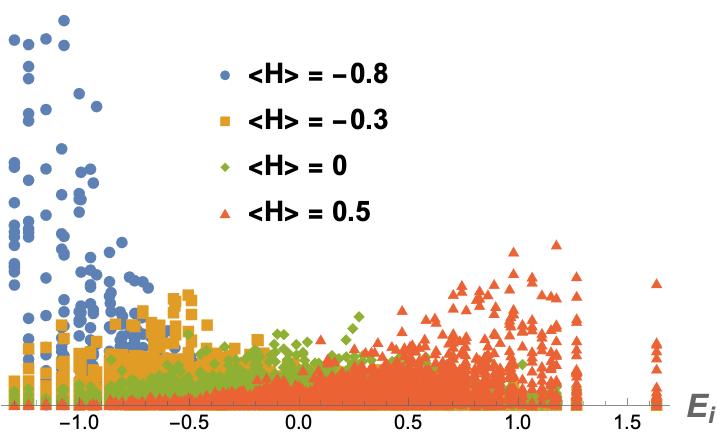}
		\label{fig: unentangled state energy distribution}
	\end{subfigure}%
	\begin{subfigure}{.5\textwidth}
		\centering
		\includegraphics[width=\linewidth]{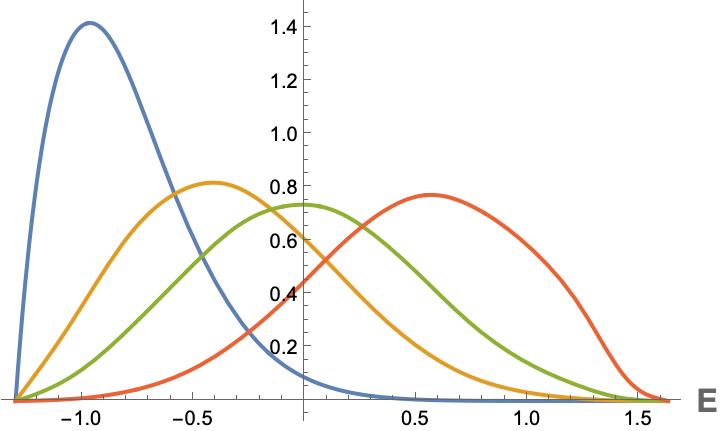}
		\label{fig: Gibbs state distributions}
	\end{subfigure}
	\caption{The left plot shows the squared magnitude of state coefficients $|c_i|^2:=|\bra{E_i}\ket{\Psi}|^2$ for random unentangled states for $L=8$.  Ten states at each of the listed energies are overlaid.  Far from being approximately microcanonical, the energy density distributions mimic those of Gibbs states.  As a qualitative comparison, the right plot shows energy density profiles obtained from Gibbs states with matching energy expectation values based on an interpolated density of states function.  The dimensionless inverse temperatures of these curves are $\beta \approx 4.6, 1.3, 0.0, -1.9$ for blue, orange, green, and red.}
	\label{fig: energy distribution of states}
\end{figure}

It is interesting to note that the random unentangled states are \textit{not} the paradigmatic states of narrow energy support often used in studies of ETH and thermalization.  States with narrow energy support are particularly natural to consider in quench scenarios, where one starts from a system eigenstate and then moves out of equilibrium through a sudden modification of Hamiltonian parameters.  But ETH is predictive for a much broader set of states.  The fact that thermalization occurs for the states considered here might at first seem surprising, but the reason in simple.  Having randomly generated unentangled states and then selected those falling at a particular energy, the resulting energy distribution tends to mimic that of a Gibbs distribution, which is qualitatively illustrated in figure \ref{fig: energy distribution of states}.  This figure compares energy distributions from a selection of random unentangled states to energy density distributions of a corresponding Gibbs state.\footnote{The smooth Gibbs state energy distributions in the right hand side of figure \ref{fig: energy distribution of states} are obtained by assuming a Wigner semicircle density of states, which is applicable to GOE ensemble Hamiltonians, scaled to the same energy range as the qubit Hamiltonian.  The actual density of states for the qubit Hamiltonian in fact has more features than the ideal Wigner semicircle, thus the comparison between left and right figures is merely qualitative.}  It is clear that the energy distribution of these states is far from a microcanonical state, and much closer to a Gibbs state.  

\begin{figure}
	\centering
	\begin{subfigure}{.48\textwidth}
		\centering
		\includegraphics[width=\linewidth]{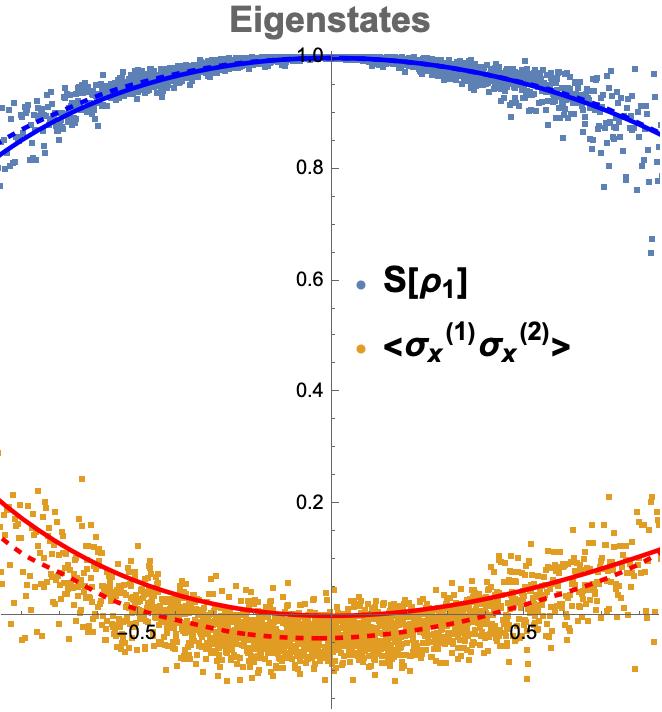}
		\label{fig: eigenstates sigxsigx}
	\end{subfigure}%
    \hspace{.01 \linewidth}
	\begin{subfigure}{.48\textwidth}
		\centering
		\includegraphics[width=\linewidth]{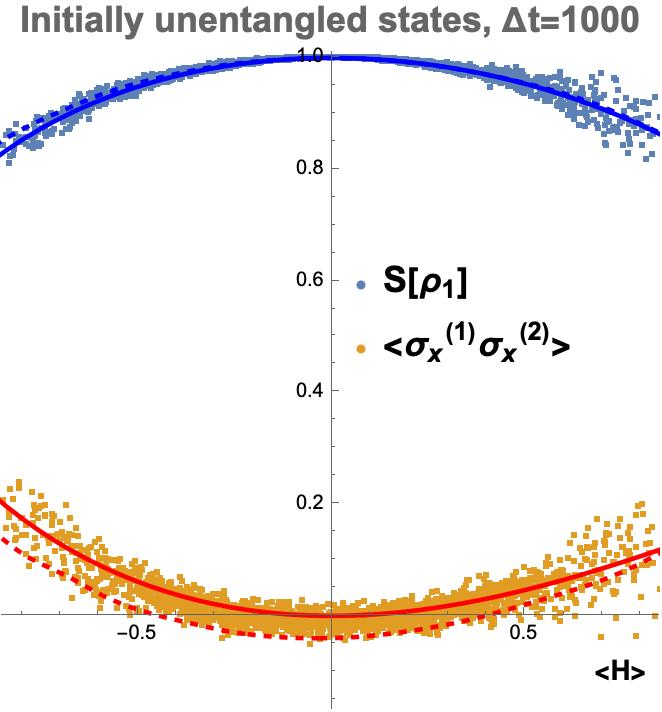}
		\label{fig: unentangled sigxsigx t=1000}
	\end{subfigure}
	\caption{The left plot shows $\langle\sigma_x^{(1)}\otimes \sigma_x^{(2)}\rangle$ and $S(\rho_1)$ for a selection of eigenstates in a system of $L=13$ (every fourth eigenstate is shown in the range $-0.85 < \langle H \rangle < 0.85$).  The right plot shows the same quantities in states across the same energy range that were initially unentangled and evolved for $\Delta t= 1000$. The solid lines shows expected values of these quantities in a Gibbs ensemble, whereas the dashed lines show the predictions of a microcanonical ensemble of width $\pm 0.1$.  Since the microcanonical ensemble simply averages over eigenstates in a small window, the dashed line of course fits better on the left plot.  But the thermalization of initially-unentangled states more closely follows the Gibbs prediction because their energy distributions are much closer to that of a Gibbs distribution.}
	\label{fig: micro versus Gibbs}
\end{figure}

Considering the discussion in section \ref{subsec: generic ETH in theory}, the equilibrium values for these random unentangled states should match the predictions of a Gibbs ensemble more closely than a microcanonical ensemble.  For local operators, the difference between ensemble predictions is small, but in some cases noticeable for the system sizes we consider.  To verify that these states thermalize more closely to Gibbs state predictions than microcanonical predictions, it is necessary to find an operator which strongly distinguishes these two.  Considering again the discussion in section \ref{subsec: generic ETH in theory}, this entails finding an operator with thermal expectation value of non-negligible concavity with respect to thermal parameters.  Figure \ref{fig: micro versus Gibbs} compares the two ensemble predictions for the two site operator $\sigma_x^{(1)}\otimes \sigma_x^{(2)}$.  As a two-site operator, this operator thermalizes less well than single site operators, but the difference between microcanonical and Gibbs state expectations values can be distinguished, and the thermalization of random unengtangled states clearly shifts toward the Gibbs distribution.

\section{Thermalization with conserved charge}
\label{sec: Thermalization with conserved charge}
In this section we discuss ETH in systems with a quasilocal conserved charge.  The meaning of quasilocal here is that it is constructed as a sum of local terms, in contrast with the parity operator $P$, for instance. We will begin in \ref{subsec: The qutrit Hamiltonian} by introducing a model, the qutrit Hamiltonian, which is built off of the qubit Hamiltonian of the previous section but admits a single such conserved charge.  In section \ref{subsec: Charge sectors of the qutrit Hamiltonian} we check that within charge sectors, the resulting Hamiltonian is chaotic.  In section \ref{subsec: Dynamics of states within Q sector}, we will consider thermalization of states within fixed charge sectors. In section \ref{subsec: generic ETH in Qutrit system}, we consider states with broad support across charge sectors, illustrating generic ETH for this system. 

\subsection{The qutrit Hamiltonian}
\label{subsec: The qutrit Hamiltonian}
Here we introduce the model we use to study thermalization in the presence of conserved quasilocal charge.  Essentially, we will take the chaotic qubit Hamiltonian of section \ref{sec: qubit spinchain models} and embed it within a qutrit model, extending the state space of each lattice site by one state.  The local support on this additional state space will be interpretable as a local charge.  Additional interaction terms are then added which allow this charge to spread locally, but maintain global conservation.  The same interaction terms restore chaotic dynamics within individual subsectors of total charge.  We will then explore the interplay of the additional charge with the dynamics of thermalization in the following sections. This method of introducing a conserved charge allows for the energy, the local charge density, and the local entanglement entropy to be tuned independently over a wide parameter range, all while keeping the total size of the Hilbert space manageable for exact diagonalization.  This allows substantial control in the initialization of states with which to study thermalization.

To build the model, consider the following $L$-qutrit Hamiltonian,

\begin{align}\label{eq: qutrit hamiltonian}
		H &= J\sum_{r=1}^{L-1}\lambda_{3}^{(r)} \lambda_{3}^{(r+1)}
		+h_1\sum_{r=1}^{L}\lambda_{1}^{(r)} 
		+h_2\sum_{r=1}^{L}\lambda_{2}^{(r)} 
		+h_3\sum_{r=1}^{L}\lambda_{3}^{(r)}  + a\Delta Q ,\\
		\label{eq: charge spread op}
		\Delta Q &= \sum_{r=1}^{L-1} \sum_{i=1}^{4} c_{i}^{(r)}  \Delta q_{i}^{(r,r+1)}
\end{align}
where the operators $\lambda_{i}$ for $i=1, ..., 8$ are the $3\times3$ Gell-Mann matrices,\footnote{We use the following basis for Gell-Mann matrices:
\begin{align*}
&\lambda_{1} = \begin{pmatrix}0&1&0\\1&0&0\\0&0&0\end{pmatrix} &
&\lambda_{2}= \begin{pmatrix}0&-i&0\\i&0&0\\0&0&0\end{pmatrix} &
&\lambda_{3}= \begin{pmatrix}1&0&0\\0&-1&0\\0&0&0\end{pmatrix} &\\
&\lambda_{4} = \begin{pmatrix}0&0&1\\0&0&0\\1&0&0\end{pmatrix} &
&\lambda_{5}= \begin{pmatrix}0&0&-i\\0&0&0\\i&0&0\end{pmatrix} &
&\lambda_{6}= \begin{pmatrix}0&0&0\\0&0&1\\0&1&0\end{pmatrix} &\\
&\lambda_{7}= \begin{pmatrix}0&0&0\\0&0&-i\\0&i&0\end{pmatrix} &
&\lambda_{8}= \frac{1}{\sqrt{3}}\begin{pmatrix}1&0&0\\0&1&0\\0&0&-2\end{pmatrix}.&  &\\
\end{align*}
}
and $\lambda_{i}^{(r)}$ is understood as the $i$'th Gell-Mann matrix acting at the $r$'th qutrit site.

First let us consider equation \eqref{eq: qutrit hamiltonian} with the coefficient $a$ set to zero so that the last term can be ignored (this term will be explained shortly). The remaining Hamiltonian can be thought of as an embedding of a qubit Hamiltonian into an enlarged Hilbert space.  Specifically, each local qubit Hilbert space $\mathcal{H}^{(2)}$ has been extended via $\mathcal{H}^{(2)}\rightarrow \mathcal{H}^{(2)}\oplus \mathcal{H}^{(1)}$, and the qubit Hamiltonian is then extended through the mapping $(\sigma_{x}, \sigma_{y}, \sigma_{z})\rightarrow (\lambda_{1}, \lambda_{2}, \lambda_{3})$. We will take the coefficient values to be those identified as chaotic for the qubit Hamiltonian \eqref{eq: qubit hamiltonian}: $h_1=1.05, h_2=0$, and $h_3=0.5$.

Note that regardless of the coefficient values, the Gell-Mann matrices employed in \eqref{eq: qutrit hamiltonian} each commute with the single-qutrit operator 
\begin{equation}
    q=\begin{pmatrix}
    0&0&0\\ 0&0&0\\ 0&0&1
    \end{pmatrix}.
\end{equation}
This becomes a local charge operator.  For a given qutrit, this operator captures by projection the amount of support placed on the extension, the additional $\mathcal{H}^{(1)}$ part of the qutrit Hilbert space.  We will sometimes refer to this as the ``third slot'' value, based on its placement in a computational-basis vector.\footnote{We could equally consider the traceless operator $q=\begin{pmatrix}
    -1/3&0&0\\ 0&-1/3&0\\ 0&0&2/3
\end{pmatrix}$, but defining $q$ to simply count the support on the third qutrit slot has intuitive advantages.}  Obviously, for a single qutrit the local charge eigenvalues are $0$ and $1$ (the former being doubly degenerate), and the expectation value of the charge on a single qutrit pure state is the squared magnitude of the third slot coefficient.  The total charge operator is then simply the sum of local charges at each site:
\begin{equation}
	Q=\sum_{r=1}^L q^{(r)}.
\end{equation}

As it stands (with $a$=0), the Hamiltonian \eqref{eq: qutrit hamiltonian} conserves not only the total charge but also the local charge at every site, $q^{(r)}$ for $r\in{1,...,L}$.  This results in large spectrum degeneracies, even within subsectors of definite total charge.  The one exception is the zero-charge subsector, consisting of states with no support on the third slot of any qutrit.  Restricting to this subspace, the spectrum and dynamics is completely equivalent to that of the original $L$-\textit{qubit} system, which we have taken to correspond to the Hamiltonian \eqref{eq: qubit hamiltonian}.  In a rather direct way, increasing the amount of local charge at any site(s) reduces the relative importance of the chaotic dynamics between local $\mathcal{H}^{(2)}$s by moving some support to the inactive $\mathcal{H}^{(1)}$ component. The limit of maximal charge corresponds to a single eigenstate, with all support on the third slot of each qutrit. 

The asymmetry in the size of the state space at each definite charge is worth noting.  As just described, the minimal charge sector ($Q=0$), is a subspace of dimension $2^L$, while the maximally charged sector ($Q=L$) is a unique state. More generally, for any total charge eigenvalue $n\in (0,1,...,L)$, the state space is of dimension $|Q_n|= 2^{L-n} \binom{L}{n}$.  A \textit{typical} state, randomly selected over a Haar measure on the full Hilbert space, has charge $\expval{Q}=L/3$, corresponding to support distributed evenly across the three-state Hilbert space of each lattice site. 

By extending the Hamiltonian onto the additional $\mathcal{H}^{(1)}$ in such a trivial manner, with no coupling to the $\mathcal{H}^{(2)}$ part, we have introduced a large amount of degeneracy into the spectrum, as can be seen in  figure \ref{fig: qutrit histograms without spread}.  The term in \eqref{eq: qutrit hamiltonian} that we've ignored by setting $a=0$ is included partially to alleviate this degeneracy. If we now let $a=1$, this activates what we refer to as the charge-spreading operator $\Delta Q$ in equation \eqref{eq: charge spread op}.  Let us now describe how this operator is constructed.  

We want the total charge $Q$ to be conserved, but not the charge at every site.  We instead stipulate that charge may spread locally, transported between neighboring sites, resulting in an overall diffusion effect.  We therefore consider generic two-qutrit Hermitian operators which conserve the sum of the two charges, but not the individual charges.  It is easy to show that there are eight independent hermitian operators meeting these criteria.  If we further require that the Hamiltonian retain the time reversal invariance property that $H=H^\star$, then half of these Hermitian operators are excluded. We label the remaining four operators $\Delta q_i$ for $i\in{1,...,4}$.  \footnote{Explicitly, these can be written in terms of Gell-Mann matrices as 
\begin{align*}
&\Delta q_1 = \lambda_4\otimes\lambda_4+ \lambda_5\otimes\lambda_5,
&\Delta q_2 = \lambda_4\otimes\lambda_6+ \lambda_5\otimes\lambda_7,\\
&\Delta q_3 = \lambda_6\otimes\lambda_4+ \lambda_7\otimes\lambda_5,
&\Delta q_4 = \lambda_6\otimes\lambda_6+ \lambda_7\otimes\lambda_7.
\end{align*}}

\begin{figure}
	\centering
	\begin{subfigure}{.47\textwidth}
		\centering
		\includegraphics[width=\linewidth]{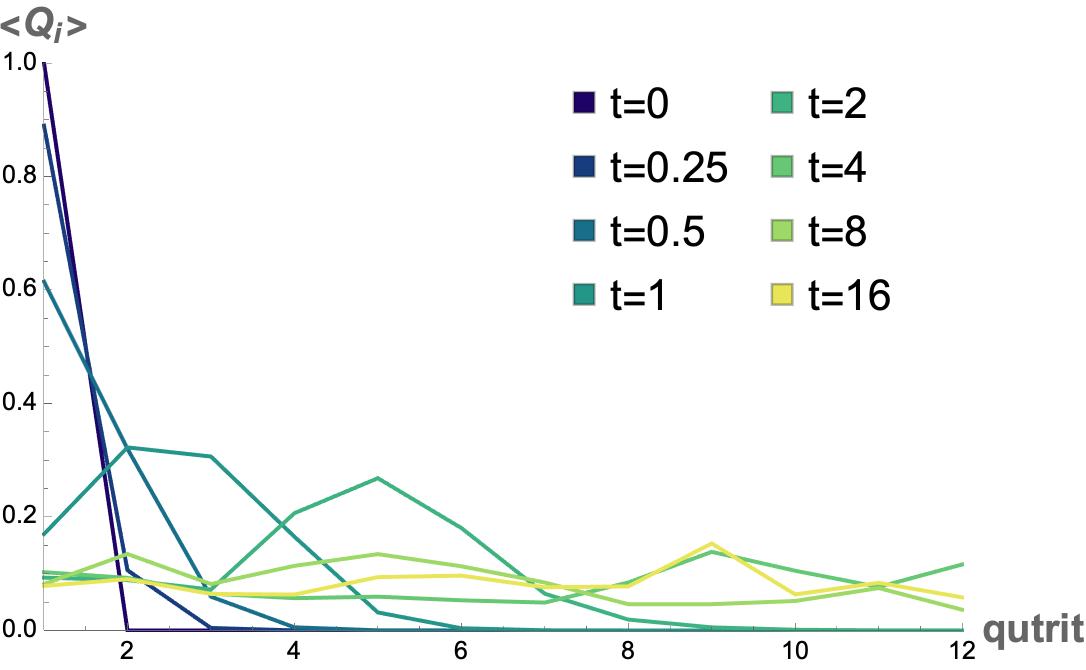}
		\label{fig: charge spread single}
	\end{subfigure}%
    \hspace{.01 \linewidth}
	\begin{subfigure}{.47\textwidth}
		\centering
		\includegraphics[width=\linewidth]{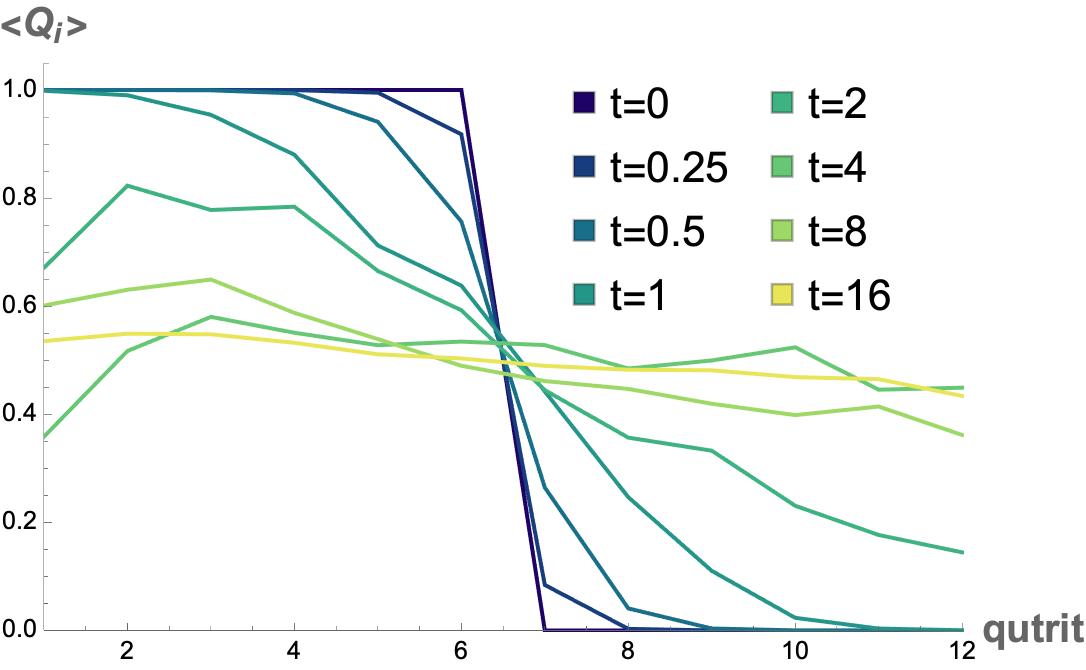}
		\label{fig:charge spread half}
	\end{subfigure}
	\caption{The effect of the charge spreading operator on the charge distribution itself is shown, starting from two different initial charge distributions. Each line represents an instantaneous distribution of charge across a 12-qutrit system, with lighter green lines representing later in time.  In the left panel, the charge distribution starts out everywhere zero except for the first qutrit, which has maximal charge.  In the right panel, the charge distribution is initially maximal on qutrits 1-6, and zero charge on qutrits 7-12.  Clearly the local charge values are no longer conserved, though the total is.  Note that the diagrams equilibrate to different values of (approximately) uniform charge density, in this case 1/12 and 1/2, respectively.}
	\label{fig: charge spreading}
\end{figure}

A general two-site charge-spreading operator may then be constructed as an arbitrary linear combination of these, and the \textit{total} charge spreading operator $\Delta Q$ consists of a sum of such charge-spreading operators placed pairwise on neighboring sites all the way down the lattice:
\begin{equation}\label{eq: charge spreading operator}
    \Delta Q := \sum_{r=1}^{L-1}\sum_{i=1}^4 c^{(r)}_i \Delta q_i^{(r)},
\end{equation}
where $\Delta q_i^{(r)}$ acts on qutrits $r$ and $r+1$, and the coefficients $c^{(i)}_r$ are independently and randomly generated.
 Concretely, we draw these from a gaussian distribution centered on 1, with spread of width 0.1, which has the desired effect of restoring chaos to individual sectors of definite total charge as we will see in the next subsection.  By construction, the inclusion of the charge spreading operator eliminates each $q^{(r)}$ as a local conserved operator. The dynamical affect of the charge spreading term on distributions of local charge is illustrated in figure \ref{fig: charge spreading}, which shows the time evolution of charge distribution from initially inhomogeneous configurations toward quasi-equilibrium configurations.

\subsection{Charge sectors of the qutrit Hamiltonian}
\label{subsec: Charge sectors of the qutrit Hamiltonian}

We now begin the analysis of the Qutrit Hamiltonian by examining the level-spacing distributions within sectors of definite total charge $Q$.

\subsubsection{Level-spacings without the charge spreading term ($a=0$)}
\label{subsubsec: levels without charge spreading term}

\begin{figure}[h]
	\centering
	\begin{subfigure}{.3\textwidth}
		\centering
		\includegraphics[width=\linewidth]{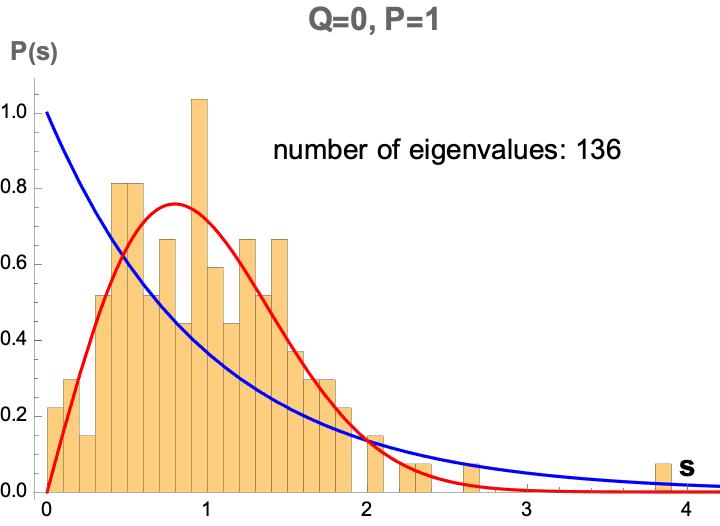}
		\label{fig: charge spread off Q0P1}
	\end{subfigure}%
    \hspace{.01 \linewidth}
	\begin{subfigure}{.3\textwidth}
		\centering
		\includegraphics[width=\linewidth]{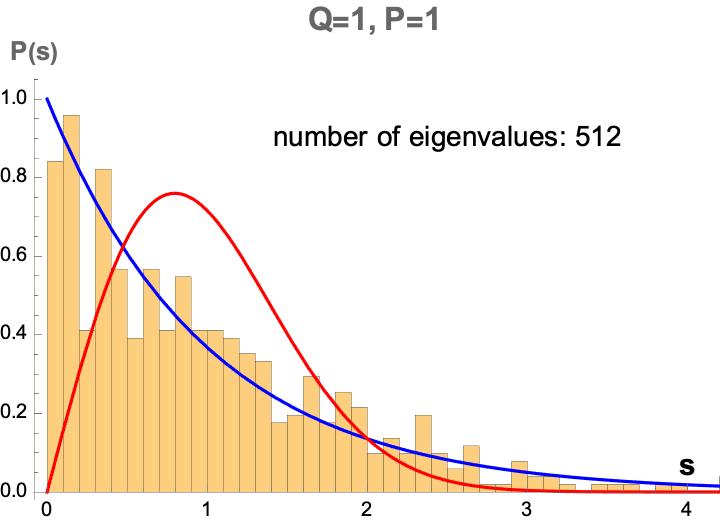}
		\label{fig: charge spread off Q1P1}
	\end{subfigure}
    \hspace{.01 \linewidth}
	\begin{subfigure}{.3\textwidth}
		\centering
		\includegraphics[width=\linewidth]{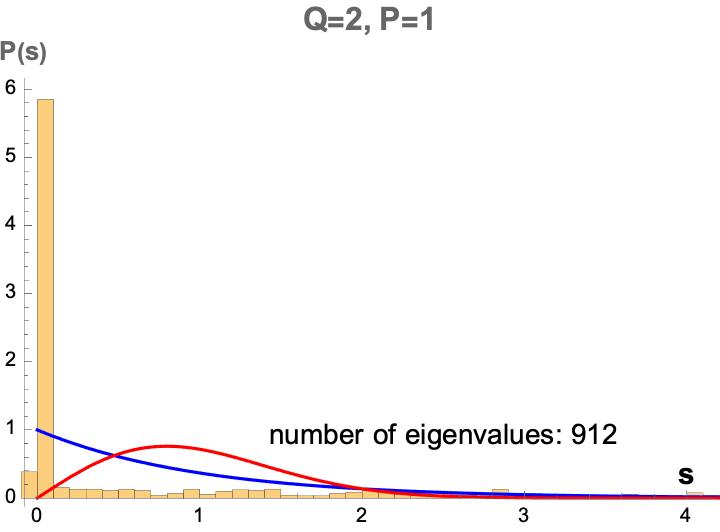}
		\label{fig: charge spread off Q2P1}
	\end{subfigure}
	\caption{Spectra are shown across several subsectors of definite total charge and definite parity for the Hamiltonian \eqref{eq: qutrit hamiltonian} with $a=0$ (or equivalently, all $c_r^{(i)}=0$).  Setting these terms to zero eliminates the charge spreading operator, resulting in a Hamiltonian with hugely degenerate spectra; the parity symmetry is exactly retained, and the local charge at every site is also a conserved quantity.  Shown here are the sectors with $P=1$ and $Q=0, 1,$ and $2$.}
	\label{fig: qutrit histograms without spread}
\end{figure}

We first briefly consider the case that $a=0$ in the qutrit Hamiltonian \eqref{eq: qutrit hamiltonian}, turning off the charge spreading operator $\Delta Q$.  In this case, there is an overall parity symmetry that inverts the order of qutrits, which again we denote $P$.  More importantly, this Hamiltonian also has an extensive number of conserved local quantities given by the individual charge operators $q^{(r)}$ at each site.    This entails that even if we restrict to sectors of definite total charge and parity we do not find chaos.  The level spacings of a few such sectors are shown in figure \ref{fig: qutrit histograms without spread}.  The $Q=0$ sector is a special case, in the sense that it is insensitive to the presence or absence of the charge spreading operator, and its spectrum is identical to that of the corresponding qubit Hamiltonian (the action restricted to the $\mathcal{H}^{(2)}$ at each site) and hence chaotic.  The case of $Q=1$ is intermediate, more closely matching the Poisson distribution.  All higher $Q$ sectors merely resemble the $Q=2$ sector, with degeneracies dominating the distribution.

\subsubsection{Level spacings with the charge spreading term ($a=1$)}
\label{subsubsec: levels with charge spreading}

\begin{figure}[H]\centering
	\begin{subfigure}{.3\textwidth}\centering
		\includegraphics[width=\linewidth]{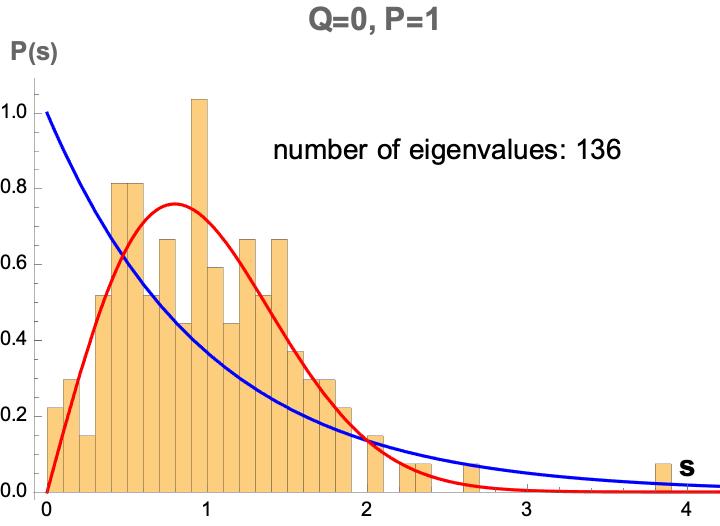}
	\end{subfigure}\hspace{.01 \linewidth}
	\begin{subfigure}{.3\textwidth}\centering
		\includegraphics[width=\linewidth]{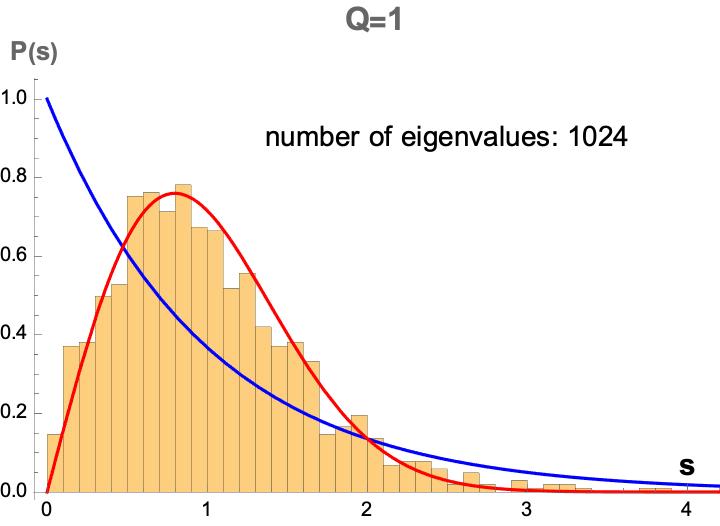}
	\end{subfigure}\hspace{.01 \linewidth}
	\begin{subfigure}{.3\textwidth}\centering
		\includegraphics[width=\linewidth]{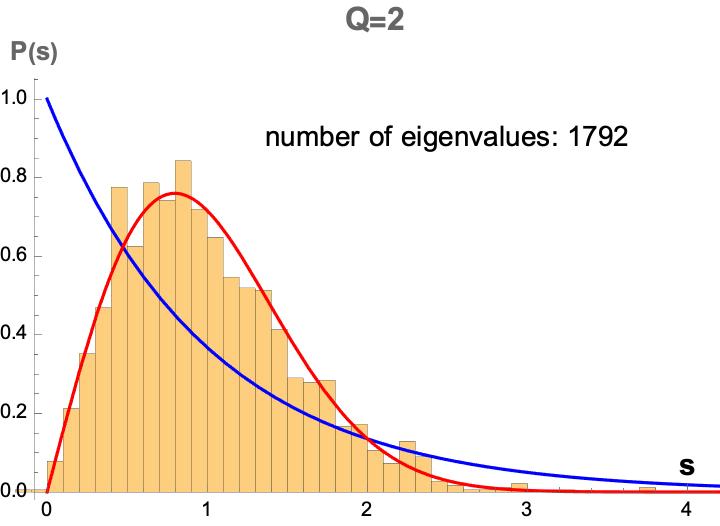}
	\end{subfigure}\hspace{.01 \linewidth}
    \begin{subfigure}{.3\textwidth}\centering
		\includegraphics[width=\linewidth]{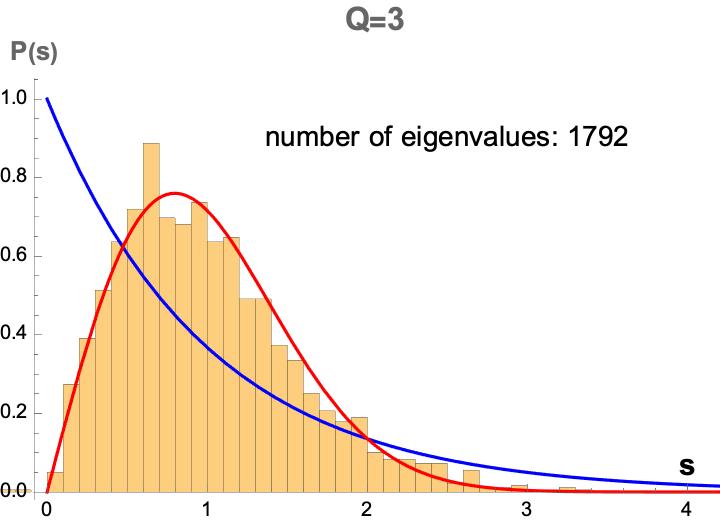}
	\end{subfigure}\hspace{.01 \linewidth}
	\begin{subfigure}{.3\textwidth}\centering
		\includegraphics[width=\linewidth]{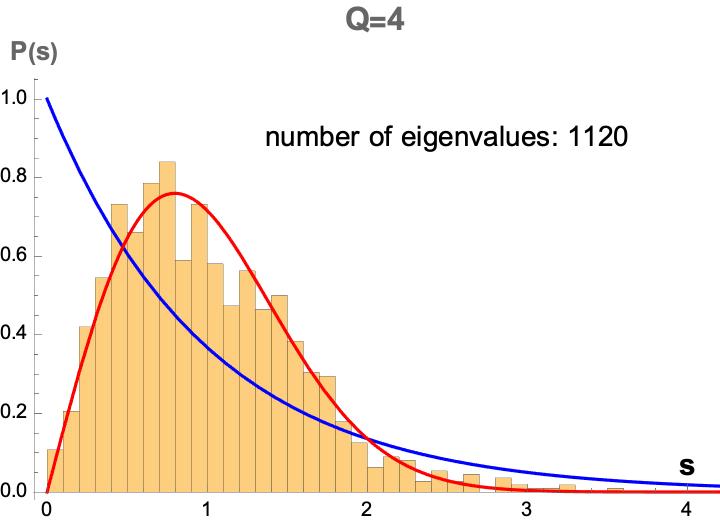}
	\end{subfigure}\hspace{.01 \linewidth}
	\begin{subfigure}{.3\textwidth}\centering
		\includegraphics[width=\linewidth]{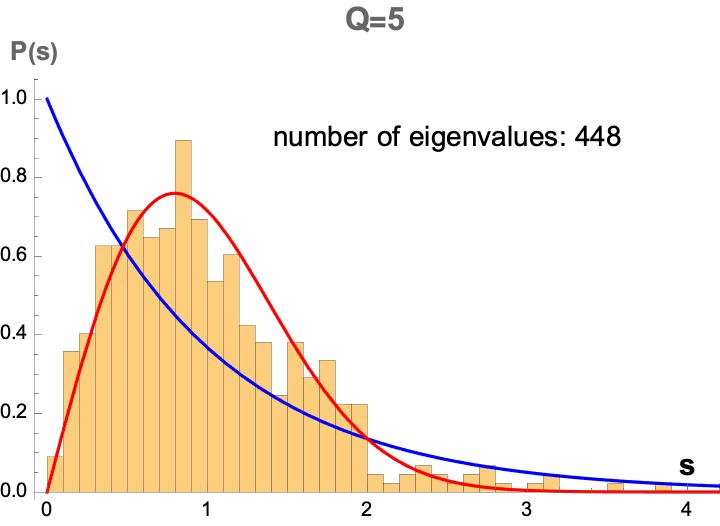}
	\end{subfigure}\hspace{.01 \linewidth}
    \begin{subfigure}{.3\textwidth}\centering
		\includegraphics[width=\linewidth]{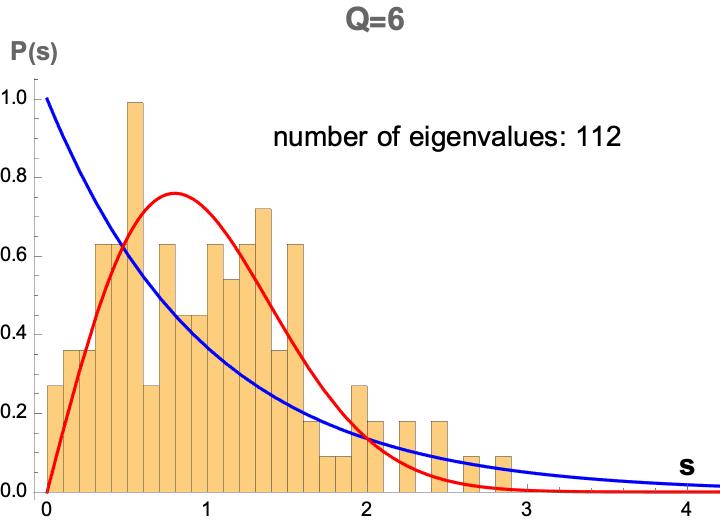}
	\end{subfigure}\hspace{.01 \linewidth}
	\begin{subfigure}{.3\textwidth}\centering
		\includegraphics[width=\linewidth]{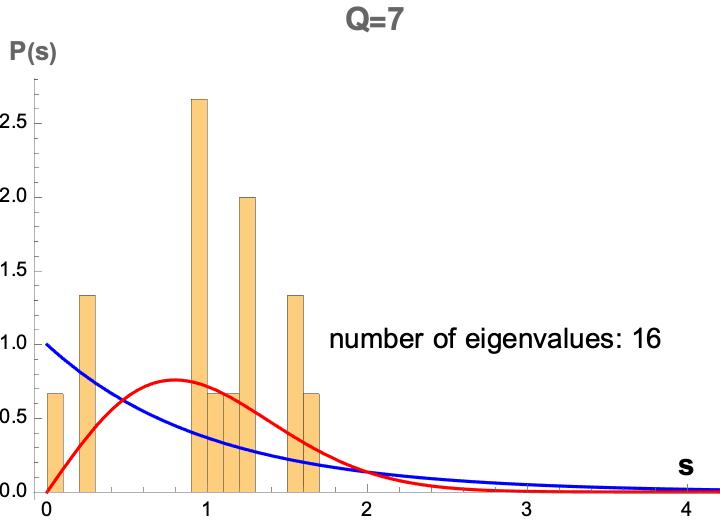}
	\end{subfigure}\hspace{.31 \linewidth}
	\caption{The level spacing distributions of total charge eigensectors within the qutrit Hamiltonian \eqref{eq: qutrit hamiltonian} with $a=1$ and coefficients $c_r^{(i)}$ in the charge spreading operator drawn independently from a normal distribution of width 0.1 centered at 1. Each panel shows a fixed charge sector of the Hamiltonian. The fit to the GOE distribution is once again apparent, although it is poorer for sectors with fewer eigenvalues. The maximal charge $Q=L$ case is not shown as the charge sector contains only a single eigenvalue.}
	\label{fig: qutrit histograms with spread}
\end{figure}

We now consider the full qutrit Hamiltonian, including the charge spreading operator by setting $a=1$.  By design, this term eliminates the local conserved quantities $q^{(r)}$.  The parity symmetry is likewise eliminated due to the randomization of coefficients $c_i^{(r)}$.  The caveat to this last statement is that the $Q=0$ subsector retains a parity symmetry, being insensitive to the charge spreading operator.  Figure \ref{fig: qutrit histograms with spread} shows level spacings for charge sectors 0 through 7 in a system of $L=8$. For the larger eigensectors, $Q=1$ through $Q=5$, the emergence of a Wigner-Dyson distribution is clear, and thus we expect states in these sectors to thermalize.
Where there are fewer eigenstates, e.g. $Q=6,7$, we see a poor fit to the Wigner-Dyson distribution due to a lack of eigenstates.  We note that even for higher values of $Q$ the distributions still more closely resemble the GOE distribution than the Poisson distribution.  Increasing the system size increases both the maximum possible conserved charge, as well as the size of individual charge sectors, with the exception of the maximal charge $Q=L$ sector, which will \textit{always} admit a unique state. 

\subsection{Dynamics of states within fixed $Q$ sector}
\label{subsec: Dynamics of states within Q sector}

Having established that the Hamiltonian is chaotic within charge sectors by finding that for some values of $Q$ the level-spacing distributions approximate a Wigner-Dyson distribution, we next consider the dynamics of states in such regimes.  
In this section we will see conformity to the expectations of ETH within individual charge sectors, with improving clarity for sectors with larger state space.  We check this by comparing the eigenstates themselves to a set of states we call ``random microcanonical'' states. 

Since the charge operator $Q$ commutes with the Hamiltonian $H$, we are able to identify simultaneous eigenstates of both operators. Hamiltonian eigenstates are divided into charge sectors accordingly.  We then build states consisting of a random superpositions of eigenstates within a small energy window $E \pm \Delta E$ and a fixed charge sector.  We call these ``random microcanonical states,''  (note that the charge `window' is precisely one sector at this point).  We compare subsystem entanglement entropies and operator expectation values in these states with those in the eigenstates themselves.  For the qutrit spinchains of size $L=9$ used in this section, generating curves for the Gibbs state predictions would be too large of a task numerically.  On the other hand, microcanonical predictions can be easily inferred visually, at least in sectors where the ETH clustering is apparent, as an average of eigenstates in a small window.

\begin{figure}[H]\centering
	\begin{subfigure}{.32\textwidth}\centering
		\includegraphics[width=\linewidth]{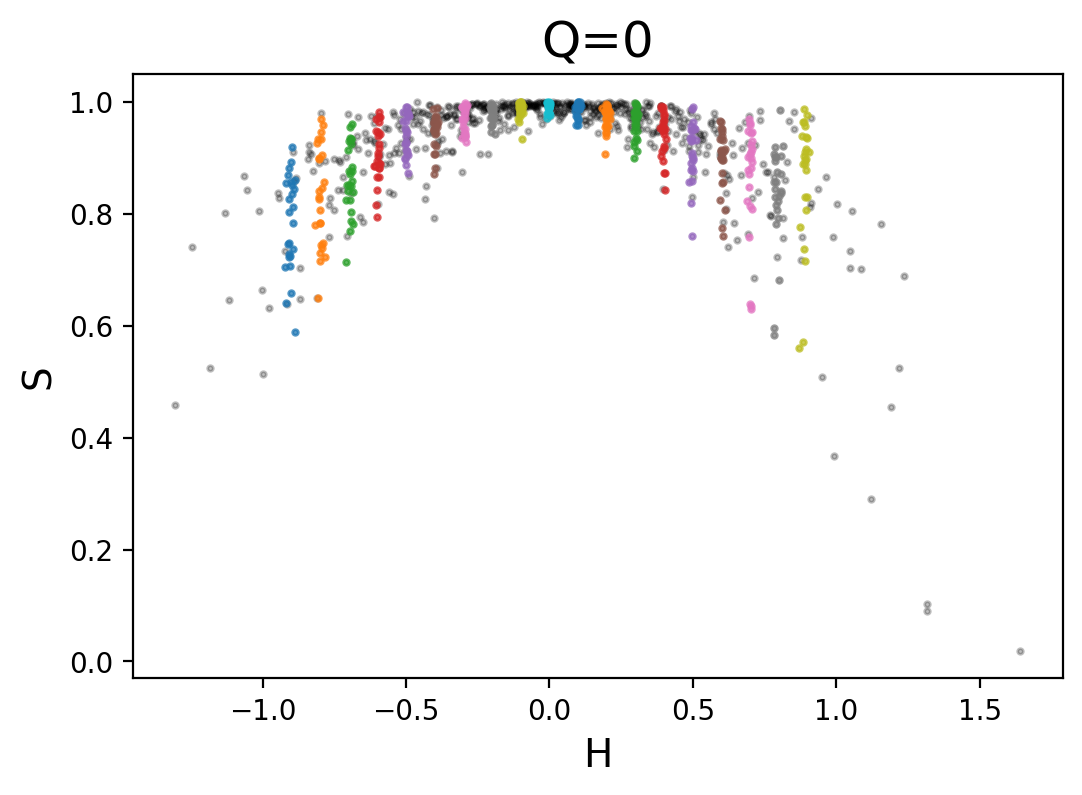}
	\end{subfigure}
	\begin{subfigure}{.32\textwidth}\centering
		\includegraphics[width=\linewidth]{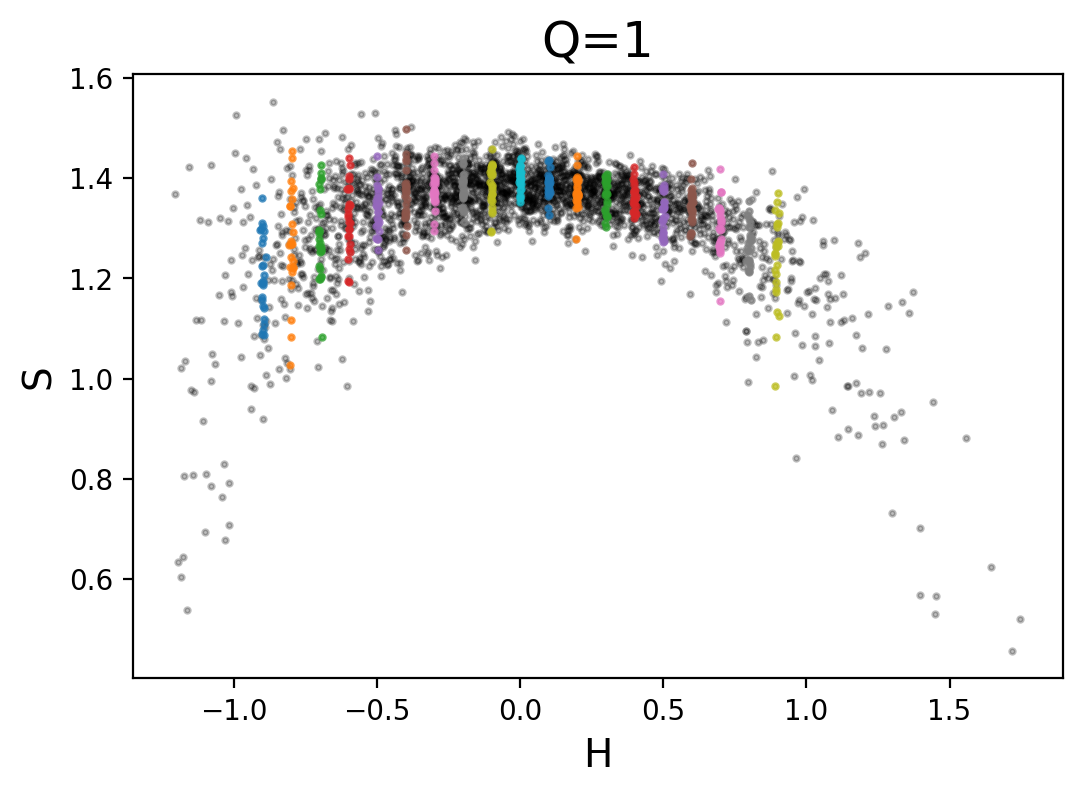}
	\end{subfigure}
	\begin{subfigure}{.32\textwidth}\centering
		\includegraphics[width=\linewidth]{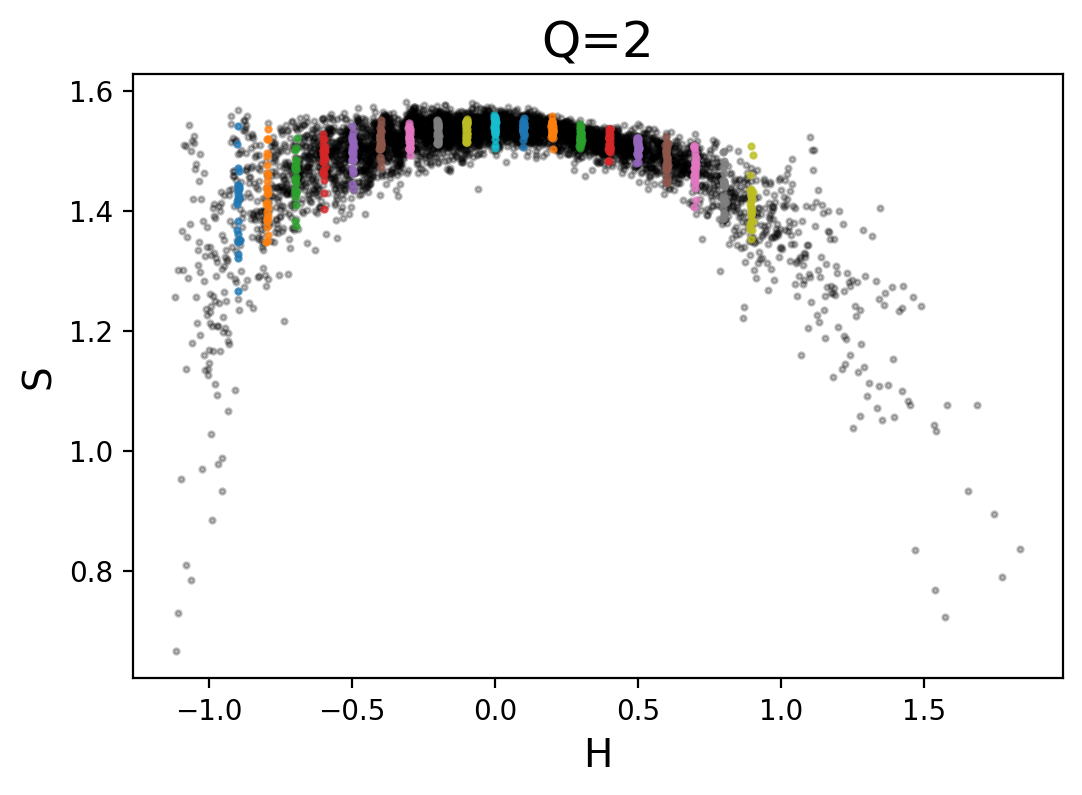}
	\end{subfigure}
    \begin{subfigure}{.32\textwidth}\centering
		\includegraphics[width=\linewidth]{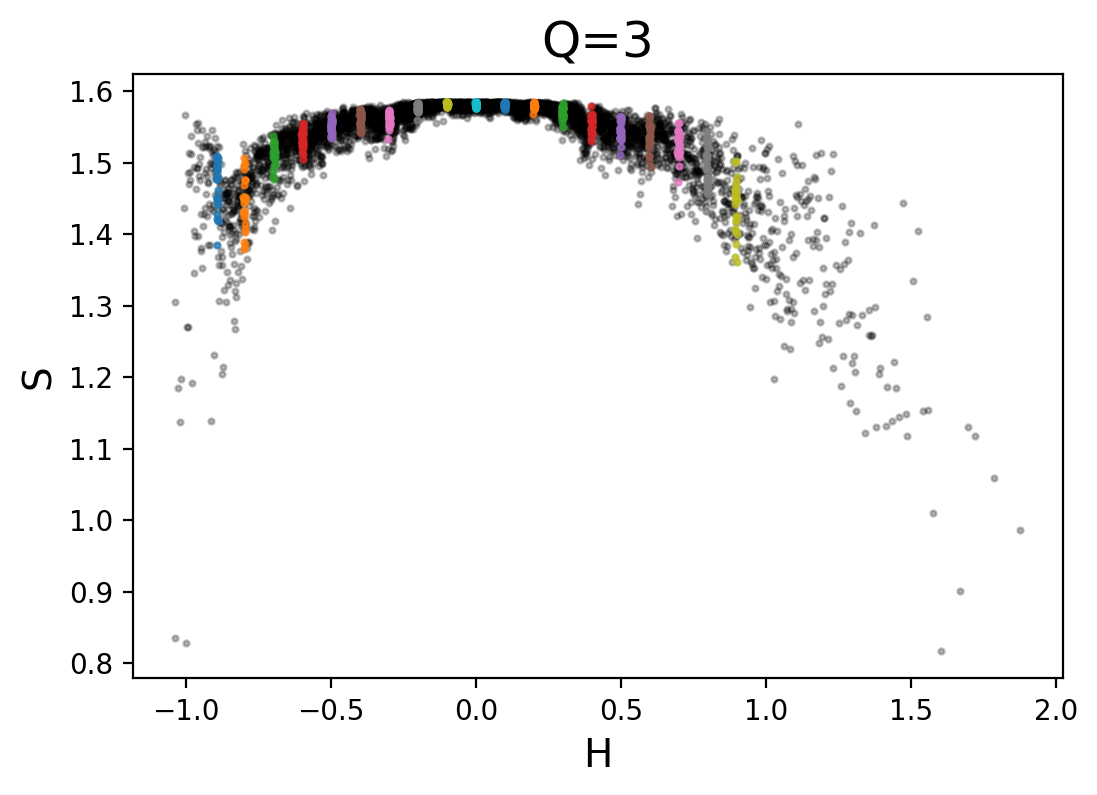}
	\end{subfigure}
	\begin{subfigure}{.32\textwidth}\centering
		\includegraphics[width=\linewidth]{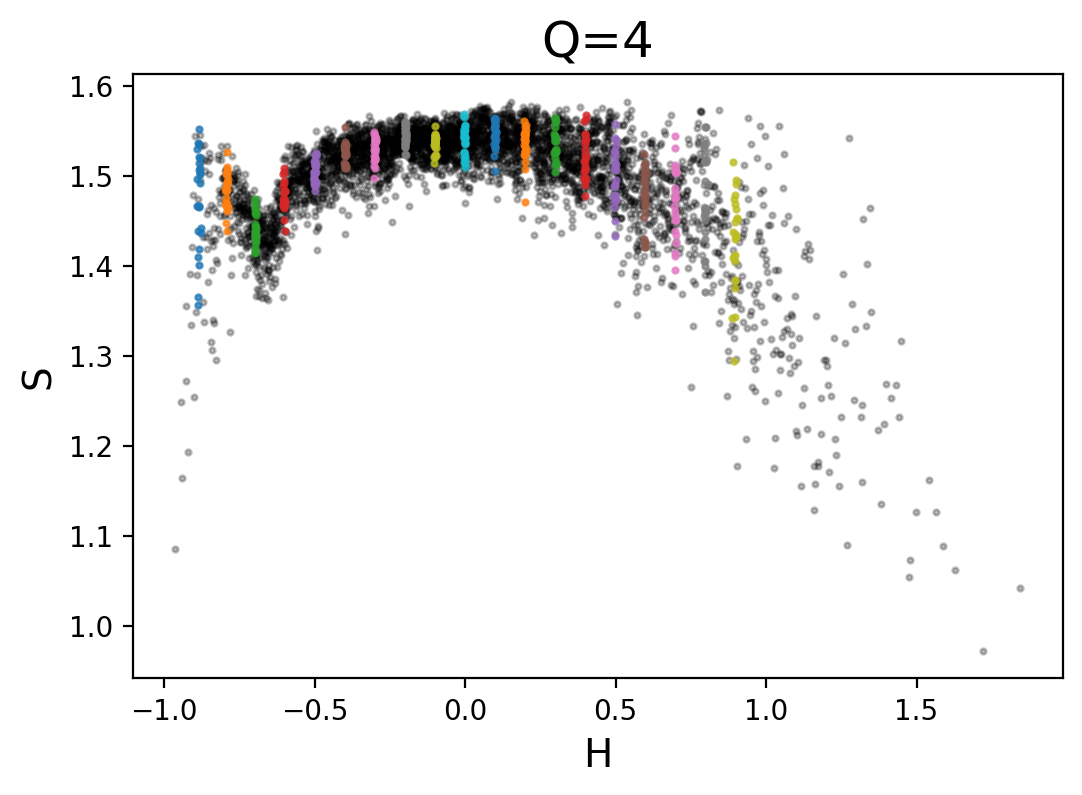}
	\end{subfigure}
	\begin{subfigure}{.32\textwidth}\centering
		\includegraphics[width=\linewidth]{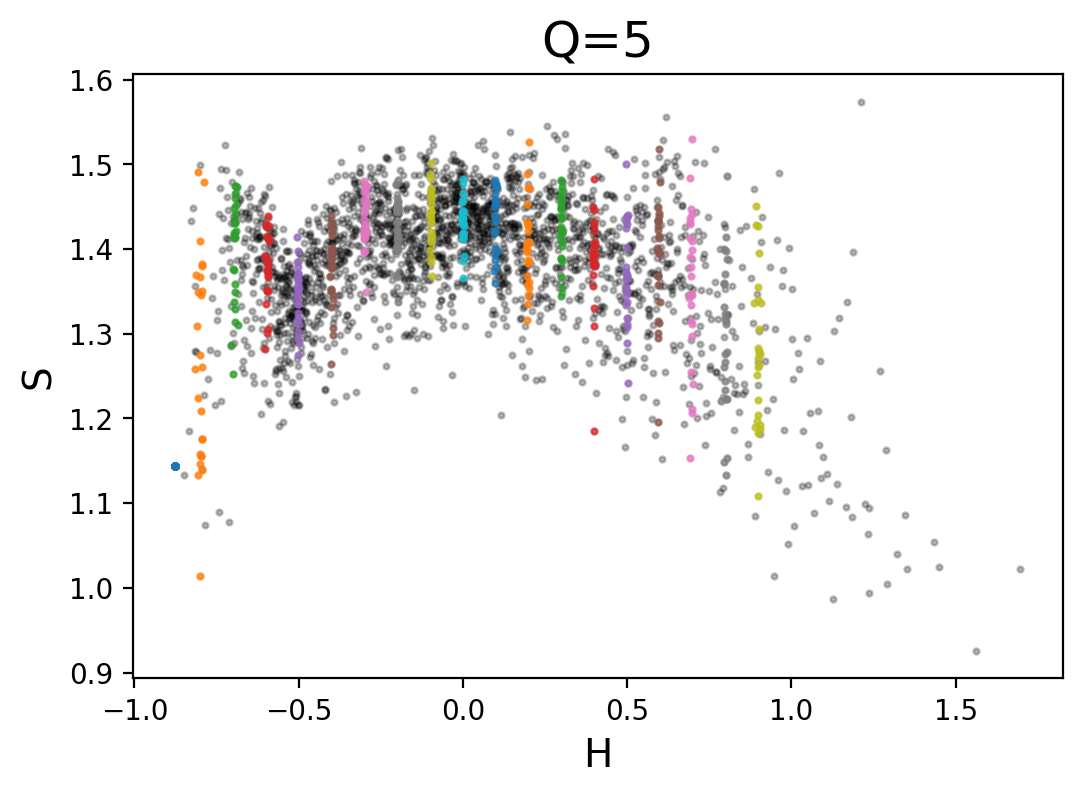}
	\end{subfigure}
    \begin{subfigure}{.32\textwidth}\centering
		\includegraphics[width=\linewidth]{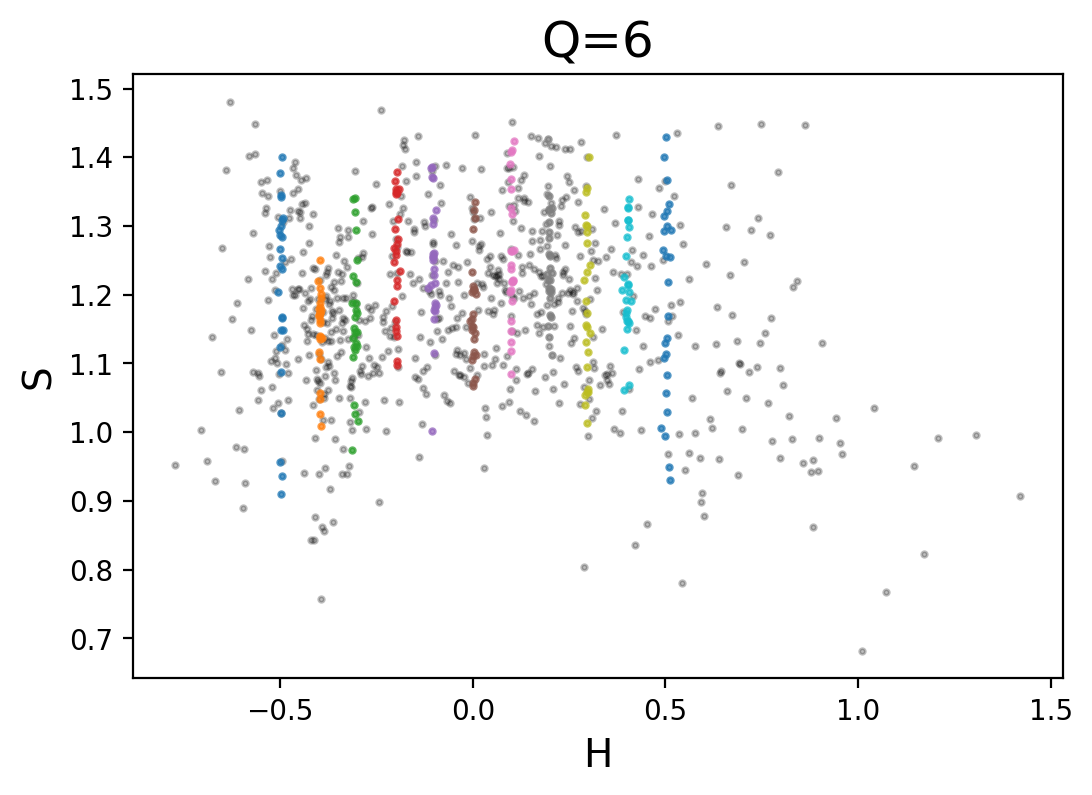}
	\end{subfigure}
	\begin{subfigure}{.32\textwidth}\centering
		\includegraphics[width=\linewidth]{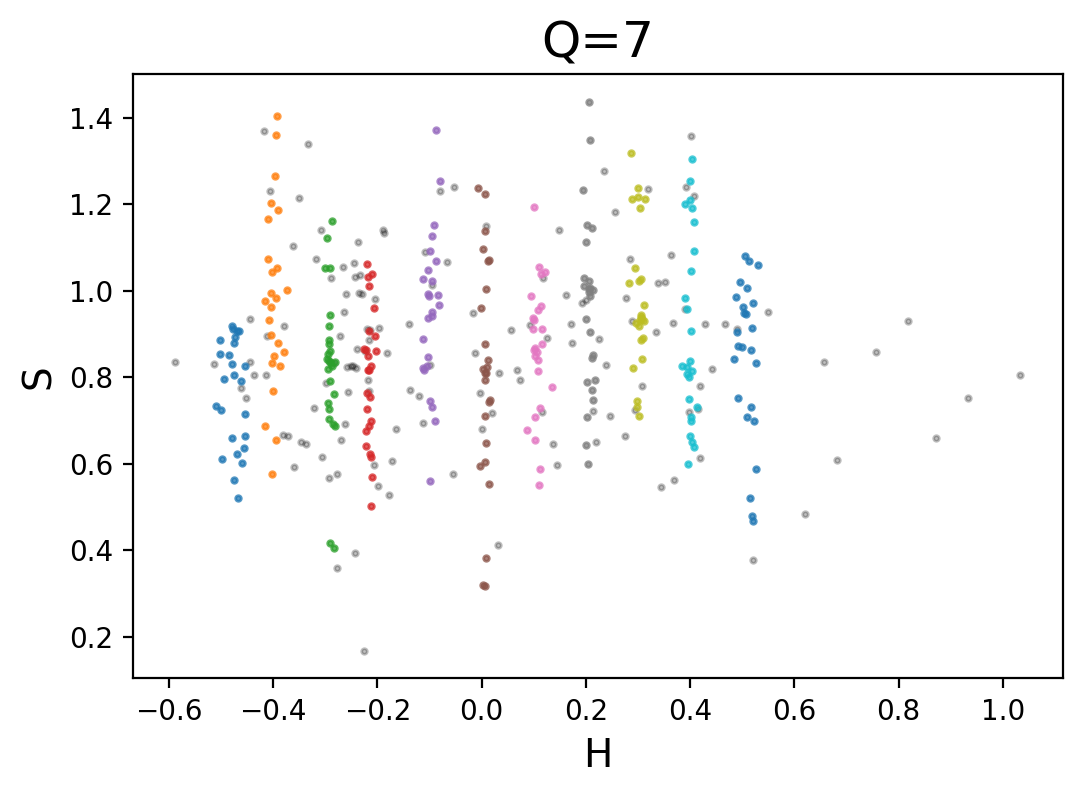}
	\end{subfigure}\hspace{.32 \linewidth}
	\caption{Each panel shows a comparison between the entropy of a single qutrit halfway down the lattice for energy eigenstates (marked by the black dots) within a fixed charge sector and for ``random microcanonical states'' (colored dots). Each vertical colored set represents the expectation value for the states that are built from a specified energy window in a fixed sector of $Q$. By construction, these expectation values lie within the range of the expectation values of the charge eigenstates themselves. }
	\label{fig: microcanonical S window}
\end{figure}

Figure \ref{fig: microcanonical S window} shows a comparison between the central qutrit entanglement entropies in random microcanonical states to the eigenstates themselves.  The black dots represent individual eigenstates.  These are most densely concentrated around $H=0$, and for values of $Q$ closer to $1/3$ of the maximal value of $Q$. This is particularly apparent in the $Q=3$ plot. In the charge sectors with the largest state space ($Q=2,3,4$) the entropies for individual eigenstates group together in a manner that can be approximated by a smooth function of energy, plus fluctuations that increase in size toward the edges of the spectrum.  By contrast, there is hardly any such discernible pattern for the sectors of smallest size ($Q=6,7$). The colorful dots are random microcanonical states, built within a fixed energy window $E\pm \Delta E$, where the values of $E$ are incremented by 0.1 and $\Delta E$ = 0.05. Each set of similarly-colored points denotes a different value of $\langle H\rangle$ while each dot represents a single generic state for a given $\langle H \rangle$ and definite $Q$. While the line for each energy window may appear to be vertical, there are small deviations in a given state energy within each energy window. This is most clear for higher values of $Q$ where there were fewer states in the window to contribute to the superposition.  These states do not correspond to the eigenstates and thus change under time evolution.  When they are evolved under the qutrit Hamiltonian the individual entropies fluctuate, but the distribution of points around the eigenstate entropies stays nearly fixed. This spread therefore represents time-dependent fluctuations that are better suppressed wherever the eigenstates better conform to ETH.  For each charge sector the entropies never cross the maximal value of entropy  $S_\text{qutrit, max} = \log_2(3) \approx 1.58$. We see that for smaller charge sectors the range of entropy values uniformly decreases in comparison to neighboring, larger charge sectors. For $Q=0$ the maximal entropy is 1, this is because states with no charge live in a qubit subsystem of the qutrit Hamiltonian, so the maximal entropy is $S_\text{qubit, max} = \log_2(2) = 1$.

\begin{figure}[H]\centering
	\begin{subfigure}{.32\textwidth}\centering
		\includegraphics[width=\linewidth]{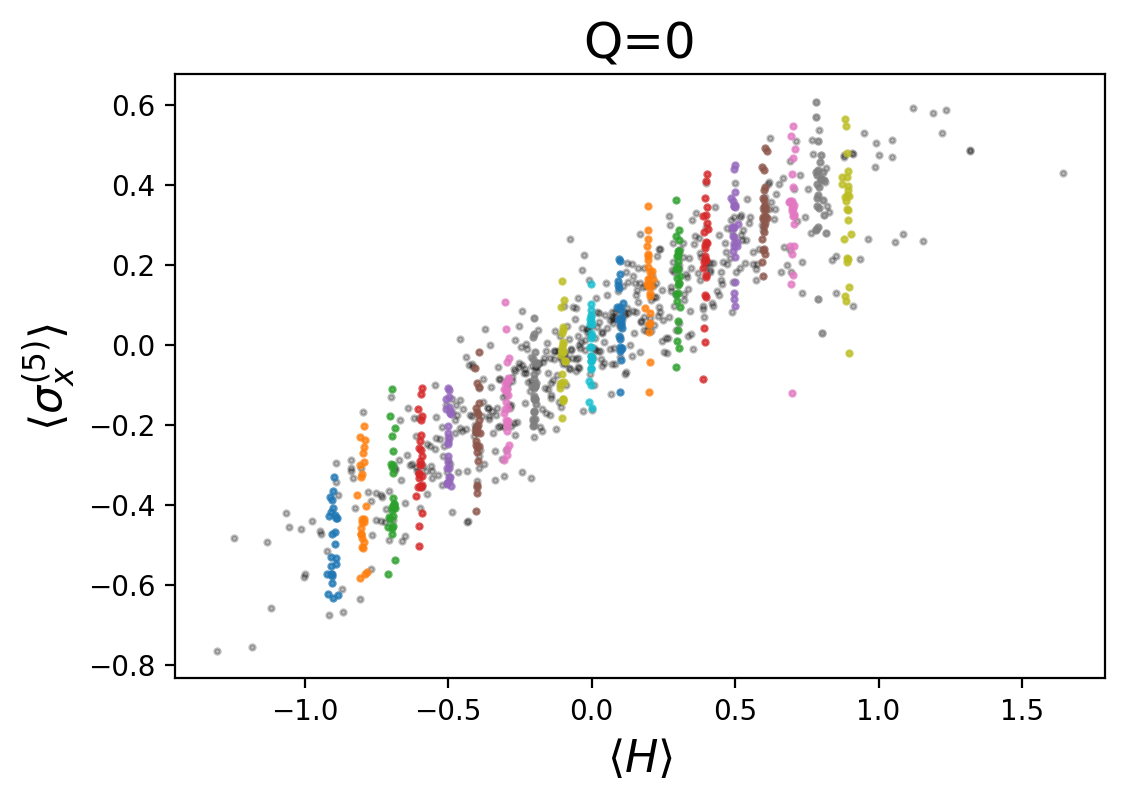}
	\end{subfigure}
	\begin{subfigure}{.32\textwidth}\centering
		\includegraphics[width=\linewidth]{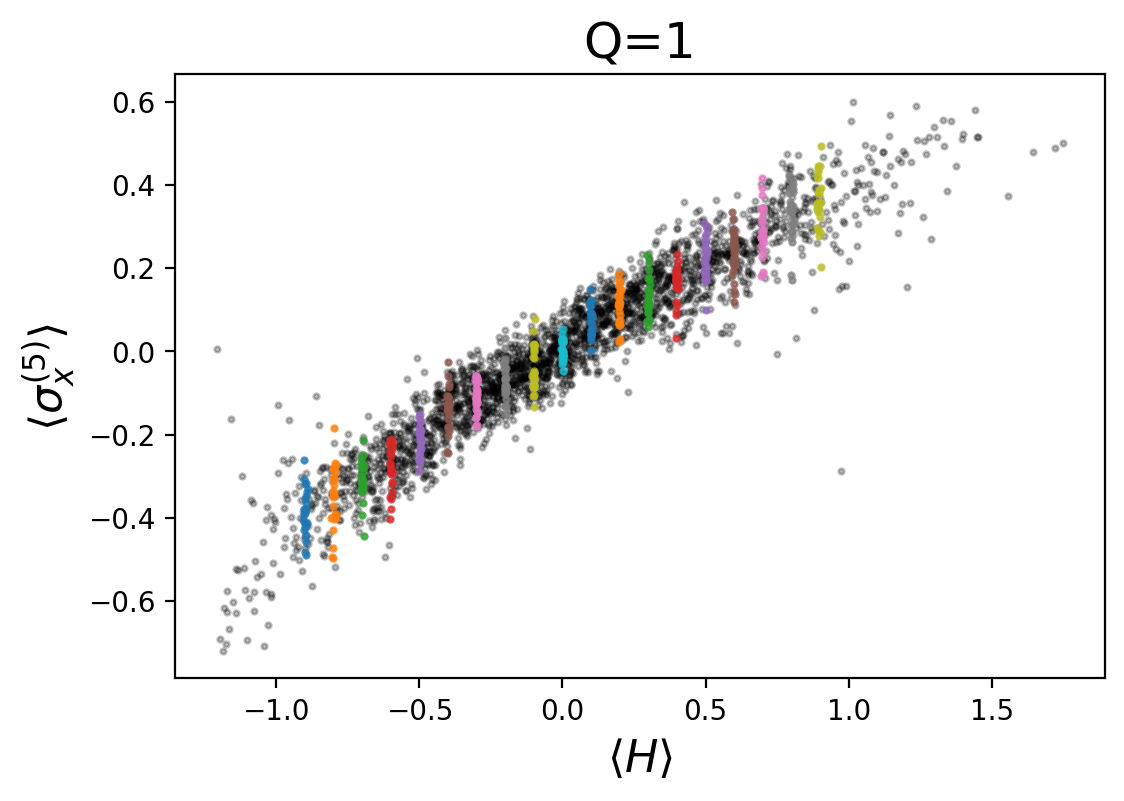}
	\end{subfigure}
	\begin{subfigure}{.32\textwidth}\centering
		\includegraphics[width=\linewidth]{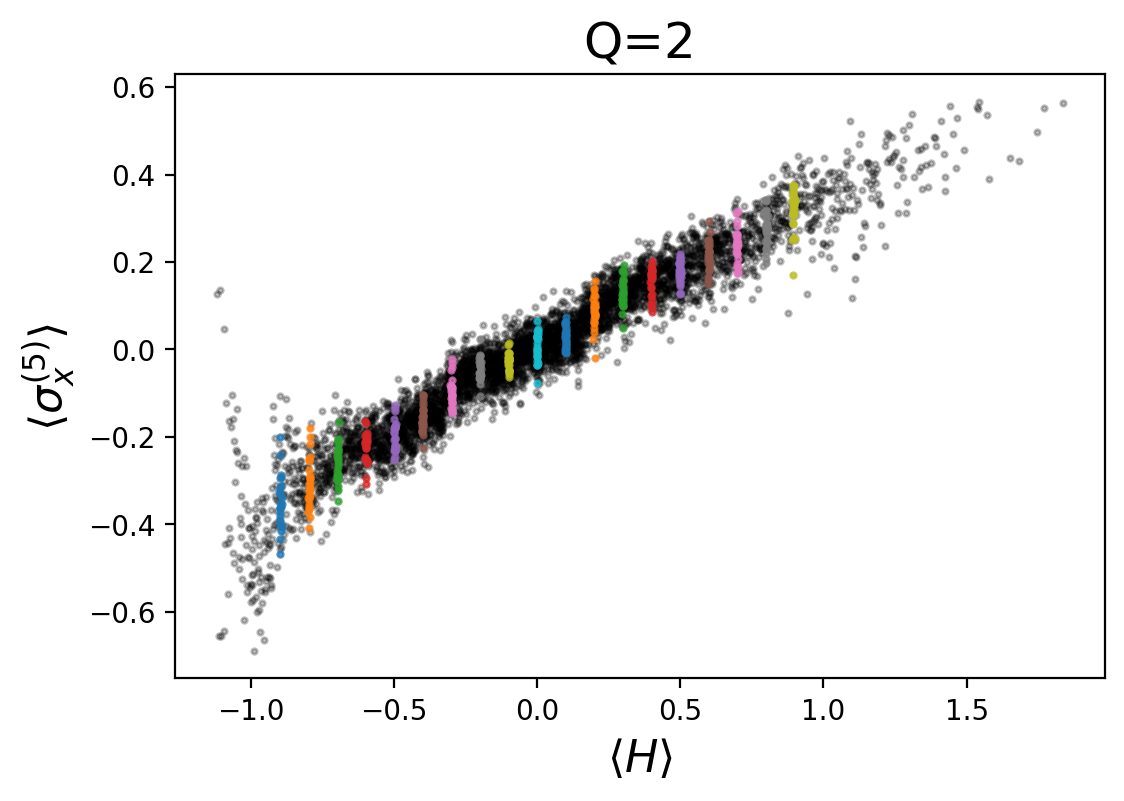}
	\end{subfigure}
    \begin{subfigure}{.32\textwidth}\centering
		\includegraphics[width=\linewidth]{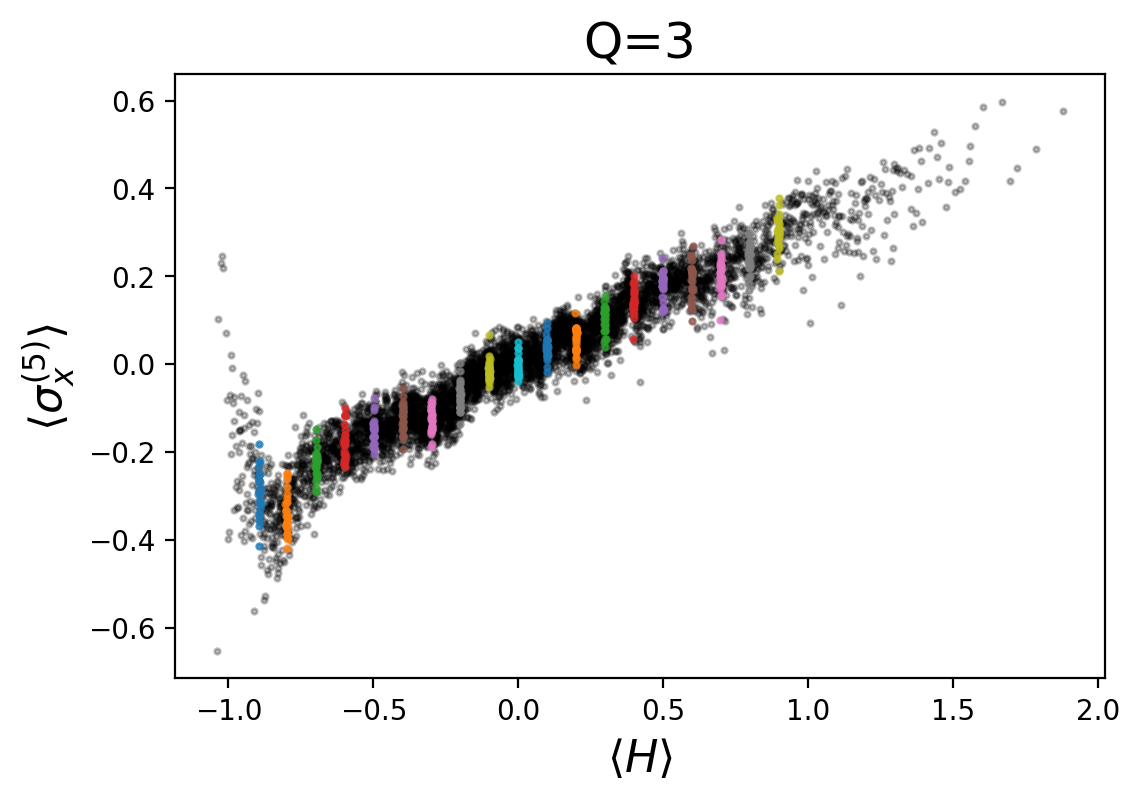}
	\end{subfigure}
	\begin{subfigure}{.32\textwidth}\centering
		\includegraphics[width=\linewidth]{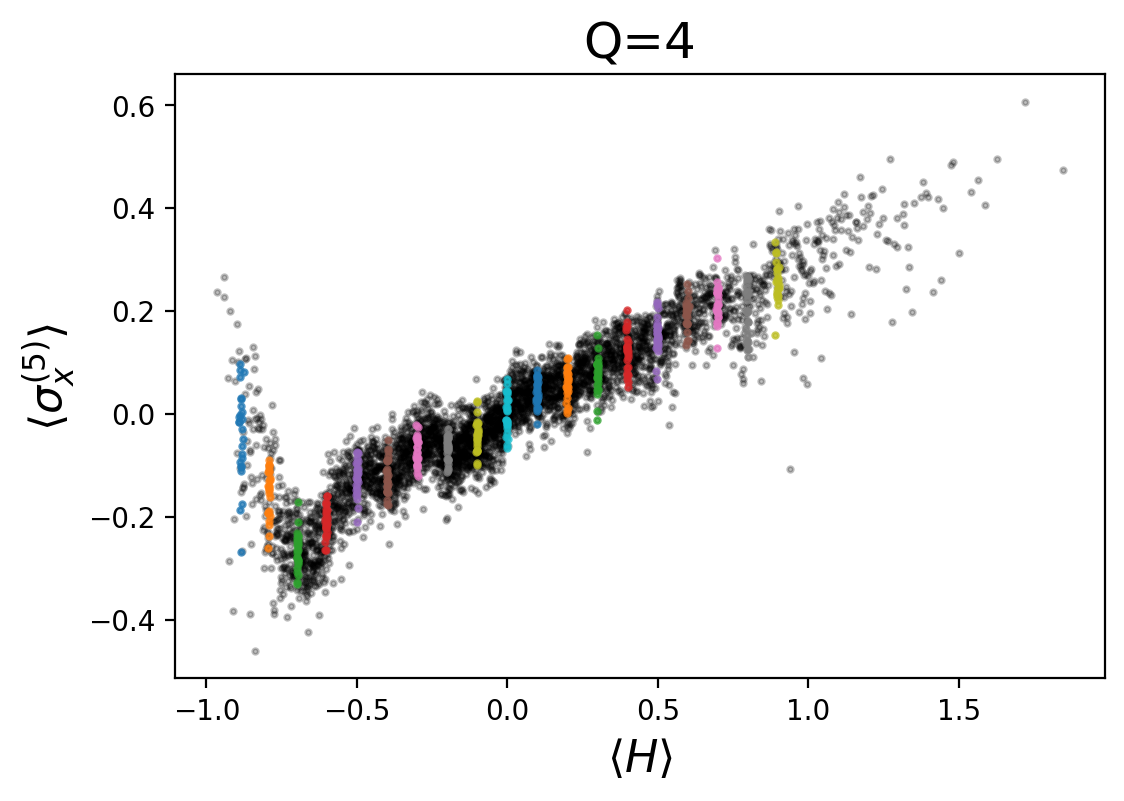}
	\end{subfigure}
	\begin{subfigure}{.32\textwidth}\centering
		\includegraphics[width=\linewidth]{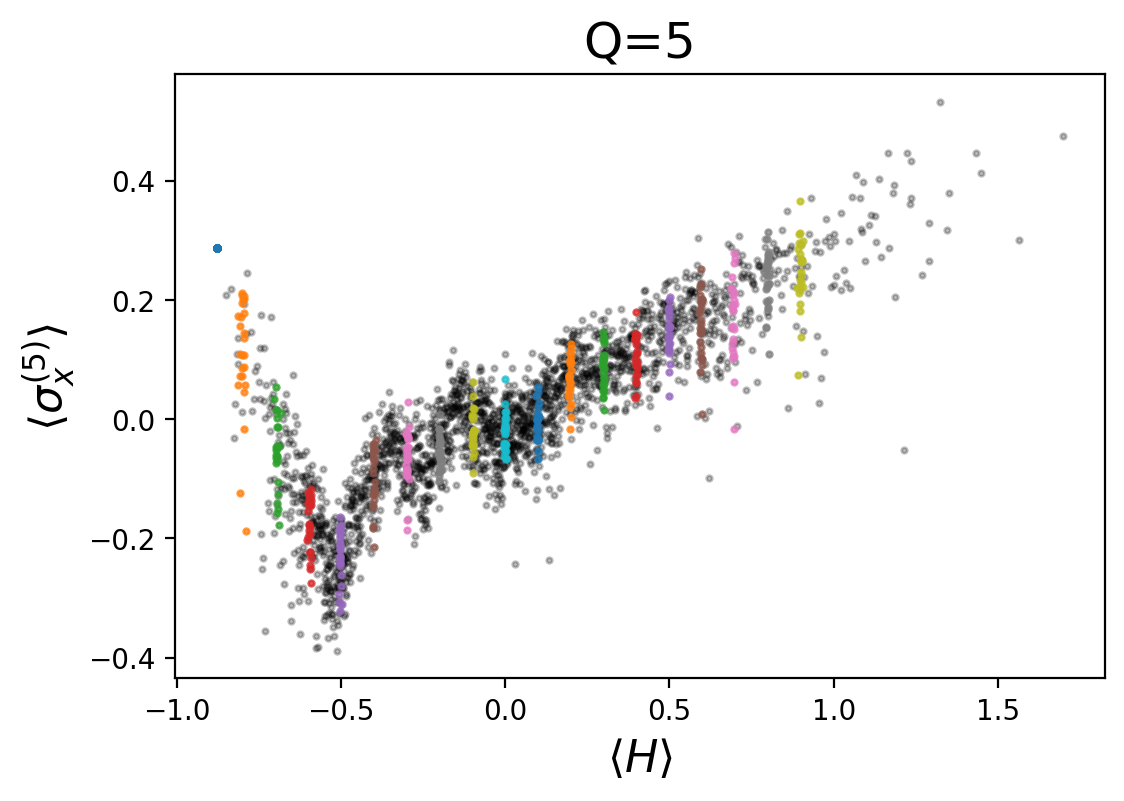}
	\end{subfigure}
    \begin{subfigure}{.32\textwidth}\centering
		\includegraphics[width=\linewidth]{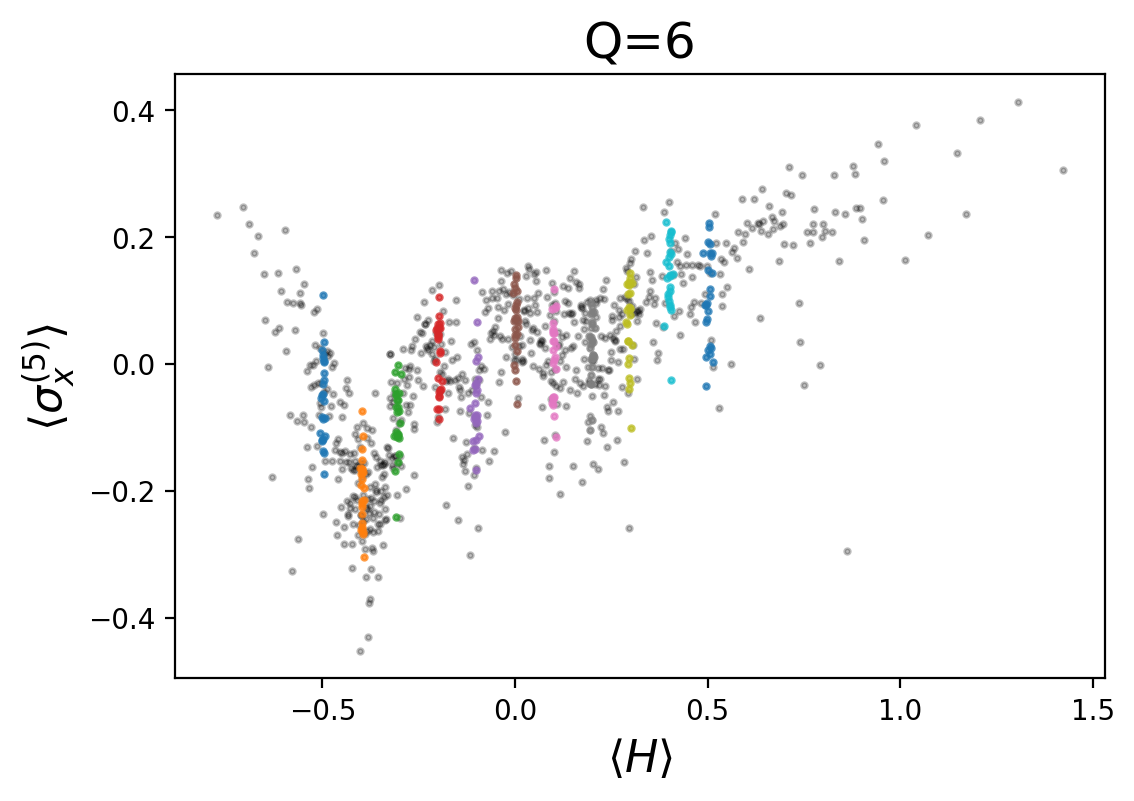}
	\end{subfigure}
	\begin{subfigure}{.32\textwidth}\centering
		\includegraphics[width=\linewidth]{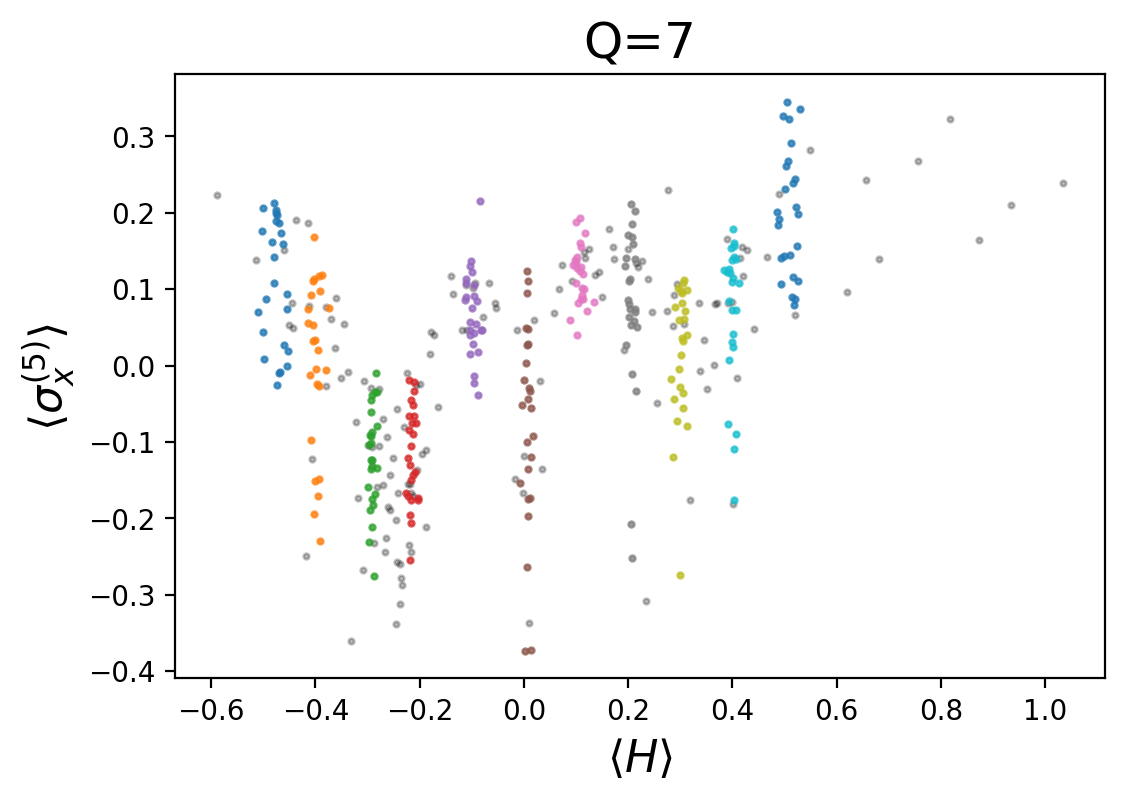}
	\end{subfigure}\hspace{.32 \linewidth}
	\caption{Each panel shows a comparison between the expectation value of a single Pauli operator $\langle \sigma_x ^{(5)}\rangle$ that acts on the qubit subsystem of the qutrit halfway down the lattice in energies eigenstates (marked by the black dots) within a fixed charge sector and for ``random microcanonical states'' (colored dots) in a fixed energy and charge window. Each vertical colored set represents the expectation value for the states that are built from a specified energy window for a fixed value of $Q$. By construction, these expectation values lie within the range of the expectation values of the eigenstates. }
	\label{fig: microcanonical pauli x window}
\end{figure}

In figure \ref{fig: microcanonical pauli x window}, we display the expectation value of a single $\lambda_1$ operator on the $5^\text{th}$ lattice site (which acts as the Pauli $\sigma_x$ operator on the qubit subsystem).  We see again that each colored band follows the expected behavior predicted by the eigenstates.  The eigenstates show expectation values clumped together, leading to the emergence of a smooth function of energy, particularly in sectors where the density of states is largest.  We do however still see a wider spread in possible expectation values for different values of $Q$. The range of $\langle \sigma^{(5)}_x \rangle$ decreases as $Q$ increases toward maximum. This is because there is less `support' on the qubit subsystem, decreasing the maximal expectation value, as more charge is added to the system. Note that the energies of the Hamiltonian have been normalized by the system size $L$, which squeezes the energy range to be order 1. 

\begin{figure}[H]\centering
	\begin{subfigure}{.32\textwidth}\centering
		\includegraphics[width=\linewidth]{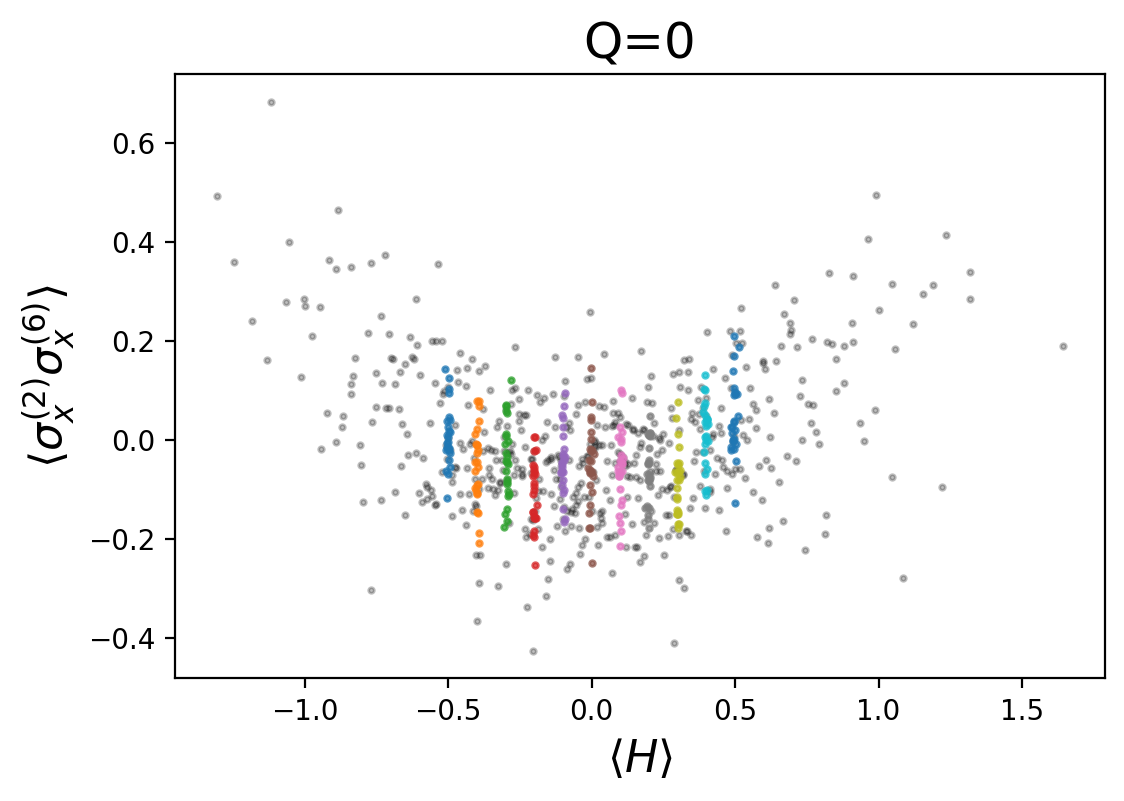}
	\end{subfigure}
	\begin{subfigure}{.32\textwidth}\centering
		\includegraphics[width=\linewidth]{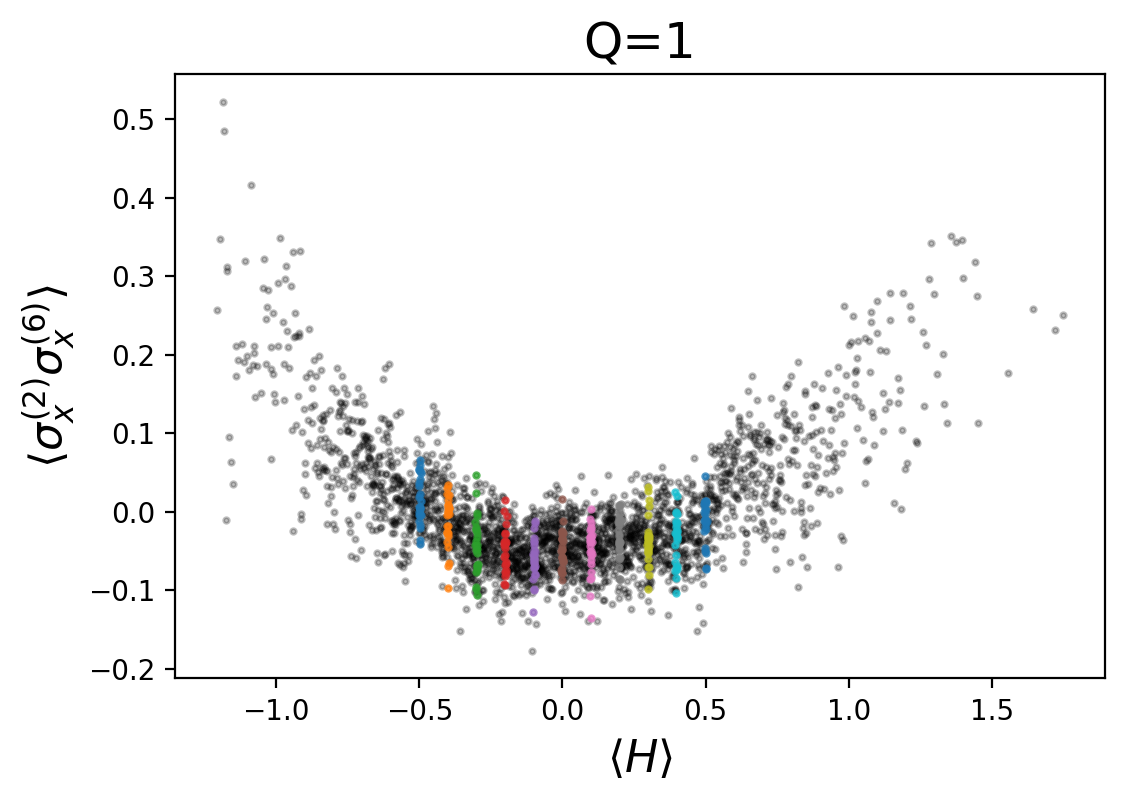}
	\end{subfigure}
	\begin{subfigure}{.32\textwidth}\centering
		\includegraphics[width=\linewidth]{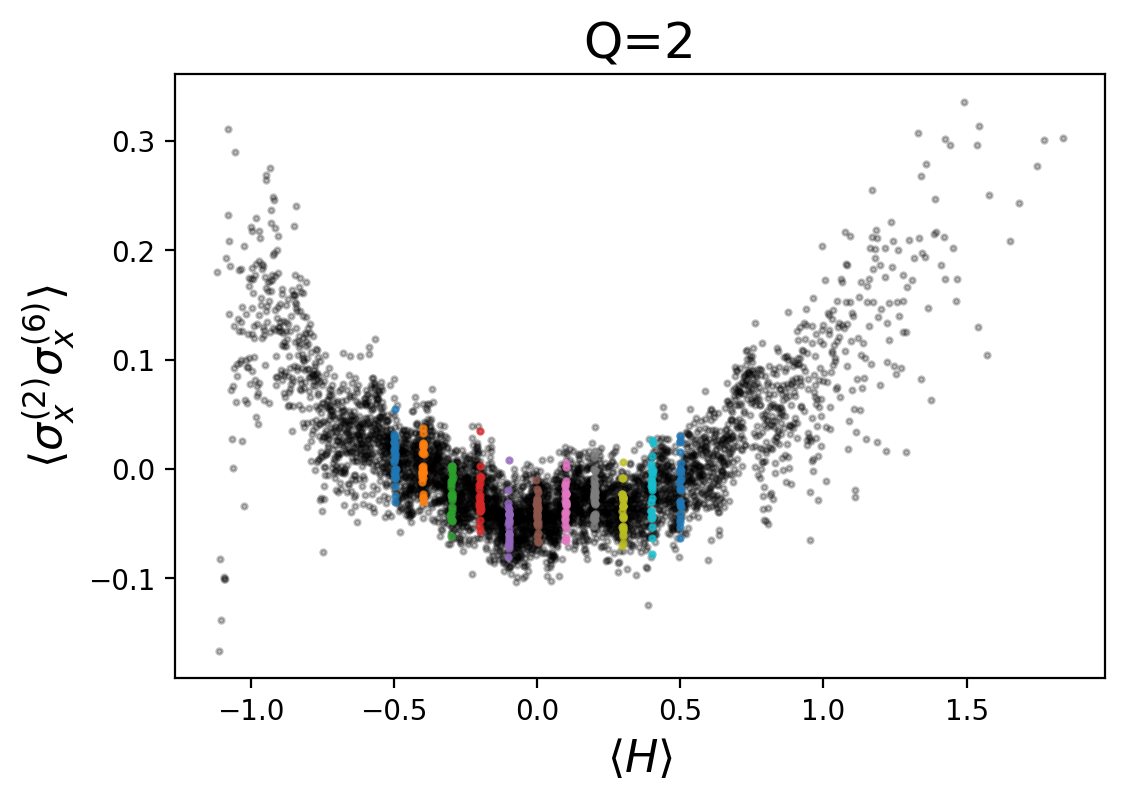}
	\end{subfigure}
    \begin{subfigure}{.32\textwidth}\centering
		\includegraphics[width=\linewidth]{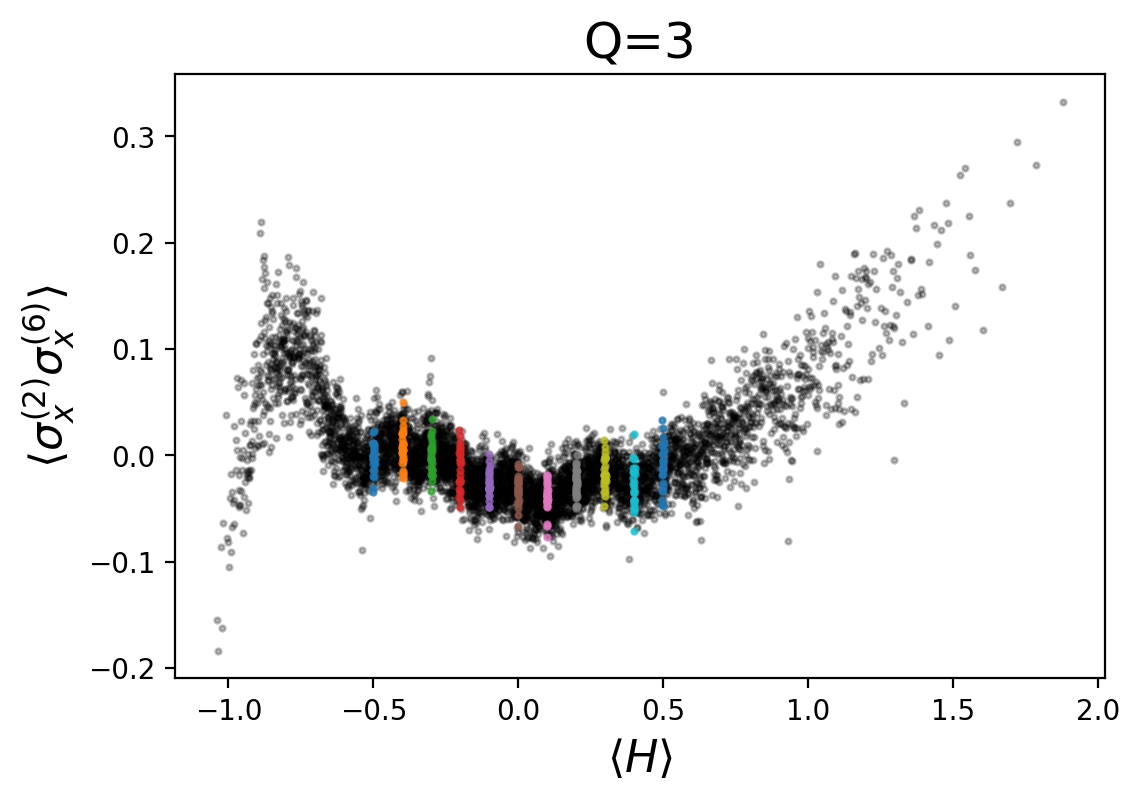}
	\end{subfigure}
	\begin{subfigure}{.32\textwidth}\centering
		\includegraphics[width=\linewidth]{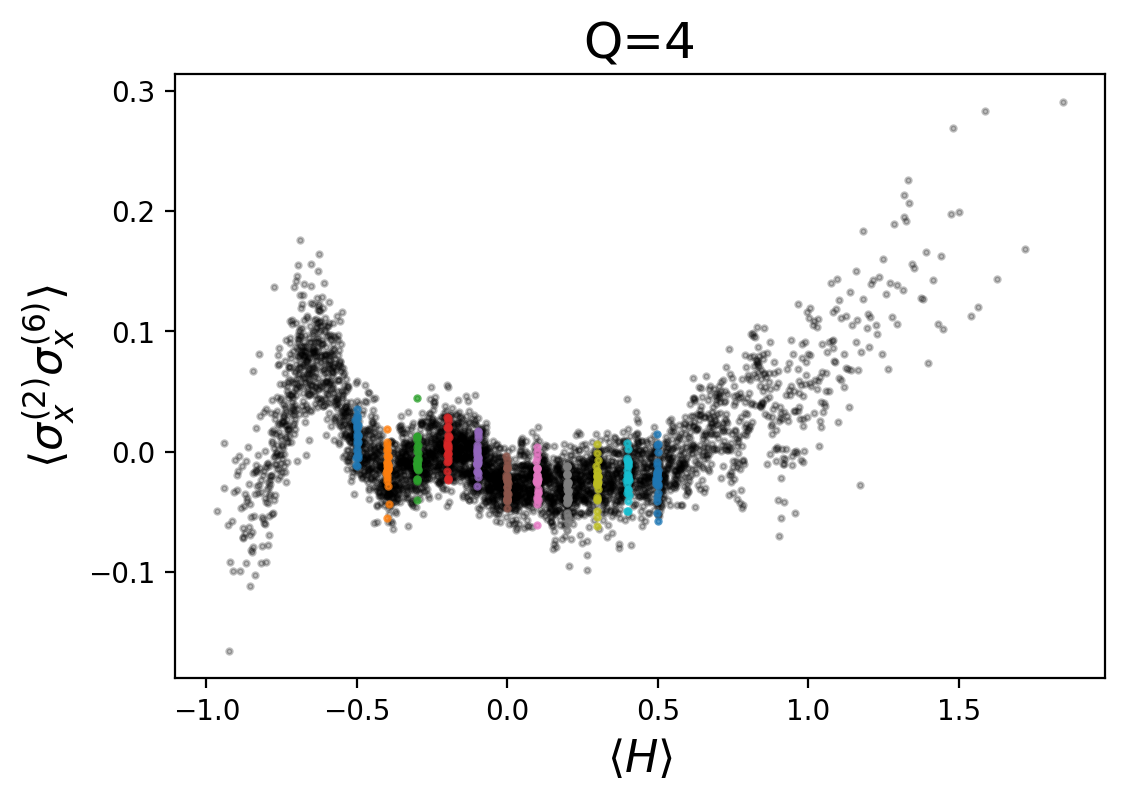}
	\end{subfigure}
	\begin{subfigure}{.32\textwidth}\centering
		\includegraphics[width=\linewidth]{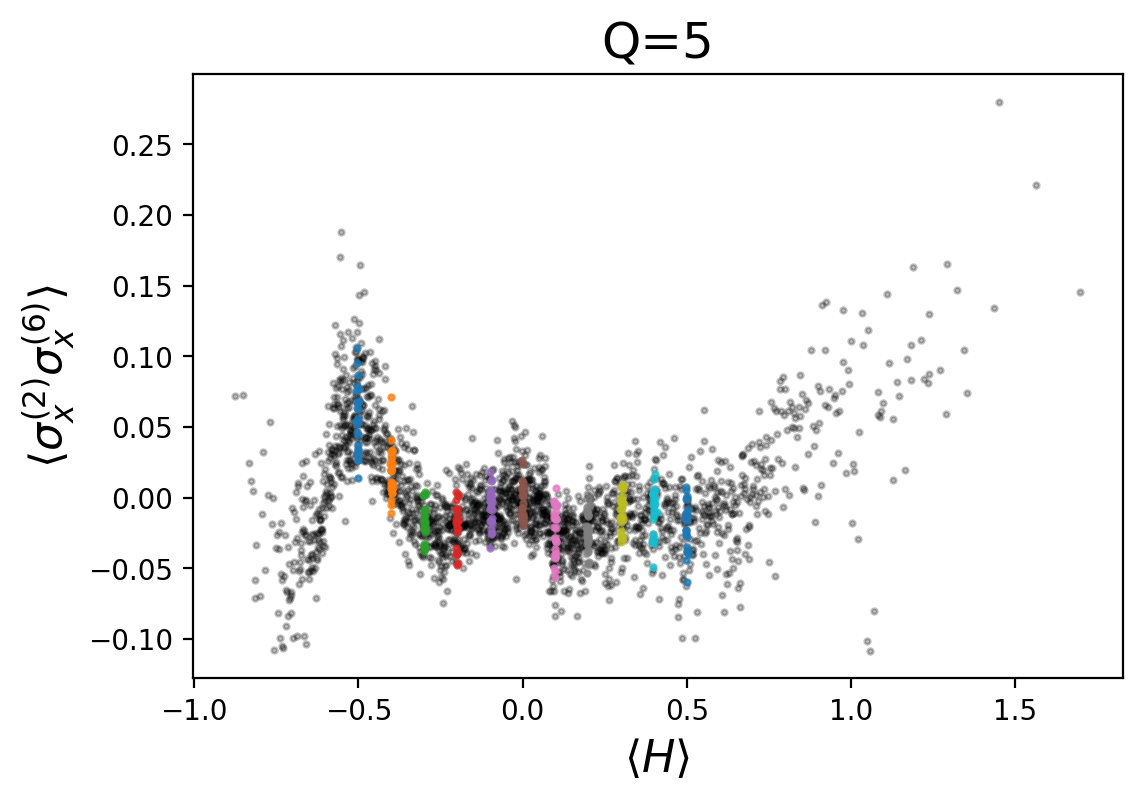}
	\end{subfigure}
    \begin{subfigure}{.32\textwidth}\centering
		\includegraphics[width=\linewidth]{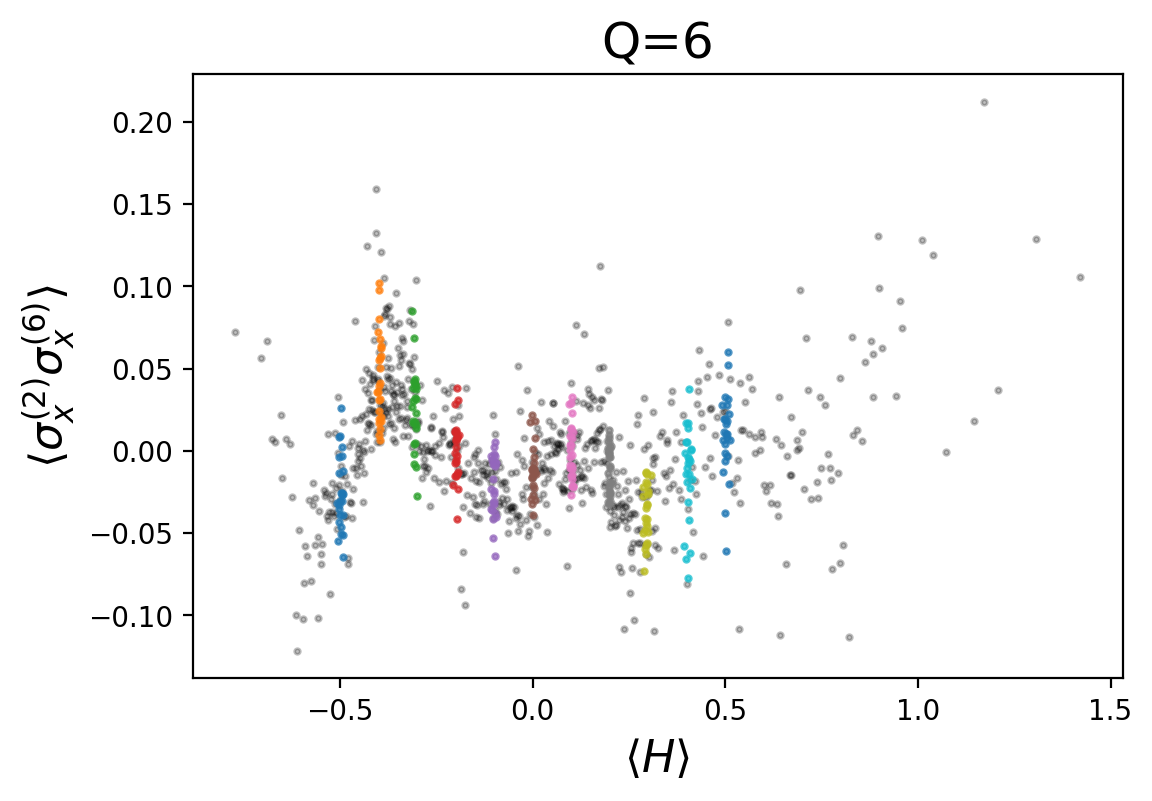}
	\end{subfigure}
	\begin{subfigure}{.32\textwidth}\centering
		\includegraphics[width=\linewidth]{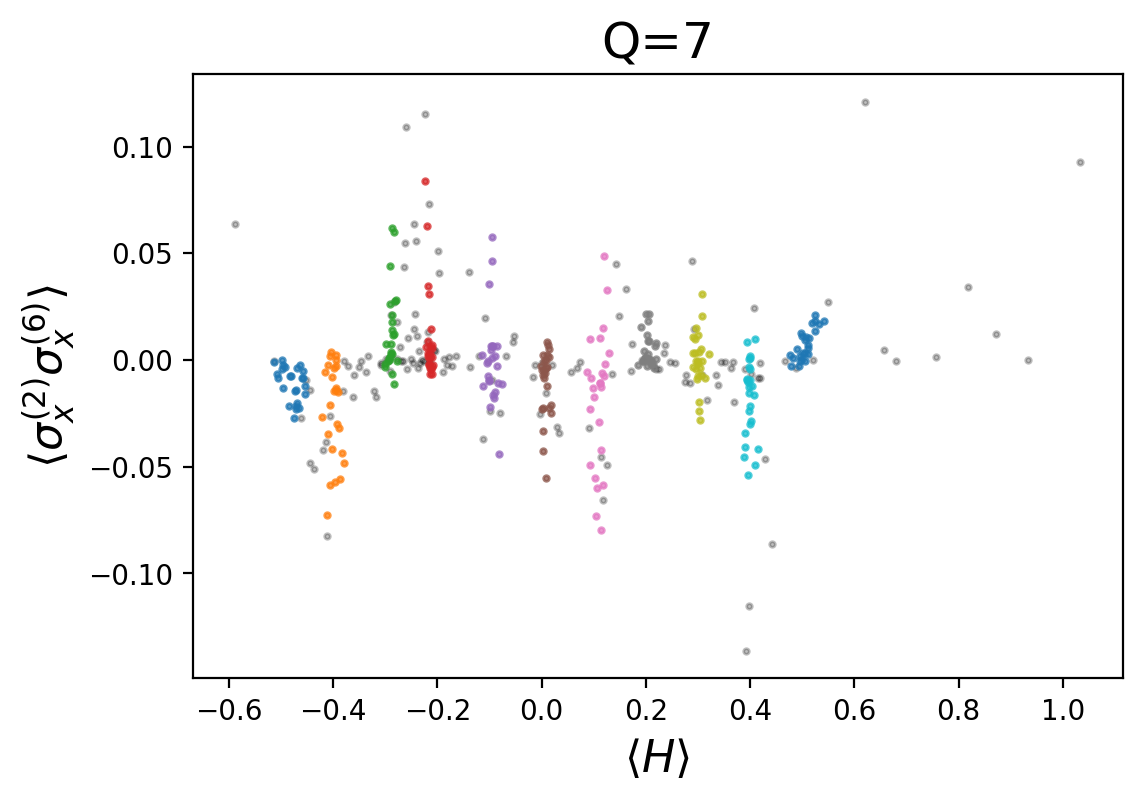}
	\end{subfigure}\hspace{.32 \linewidth}
	\caption{Each panel shows a comparison between the expectation value of a non-local product of  Pauli operators $\langle \sigma_x^{(2)} \sigma_x^{(6)} \rangle$ in energy eigenstates (marked by the black dots) within a fixed charge sector and for ``random microcanonical states'' (colored dots). Each vertical colored line represents the expectation value for the states that are built from a specified energy window for a fixed value of $Q$. By construction, these expectation values lie within the range of the expectation values of the eigenstates.}
	\label{fig: microcanonical pauli non loc x window}
\end{figure}

Figure \ref{fig: microcanonical pauli non loc x window} shows the expectation values for a non-local operator $\sigma_x^{(2)} *\sigma_x^{(6)}$ constructed from two Pauli operators acting on the qubit subsystem.  We see that even such non-local operators follow the expected pattern, with the emergence of a smooth thermal distribution. These non-local operators can thus still be thought of as `simple', hence the applicability of ETH.  In particular, this operator is nonlocal by virtue of connecting non-neighboring sites on the lattice, but it is still `$k$-local' with $k=2$ since it acts nontrivially on exactly two sites.  Both notions of locality might be expected to be relevant to the ETH form of operators, depending on the Hamiltonian in question. We do however begin to see a breakdown of ETH as the expectation values of these operators have higher variance than the single site operator, and the emergent smooth function of energy is seen to have more features.

Note that for all plots displayed in this section, in charge sectors of sufficiently large dimension, the onset of ETH is apparent through the grouping of local observables in eigenstates into a smooth function of energy, plus fluctuations which only decrease in magnitude as the system size increases.  If we compare these emergent smooth function across neighboring charge sectors, we see the emergence of smooth multivariate functions of $\langle H\rangle$ and $\langle Q\rangle$. We anticipate that for much larger systems, states drawn from a microcanonical window of both energy and charge (perhaps several neighboring charge sectors) would well thermalize to microcanonical predictions, providing a clean example of \textit{generalized} eigenstate thermalization. In the next section we will consider a natural set of states that have rather broad support across charge sectors.  We find that thermalization of these states actually improves, in the sense of having smaller temporal fluctuations, over the microcanonical states in this section as a result of this broadened support.  We take this to be an example of `generic ETH' as explained in section \ref{subsec: generic ETH in theory}.

\subsection{`Generic ETH'}
\label{subsec: generic ETH in Qutrit system}

We now turn our attention to the behavior of states outside of charge sectors. We construct special out-of-equilibrium states and study their dynamics as they evolve. Most states belonging to a specified charge sector and fixed energy window will be highly entangled. Finding an unentangled state under such considerations is akin to finding a `needle in a haystack'. Intuitively this difficulty in finding low entanglement states can be understood through an equipartition argument and most states in a Hilbert space being entangled. To study full growth and saturation of entanglement entropy in states that evolve under the qutrit Hamiltonian we construct completely unentangled states by tensoring states on individual lattice sites. We can thereby construct unentangled states which are arbitrary with one caveat; the states can be tuned to a specific charge expectation value.
We will see that these states have a broad spread in both energy and charge, but they still appear to thermalize.  We dub behavior of these states `generic eigenstate thermalization' for the qutrit system, since we have loosened the restrictions typically assumed of states following generalized ETH.

The states we consider are constructed as follows: a randomly oriented qubit is embedded into a larger Hilbert space to construct a qutrit.\footnote{We use the following orthonormal bases for qutrits
\begin{equation*}
\ket{0}= \left(\begin{array}{c} 1 \\ 0 \\0 \end{array}\right)
\quad
\ket{1} = \left(\begin{array}{c} 0 \\ 1 \\0 \end{array}\right)
\quad
\ket{2} = \left(\begin{array}{c} 0 \\ 0 \\1 \end{array}\right)
\end{equation*} 
The states in this section and all subsequent dynamics were done using the Quantum Toolbox in Python (QuTiP) \cite{Johansson_2013}.} The states are defined with a continuous parameter $f$, which determines the expectation value of the local charge operator, $\langle \hat{q}\rangle = f$. 
The state describing a single lattice site is thus given by 
\begin{equation}
\begin{aligned}
\ket{\psi_f} &= \sqrt{1-f} \left(\cos \theta \ket{0} + e^{i \phi_1} \sin \theta \ket{1}\right) +e^{i\phi_2} \sqrt{f} \ket{2}, 
\end{aligned}
\end{equation}
where $0\leq\theta\leq\pi/2$ and $ 0\leq\phi_{1,2}<2\pi$. The parameter $f$ ranges from $0$ to $1$ since the maximum charge on a single lattice site is 1. The angles $\theta$ and $\phi$ are drawn from a uniform random distribution for each lattice site. Unentangled, and as such out-of-equilibrium, states are created by taking a tensor product of these states over single lattice sites.  The energy distribution of such states will be broad, tending to center on zero except insofar as we will post-select for specific $\langle H \rangle$ (analogous to the unentangled states considered in the qubit system, see figure \ref{fig: energy distribution of states}).  The net charge expectation value, the conserved charge of the system $\langle Q\rangle $, is the sum of the lattice charges $f$. We choose to initialize states with the net charge spread equally over each lattice site in the system. Other initialized charge distributions were tested, but no quantitative effects were found on the long-term dynamics of the observables studied in this section (in line with the expectation that charge distribution thermalizes to near uniformity anyway, see figure \ref{fig: charge spreading}). 
With $f$ the same on all sites, the total charge expectation value is precisely $\langle Q\rangle = f L$.  However the amount of support in each charge sector follows a binomial distribution: letting $\Pi_q$ represent the projector onto the charge sector $Q=q$ we have
\begin{equation}\label{eq: charge distribution}
    \langle \Pi_q \rangle = \frac{L!}{(L-q)!q!}f^q(1-f)^{L-q}
\end{equation}
for $q = 0, ..., L$.  For large systems, this is well approximated by a Gaussian distribution $\mathcal{N}\left(L f, Lf (1-f)\right)$.  Thus the width of the charge distribution grows as $\sqrt{L}$.\footnote{As a fraction of the total charge range, this amounts to a narrowing as $1/\sqrt{L}$, but regardless, for finite $L$ these states are strongly distinct from a narrow microcanonical window.}  The spread across charge sectors is thus quite broad, and it might be considered surprising that such states thermalize.  We have explained our understanding of this in section \ref{subsec: generic ETH in theory}, essentially employing the standard mechanism of (generalized) ETH, but pushed to broader circumstances.

The energies of these unentangled states can be understood as the the energy due to the random orientation of the qubit subsystem in the magnetic field weighted by the charge. By construction in the $\langle Q \rangle = 0$ case, the allowed energies of states are exactly those of the qubit subsystem since $f=0$. The contribution of that sector to $\langle H \rangle _{\ket{\psi}}$ decreases with the growth of $f$, however the part of the state built from $\ket{2}$ has by definition 0 energy since it only interacts with the charge spreading term, as such the unique state with $\langle Q \rangle = L$ has $\langle H \rangle = 0$. Thus we can see how the expectation values of energies are squeezed as we increase the charge expectation value as can be seen every figure. 

\begin{figure}[H]\centering
	\begin{subfigure}{.32\textwidth}\centering
		\includegraphics[width=\linewidth]{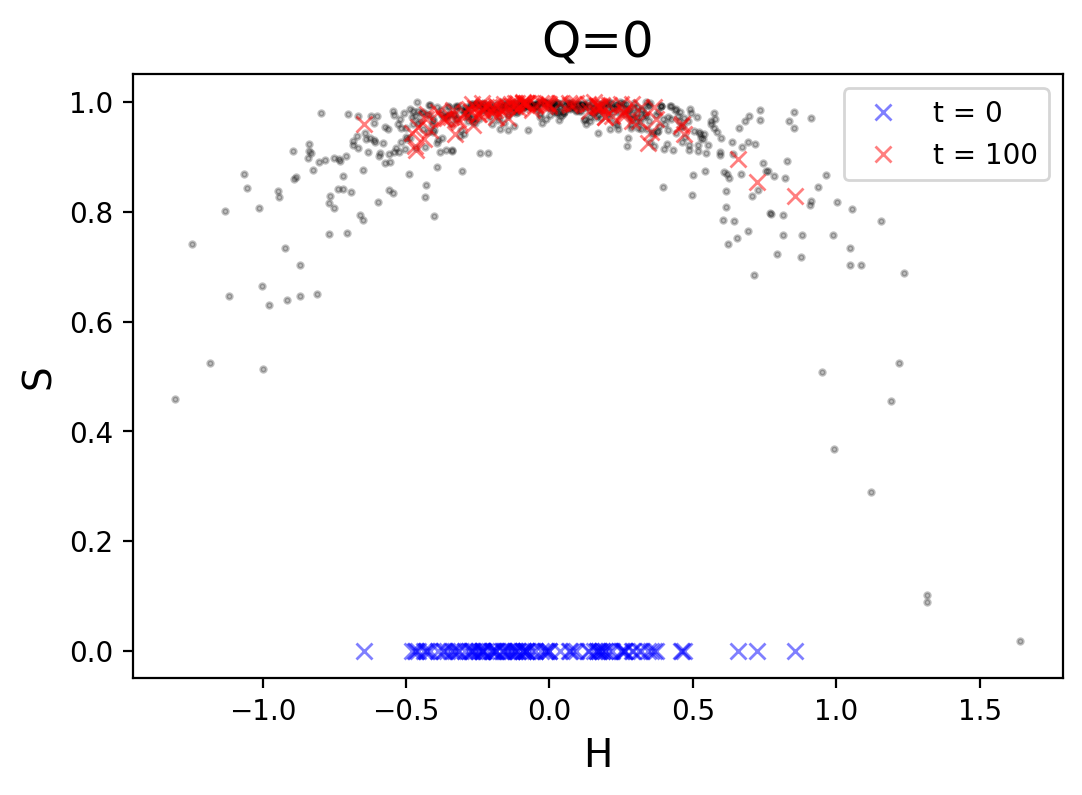}
	\end{subfigure}
	\begin{subfigure}{.32\textwidth}\centering
		\includegraphics[width=\linewidth]{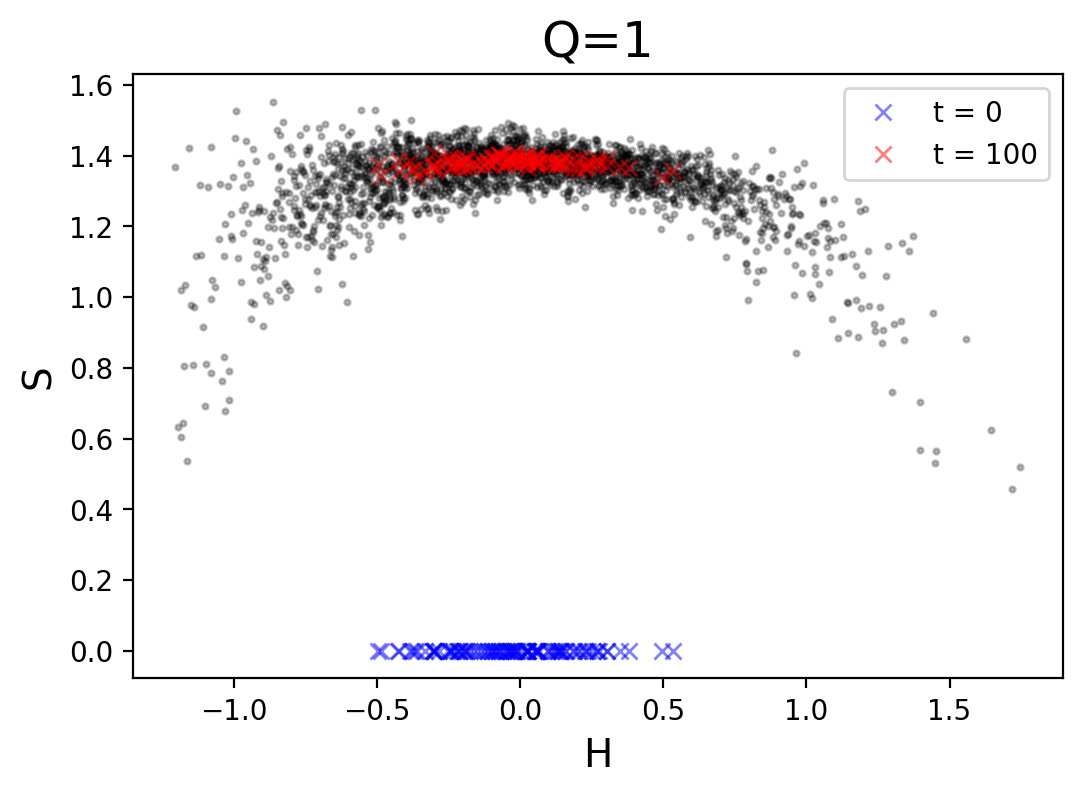}
	\end{subfigure}
	\begin{subfigure}{.32\textwidth}\centering
		\includegraphics[width=\linewidth]{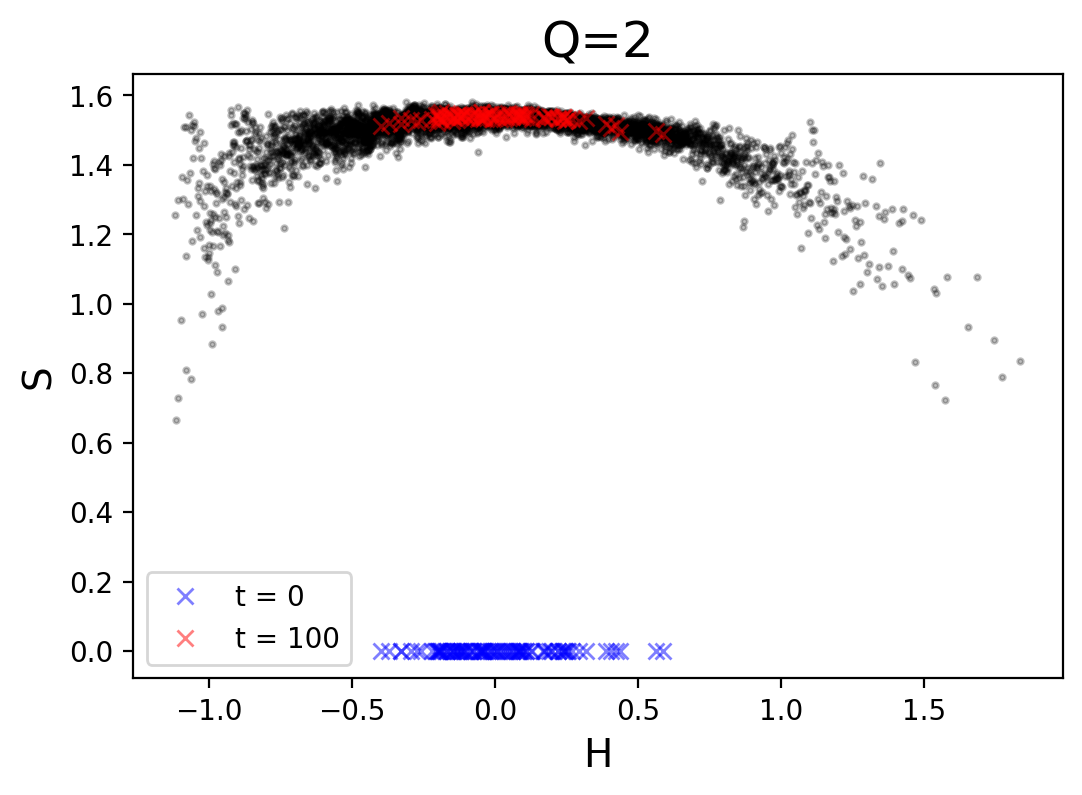}
	\end{subfigure}
    \begin{subfigure}{.32\textwidth}\centering
		\includegraphics[width=\linewidth]{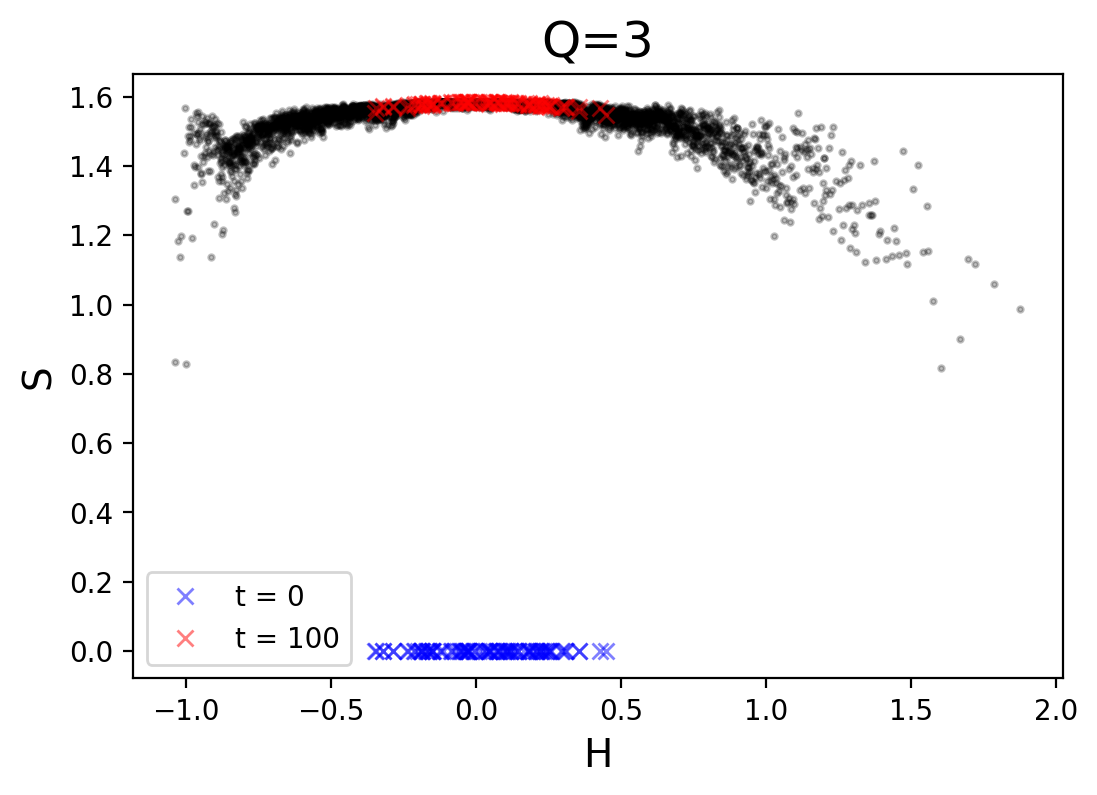}
	\end{subfigure}
	\begin{subfigure}{.32\textwidth}\centering
		\includegraphics[width=\linewidth]{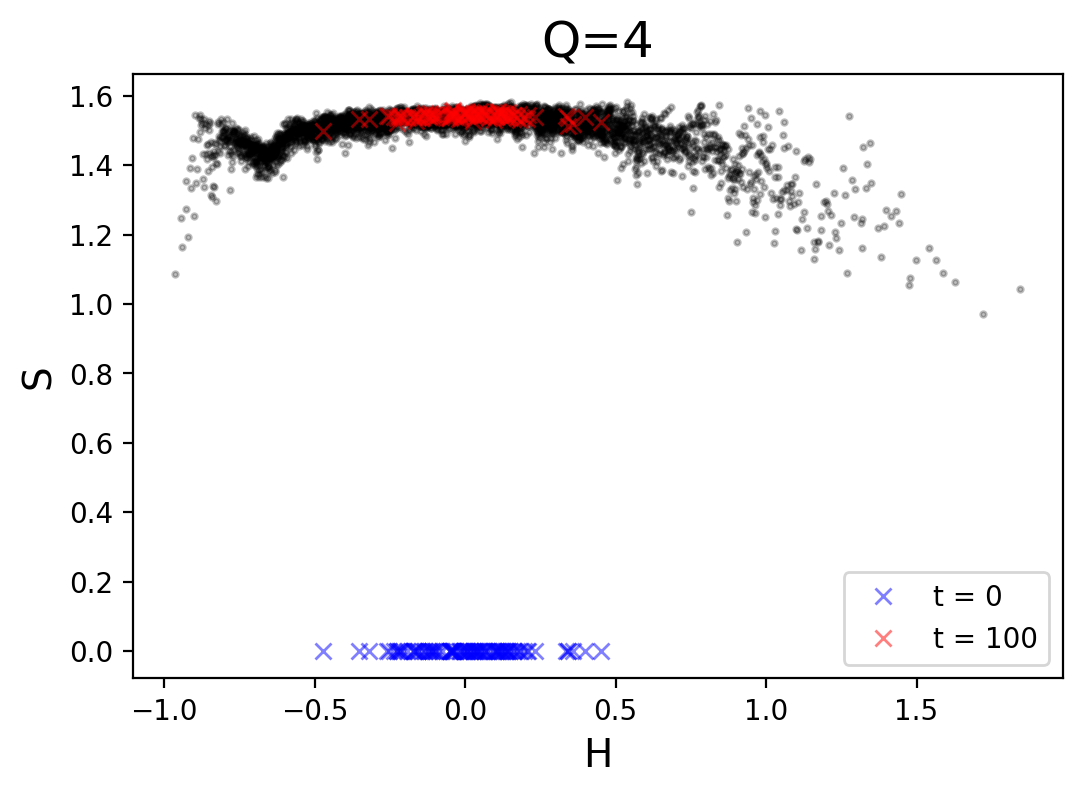}
	\end{subfigure}
	\begin{subfigure}{.32\textwidth}\centering
		\includegraphics[width=\linewidth]{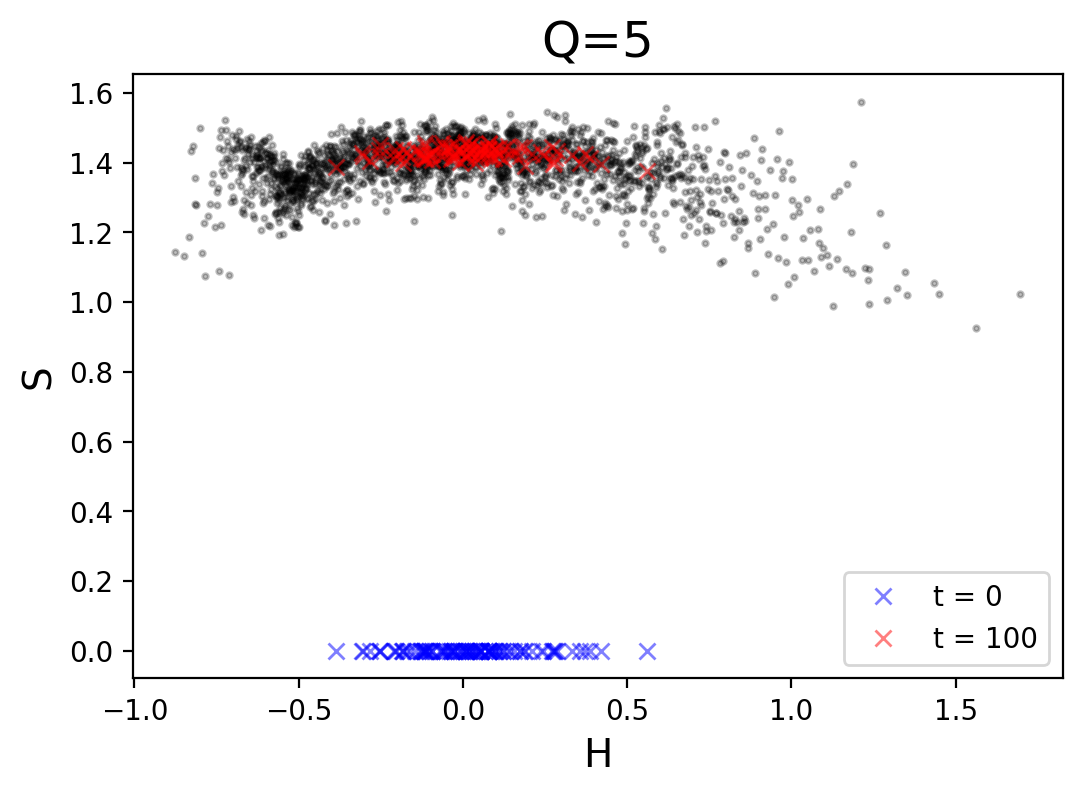}
	\end{subfigure}
    \begin{subfigure}{.32\textwidth}\centering
		\includegraphics[width=\linewidth]{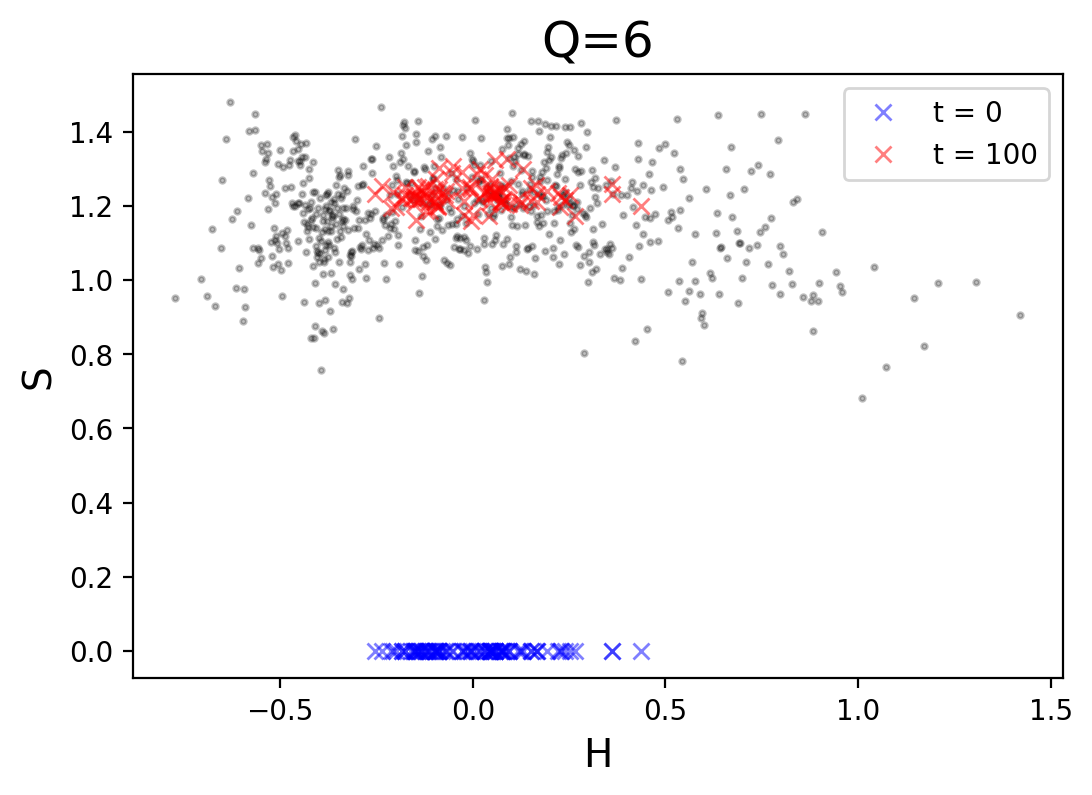}
	\end{subfigure}
	\begin{subfigure}{.32\textwidth}\centering
		\includegraphics[width=\linewidth]{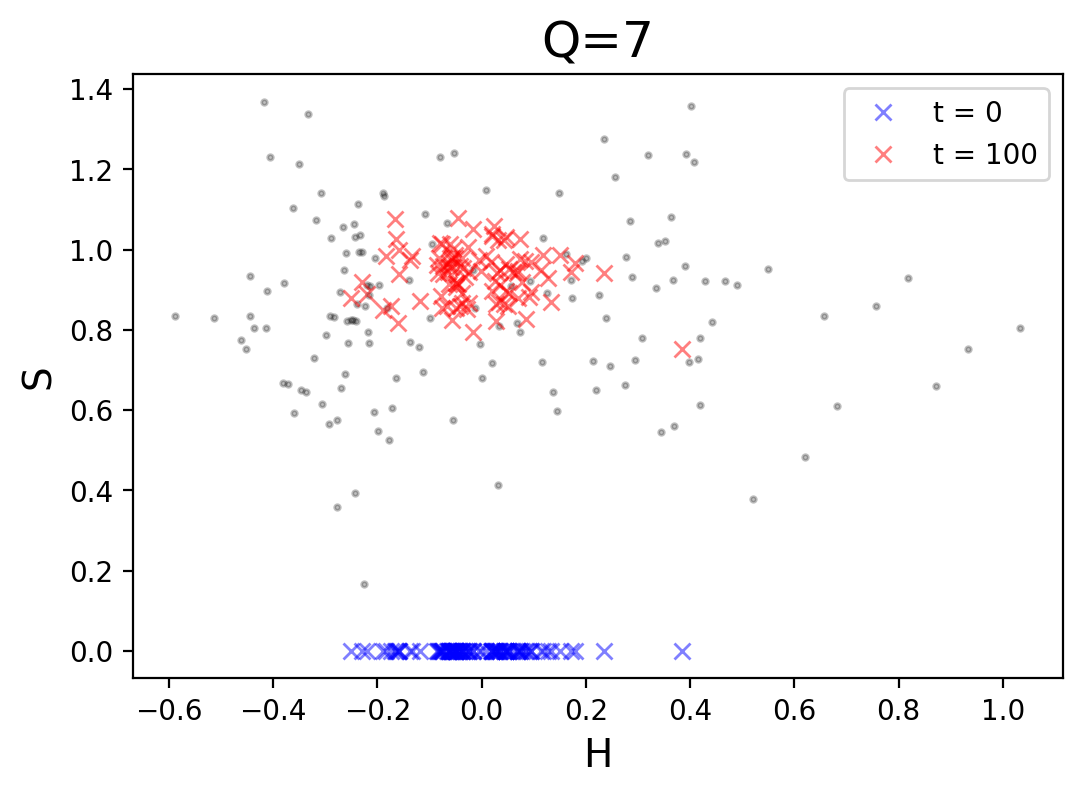}
	\end{subfigure}
    \begin{subfigure}{.32\textwidth}\centering
		\includegraphics[width=\linewidth]{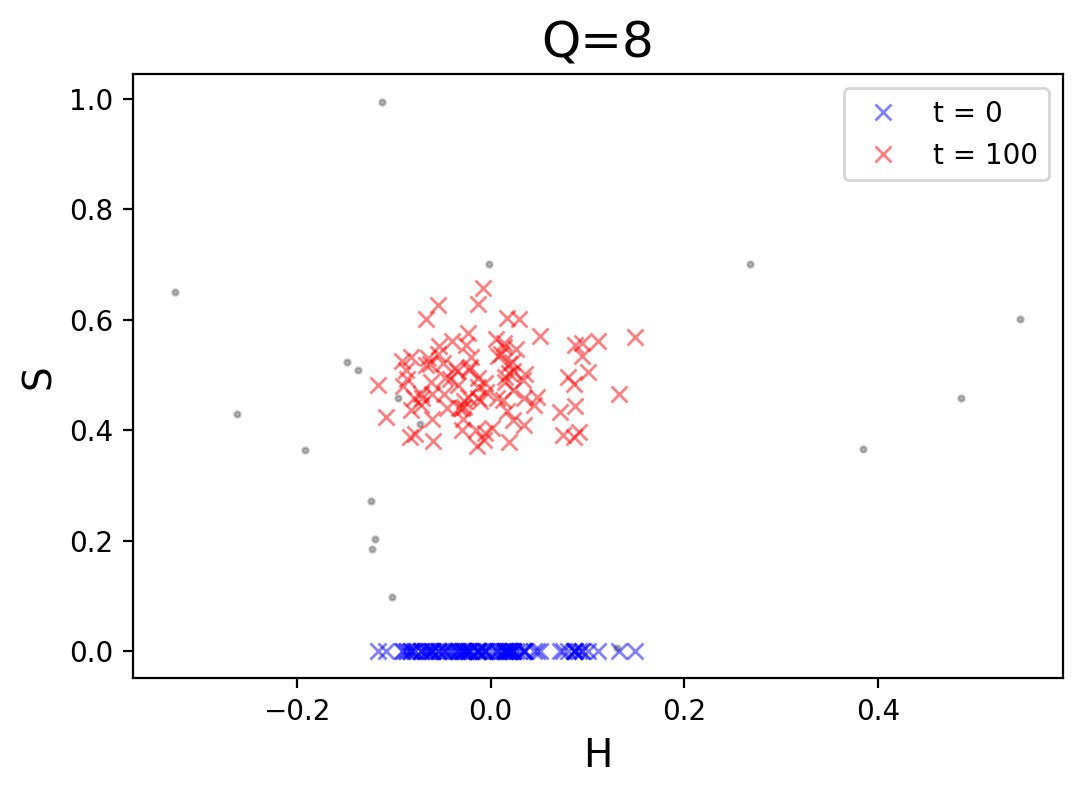}
	\end{subfigure}\hspace{.32 \linewidth}
	\caption{Each panel shows a comparison between the entropies and energies of eigenstates (marked by the black dots) within a fixed charge sector $Q$ and the time evolution of initially unentangled states with fixed $\langle Q\rangle$ (blue dots $t=0$, red dots $t=100$).}
	\label{fig: charge S}
\end{figure}

We now turn our attention to the dynamics of these unentangled states evolved by the qutrit Hamiltonian. The thermalization of out-of-equilibrium states is accompanied by a characteristic growth and saturation of entanglement entropy. First, to test if these out-of-equilibrium states do in fact thermalize we compare the distribution of entropies of time evolved states to the entropies of eigestates within a fixed charge sector. Figure \ref{fig: charge S} shows how the entropies of initially random unentangled states grow to match the distribution expected from a thermal density matrix under time evolution. As with the states of the previous section, the variance of entropy is minimized for states built from the largest charge sector, $Q=3$. We see that these states built under weaker assumptions, namely that they have a fixed charge expectation value, show similar behavior: for the charge sectors of largest dimension ($Q=2,3,4$), they follow the clear thermal pattern established by the eigenstates, whereas in sectors ($Q=6,7,8$), ETH fails and the pattern breaks down.  The initial entropy values are of course zero, and we see that across all energies after a large time step ($t=100$) the final entropy values lie in the middle of the eigenstate entropy distribution. In fact the typical temporal fluctuations, which are well represented by the vertical spread of red marks, are substantially smaller than those of the microcanonical states of the previous section.  This is understood by considering the remarks made in the penultimate paragraph of section \ref{subsec: generic ETH in theory}: the latter states superimposed a relatively small number of eigenstates, and had temporal fluctuations determined by the variance of the eigenstate expectation values themselves (see the colored dots in plots of the previous section). In a thermodynamic limit, as long as the density of states increases quickly enough, even microcanonical states should have vanishing fluctuations.  But this illustrates that the broad support in more generic states often leads to improved thermalization in finite systems. 

\begin{figure}\centering
	\begin{subfigure}{.32\textwidth}\centering
		\includegraphics[width=\linewidth]{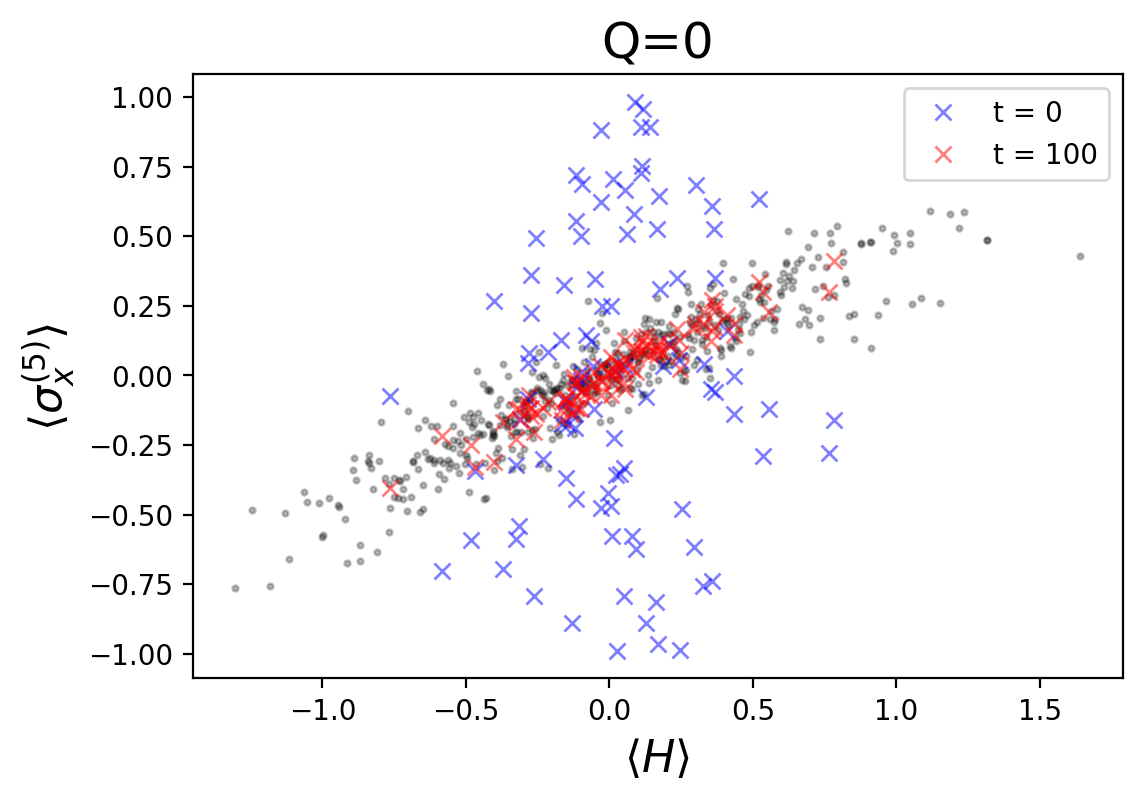}
	\end{subfigure}
	\begin{subfigure}{.32\textwidth}\centering
		\includegraphics[width=\linewidth]{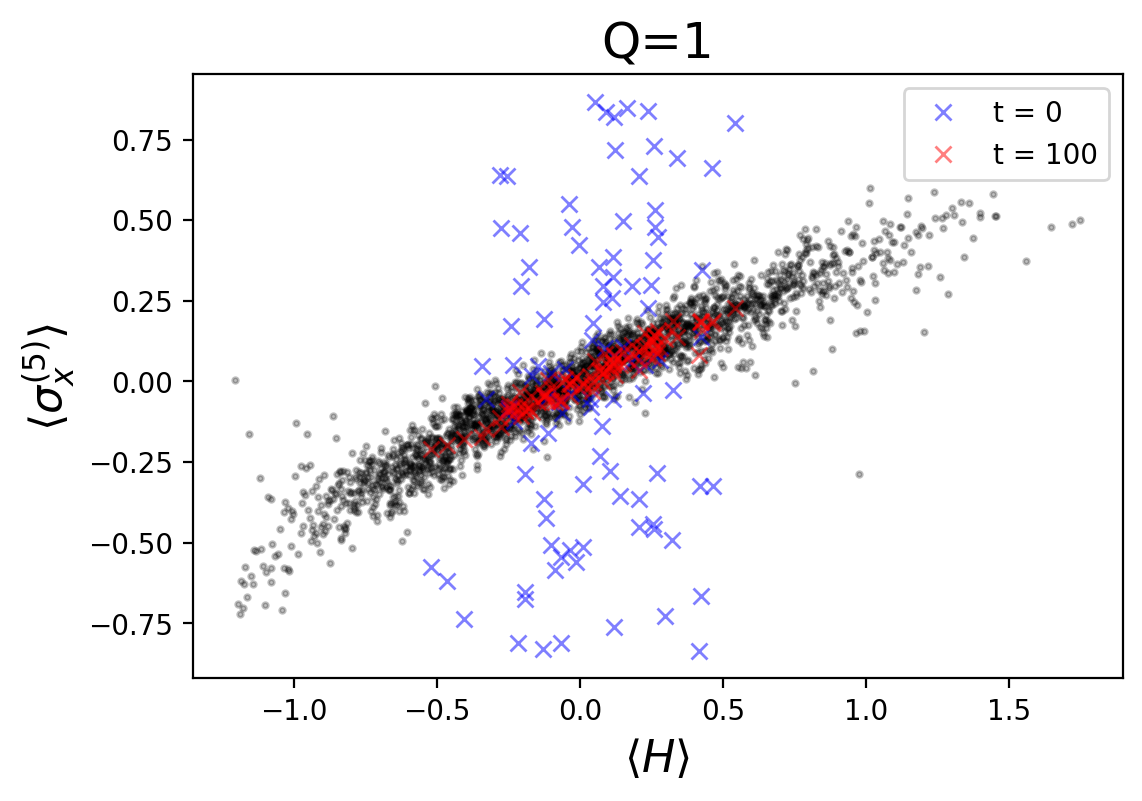}
	\end{subfigure}
	\begin{subfigure}{.32\textwidth}\centering
		\includegraphics[width=\linewidth]{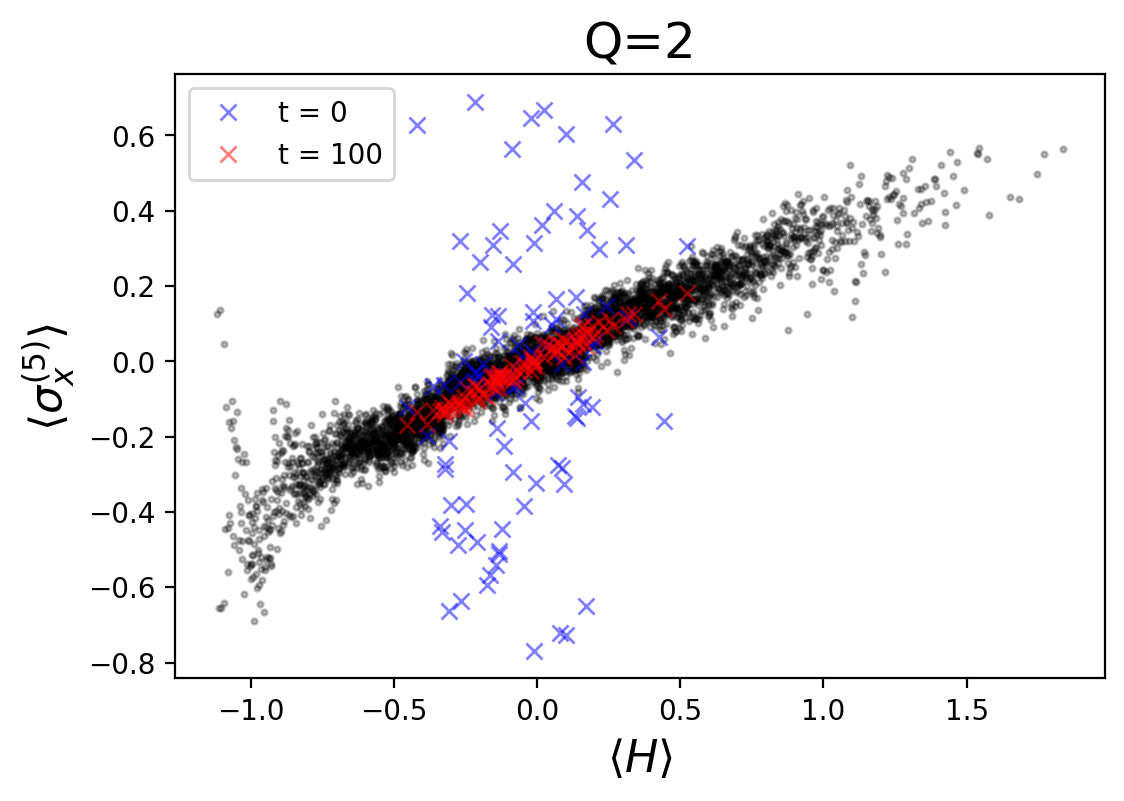}
	\end{subfigure}
    \begin{subfigure}{.32\textwidth}\centering
		\includegraphics[width=\linewidth]{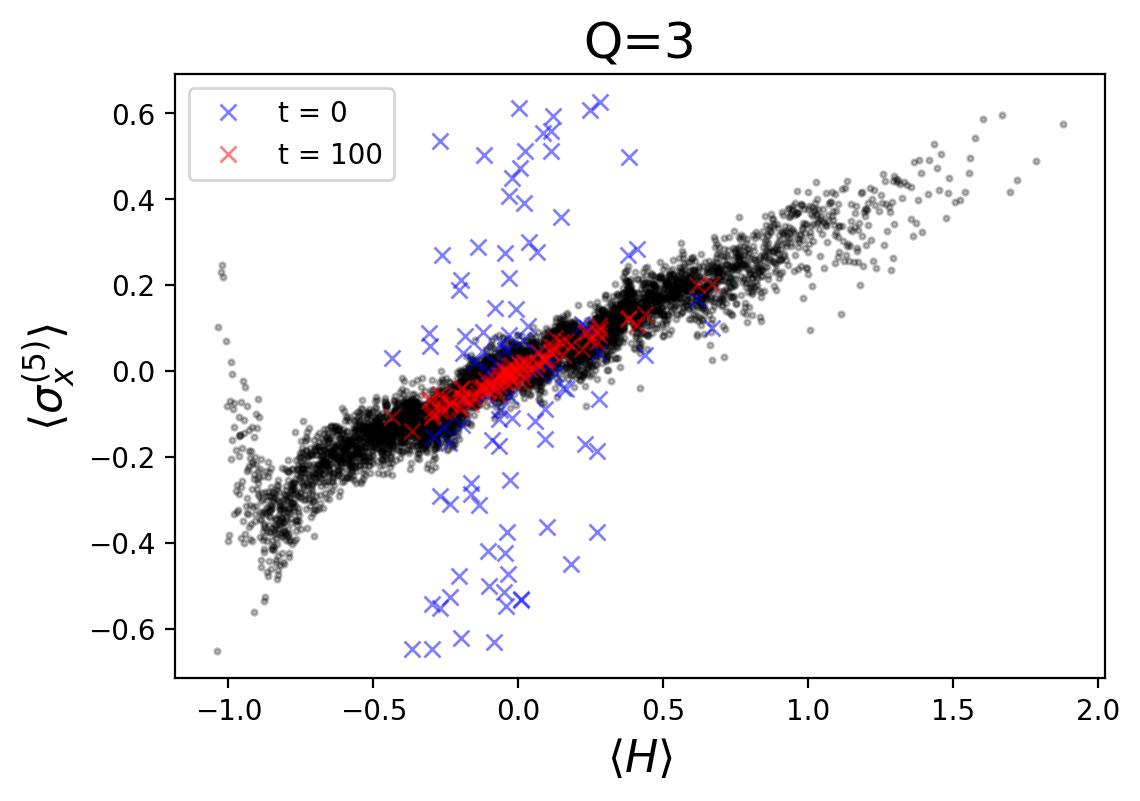}
	\end{subfigure}
	\begin{subfigure}{.32\textwidth}\centering
		\includegraphics[width=\linewidth]{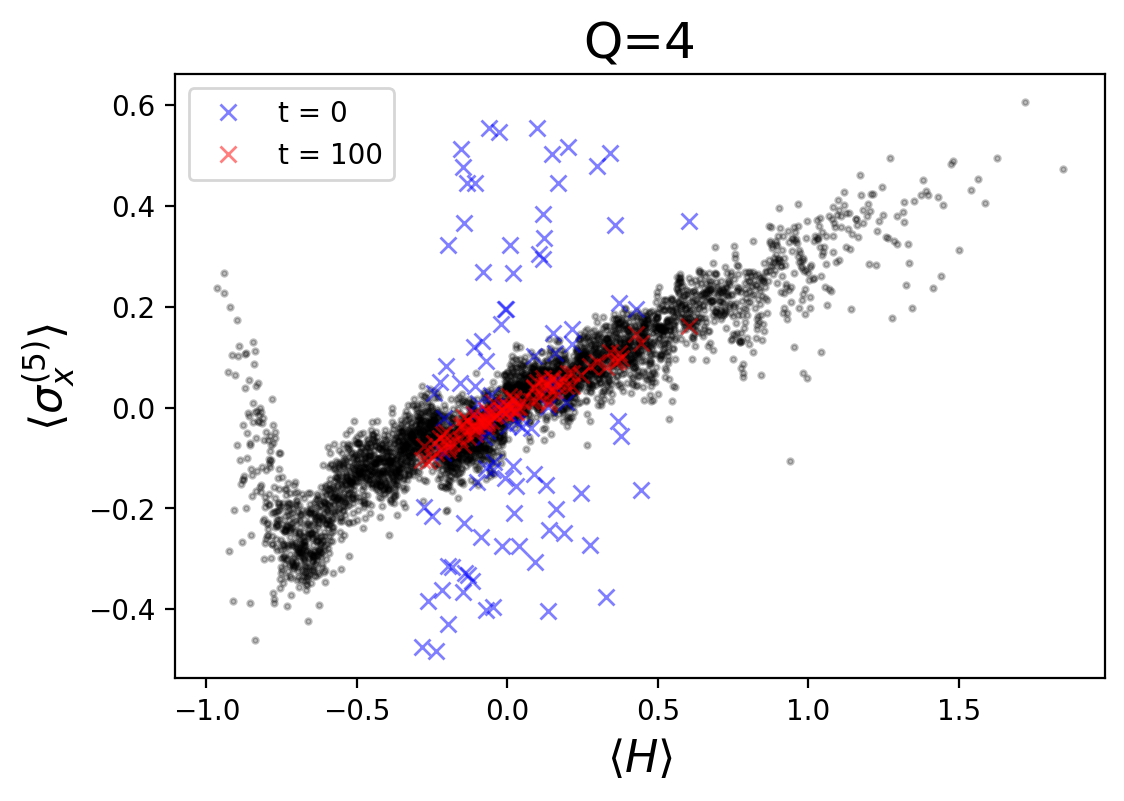}
	\end{subfigure}
	\begin{subfigure}{.32\textwidth}\centering
		\includegraphics[width=\linewidth]{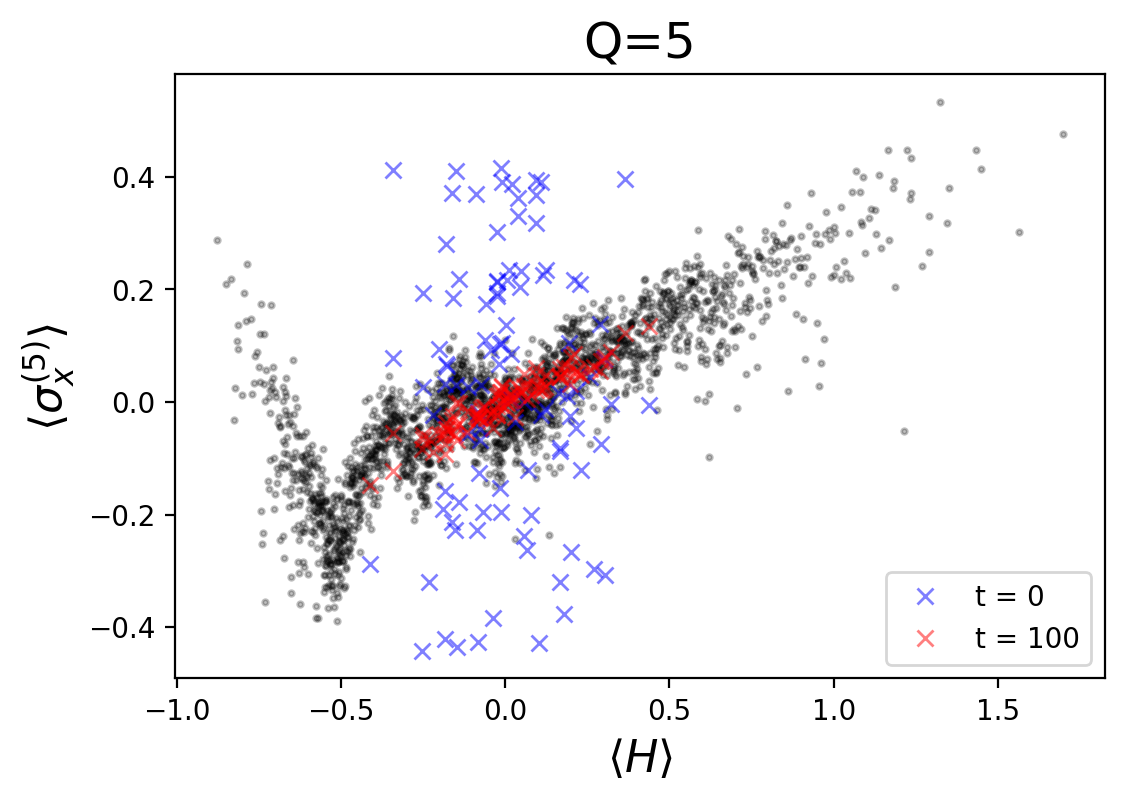}
	\end{subfigure}
    \begin{subfigure}{.32\textwidth}\centering
		\includegraphics[width=\linewidth]{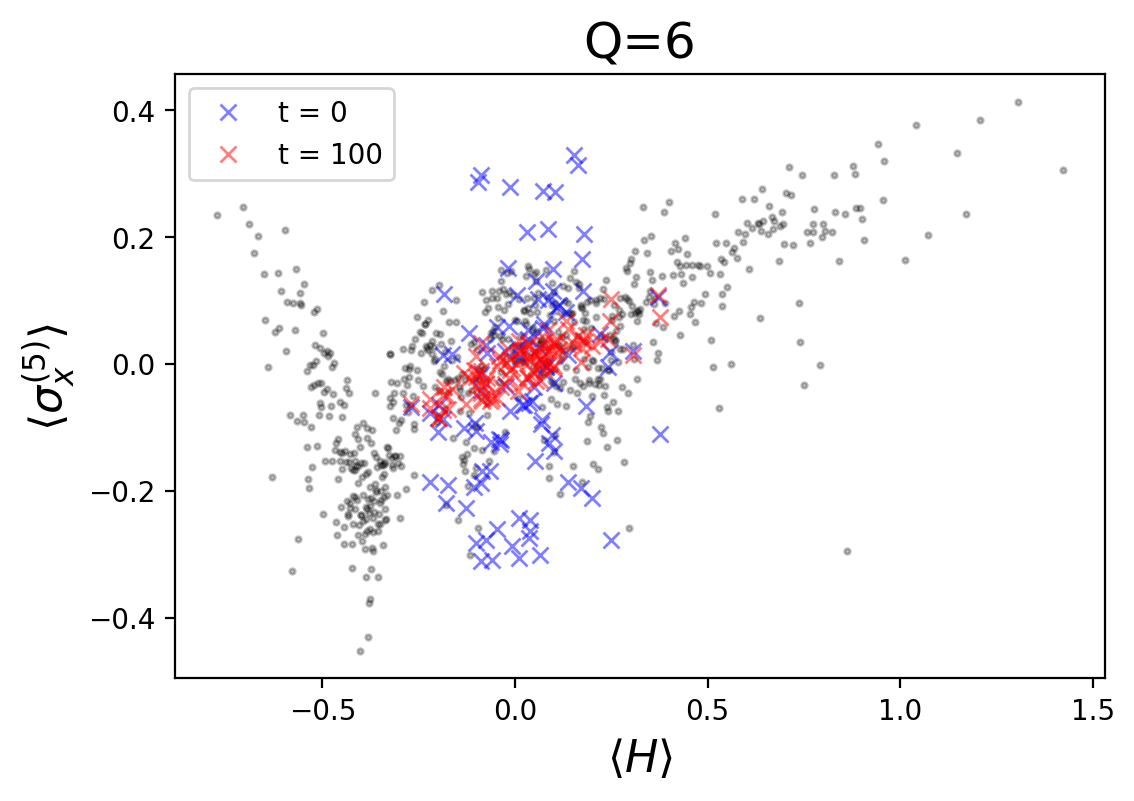}
	\end{subfigure}
	\begin{subfigure}{.32\textwidth}\centering
		\includegraphics[width=\linewidth]{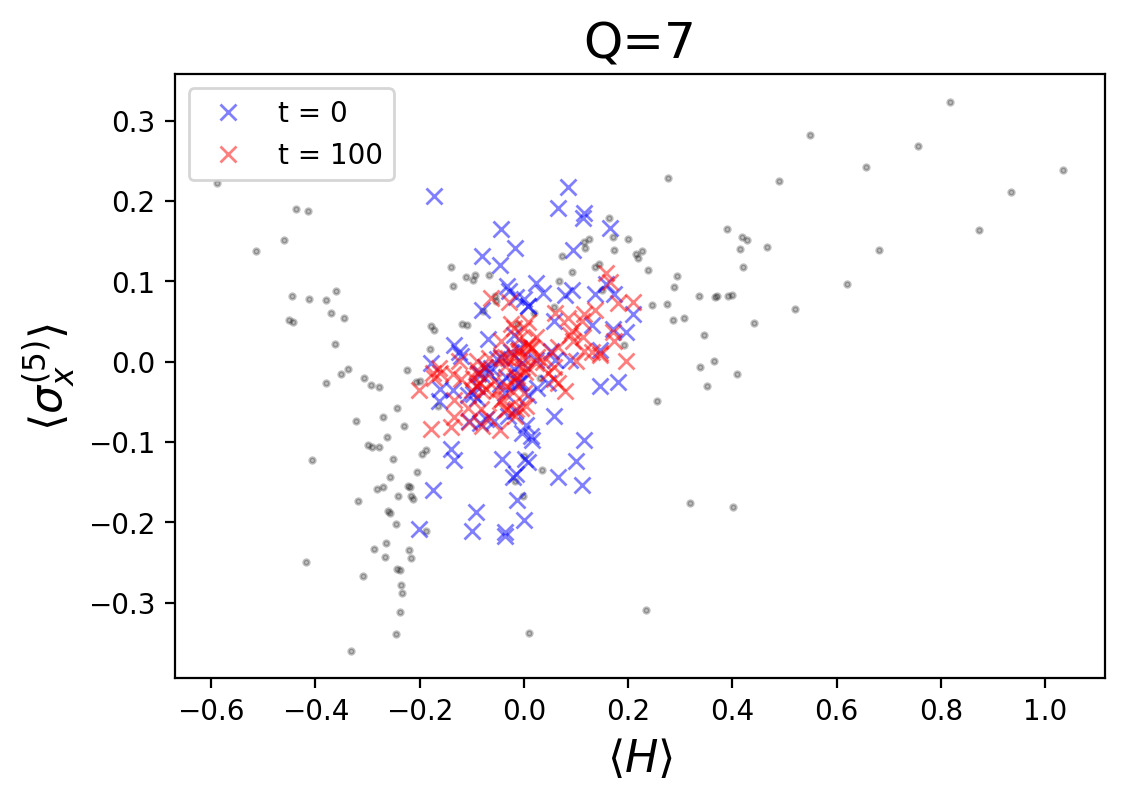}
	\end{subfigure}
 \begin{subfigure}{.32\textwidth}\centering
		\includegraphics[width=\linewidth]{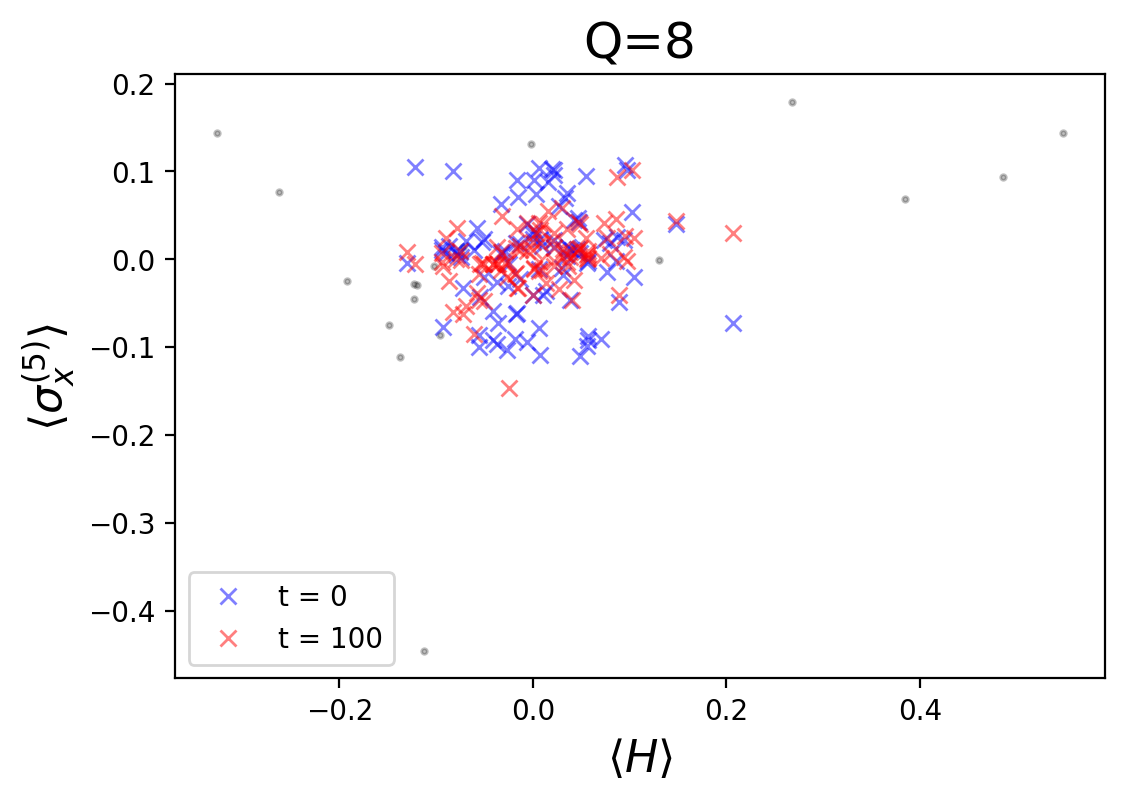}
	\end{subfigure}
	\caption{Each panel shows a comparison between the expectation value of a local Pauli operator $\langle \sigma_x \rangle$  acting on the qubit subsytem placed halfway down the lattice and energies of eigenstates (marked by the black dots) within a fixed charge sector $Q$ and the time evolution of initially unentangled states with fixed $\langle Q\rangle$ (blue dots $t=0$, red dots $t=100$).}
	\label{fig: charge pauli x}
\end{figure}

In figure \ref{fig: charge pauli x}, we additionally study the thermalization of a local Pauli operator in these states.  Much of the preceding discussion for the entropies can be repeated.  However in contrast to the initial entropy values, which by design were fixed to $S = 0$, the initial values of $\langle \sigma_x \rangle$ are completely random (seen in the blue markings) as they correspond to the orientation of the qubit subsystem at a site halfway down the spinchain. For the charge sectors with largest state space, the states again settle uniformly to thermal values (seen in the red markings), again with smaller fluctuations than microcanonical states at the same energy.  By contrast, as seen for $Q=8$, the initially randomized expectation values (shown in blue) do not differ significantly from the `thermalized' values (shown in red). This again leads to the conclusion that not every charge sector is well-thermalizing. 

\begin{figure}[H]\centering
	\begin{subfigure}{.32\textwidth}\centering
		\includegraphics[width=\linewidth]{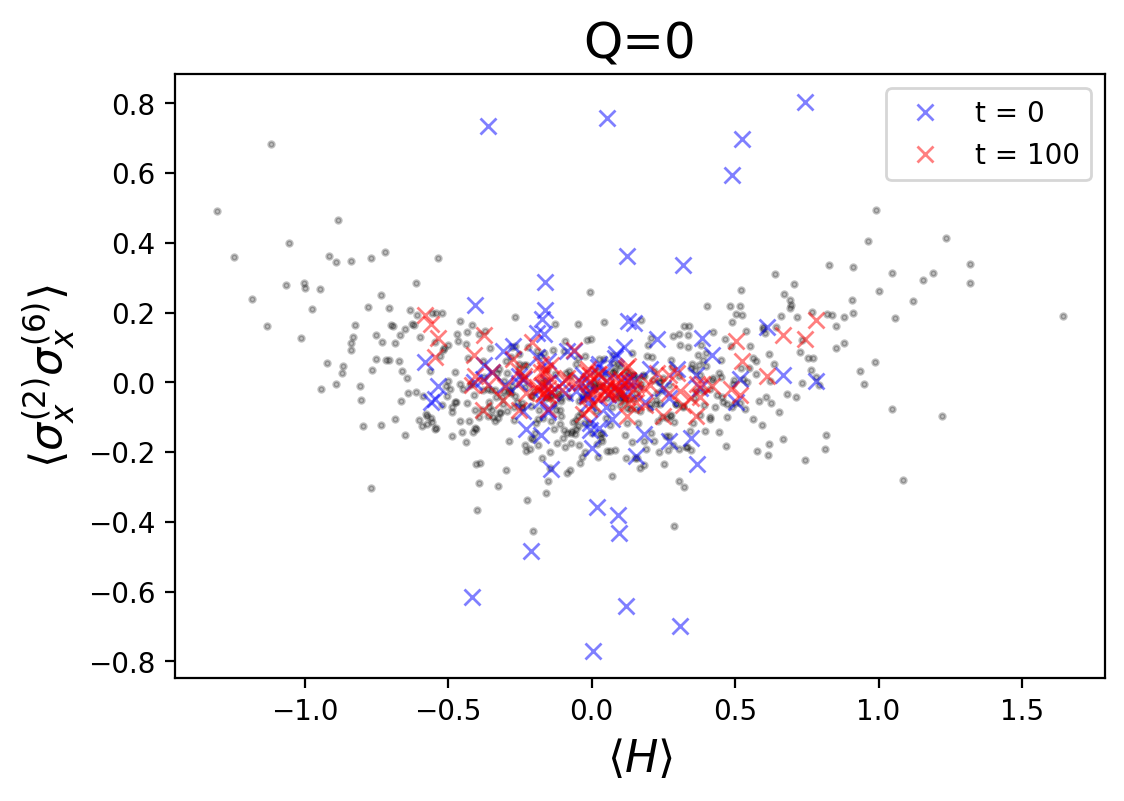}
	\end{subfigure}
	\begin{subfigure}{.32\textwidth}\centering
		\includegraphics[width=\linewidth]{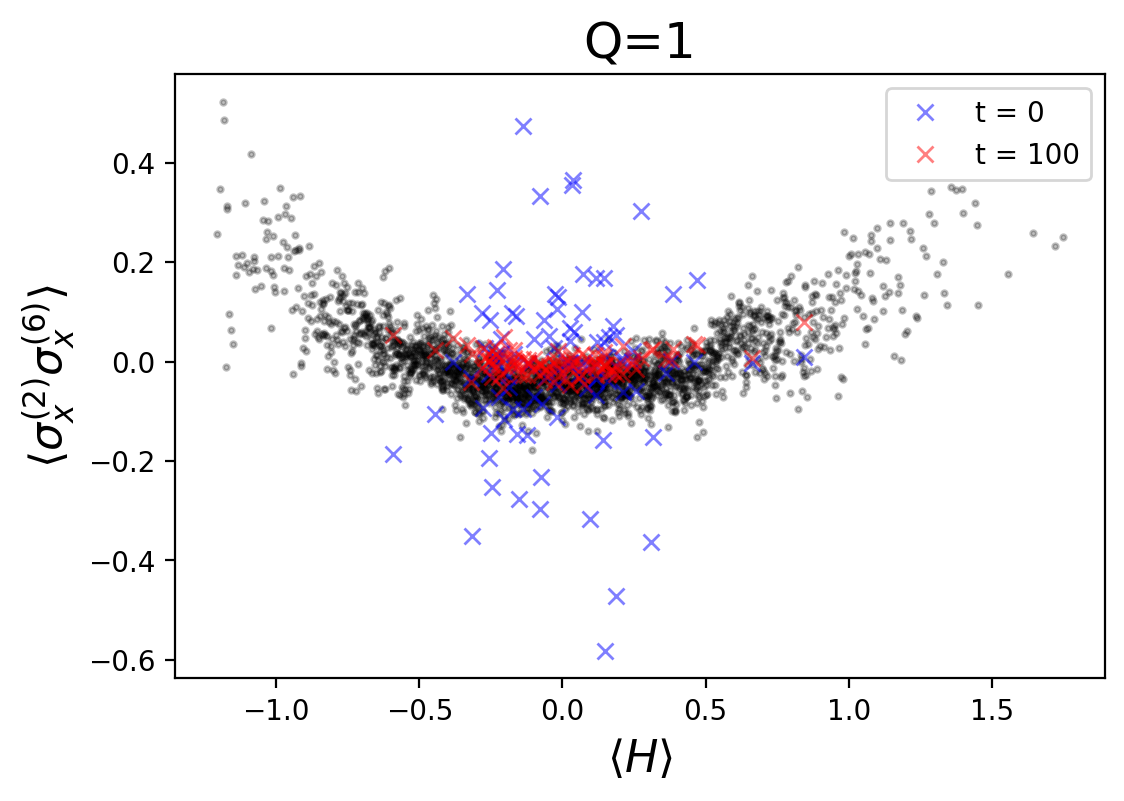}
	\end{subfigure}
	\begin{subfigure}{.32\textwidth}\centering
		\includegraphics[width=\linewidth]{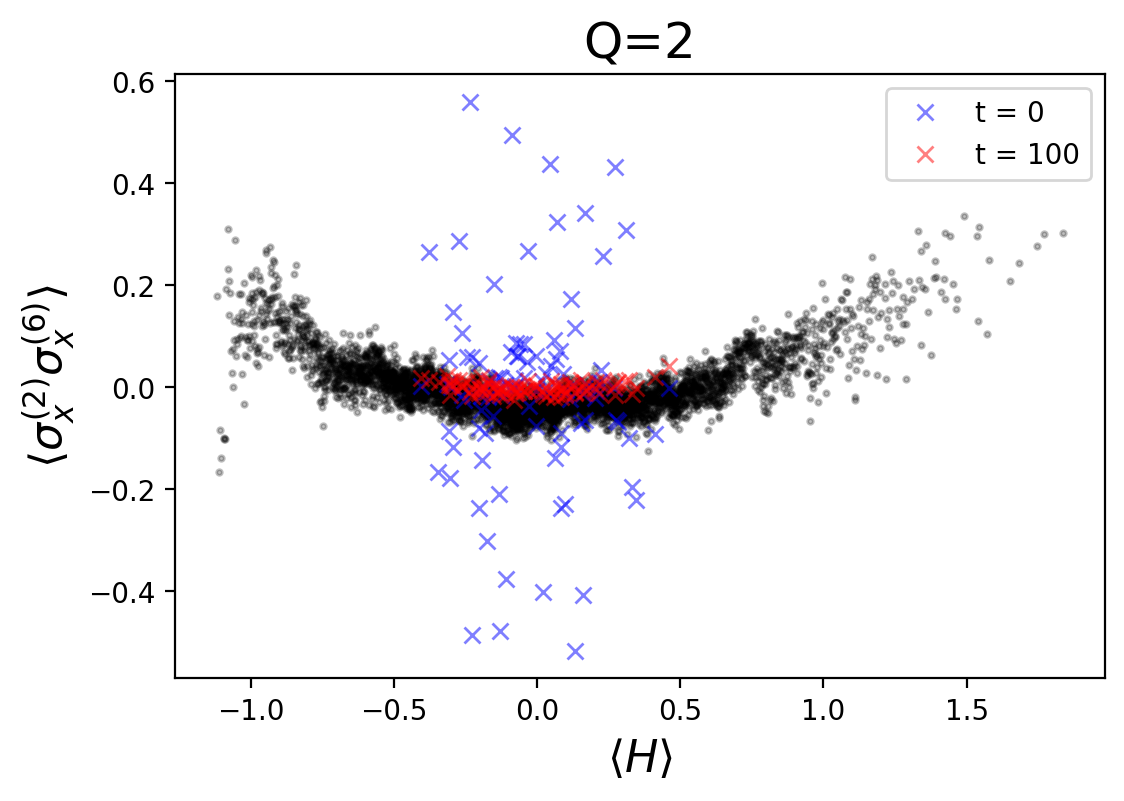}
	\end{subfigure}
    \begin{subfigure}{.32\textwidth}\centering
		\includegraphics[width=\linewidth]{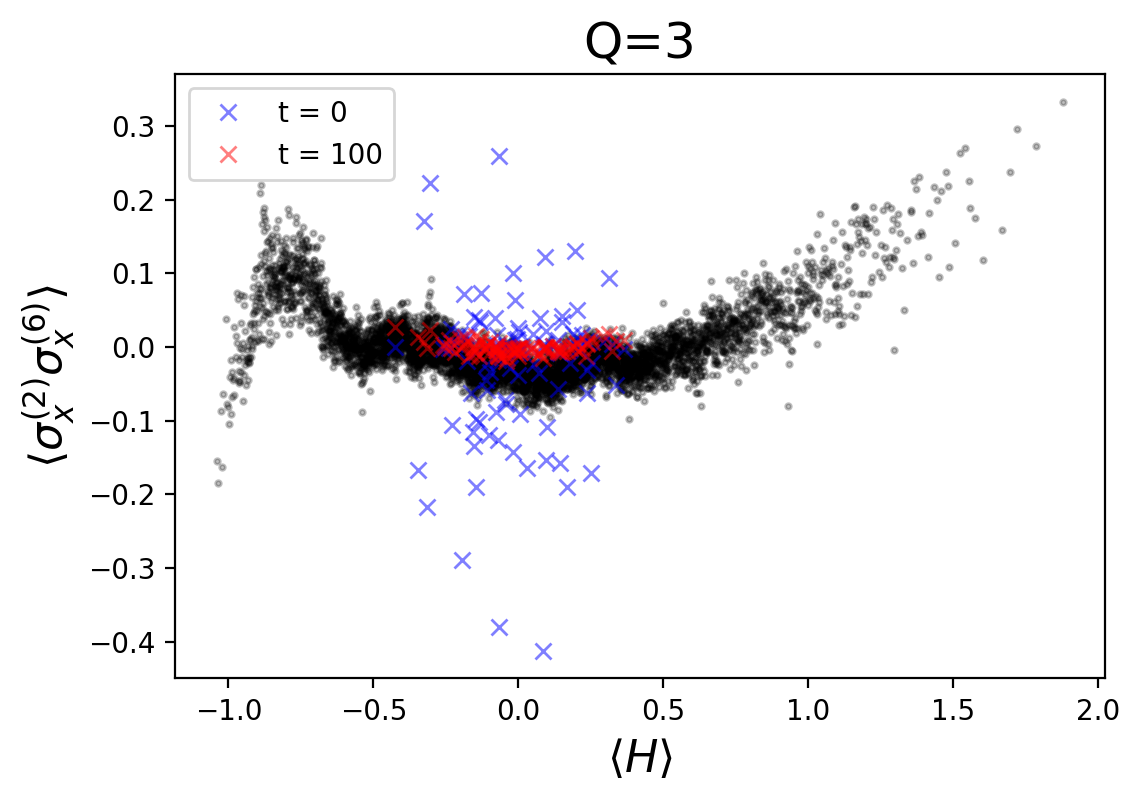}
	\end{subfigure}
	\begin{subfigure}{.32\textwidth}\centering
		\includegraphics[width=\linewidth]{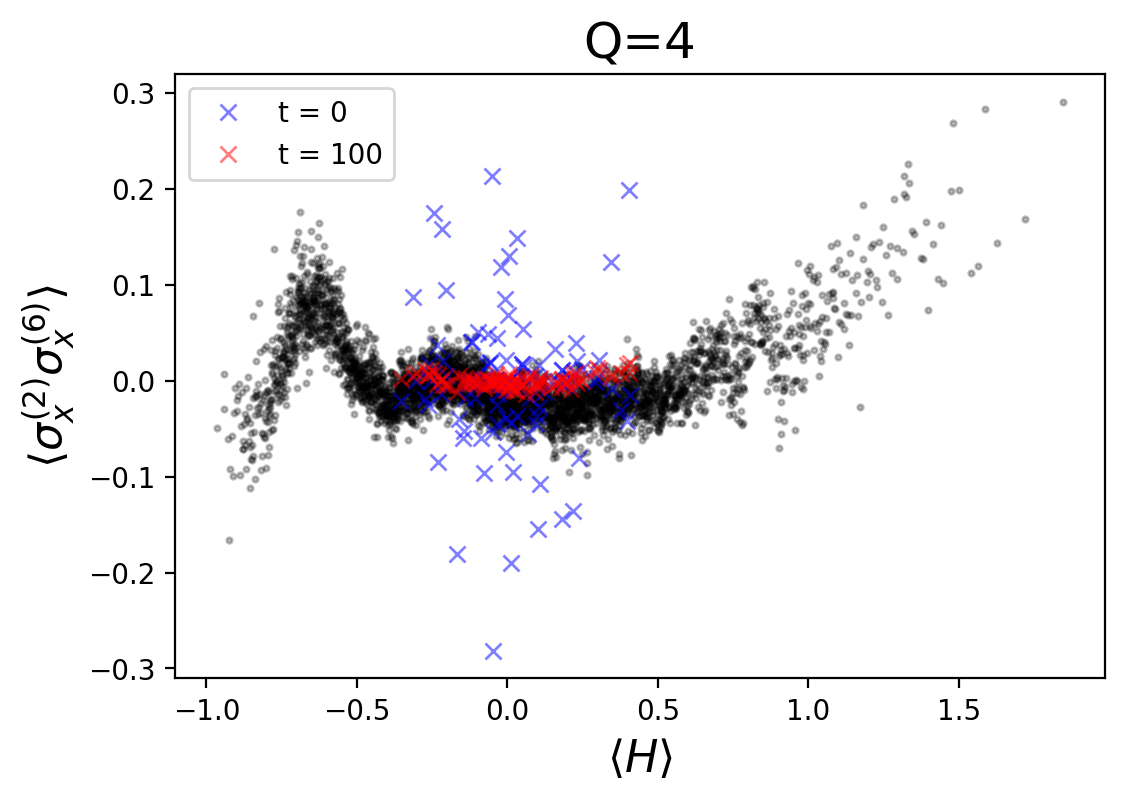}
	\end{subfigure}
	\begin{subfigure}{.32\textwidth}\centering
		\includegraphics[width=\linewidth]{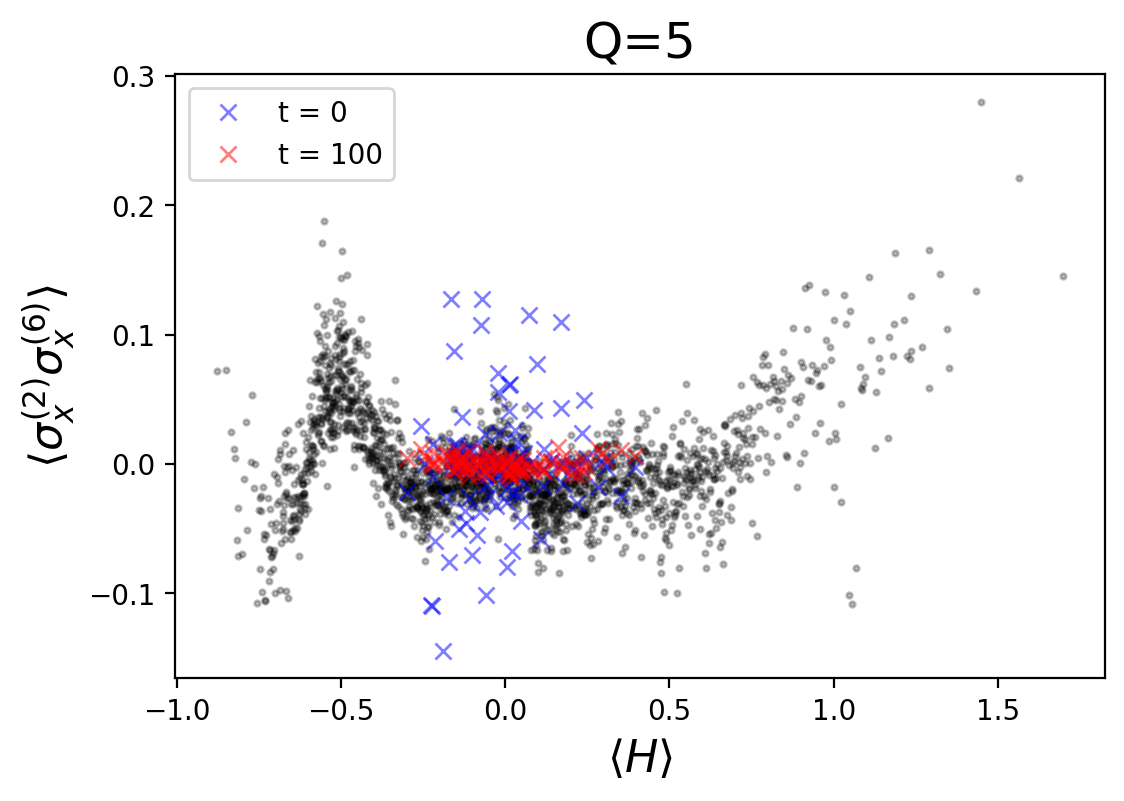}
	\end{subfigure}
    \begin{subfigure}{.32\textwidth}\centering
		\includegraphics[width=\linewidth]{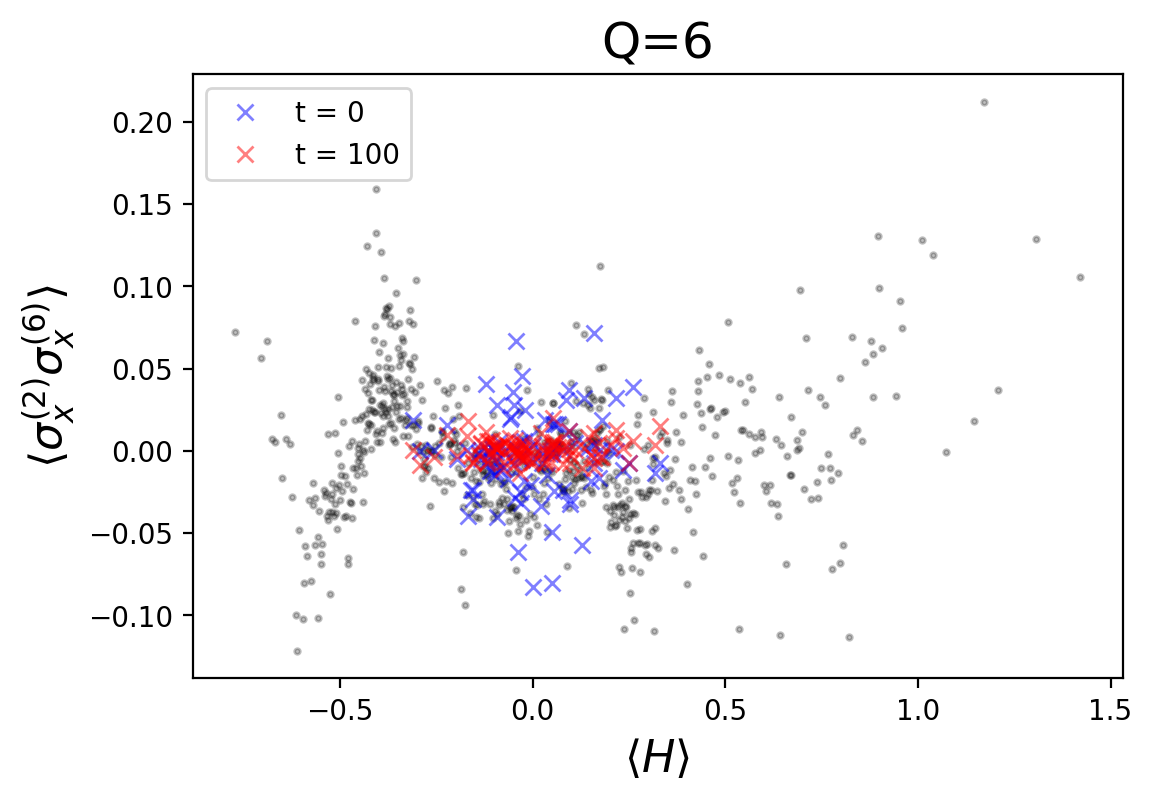}
	\end{subfigure}
	\begin{subfigure}{.32\textwidth}\centering
		\includegraphics[width=\linewidth]{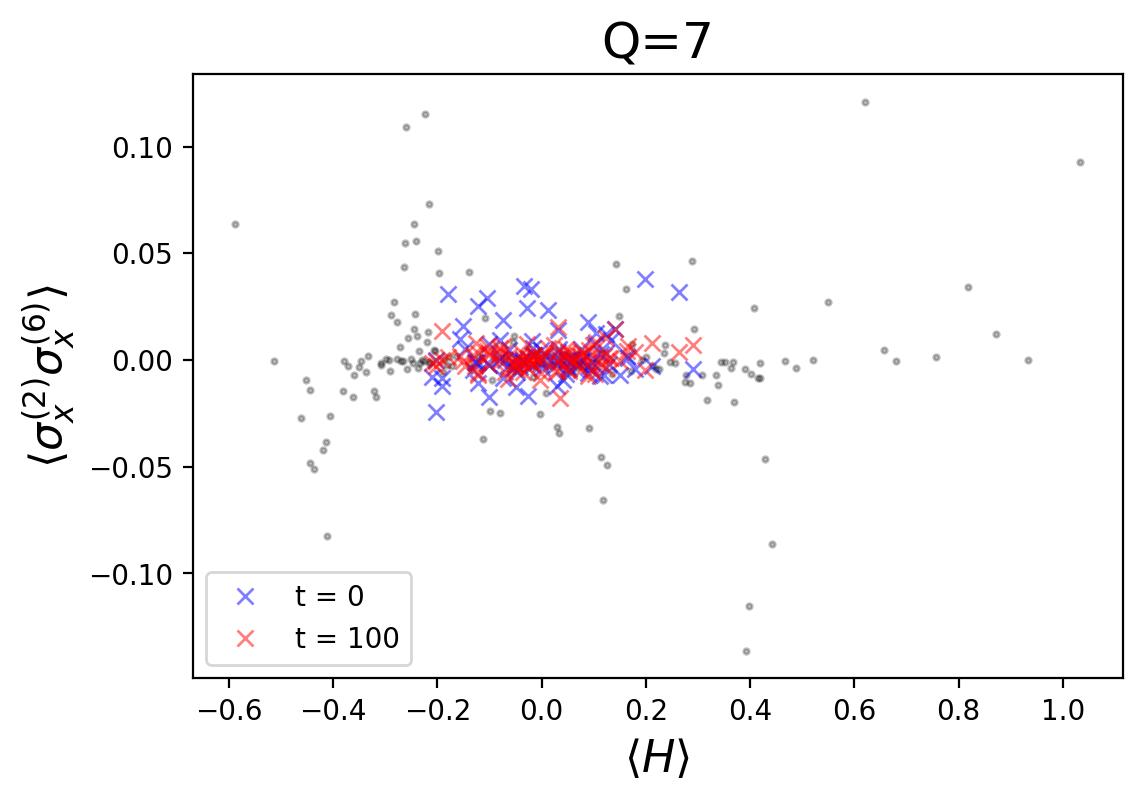}
	\end{subfigure}
 \begin{subfigure}{.32\textwidth}\centering
		\includegraphics[width=\linewidth]{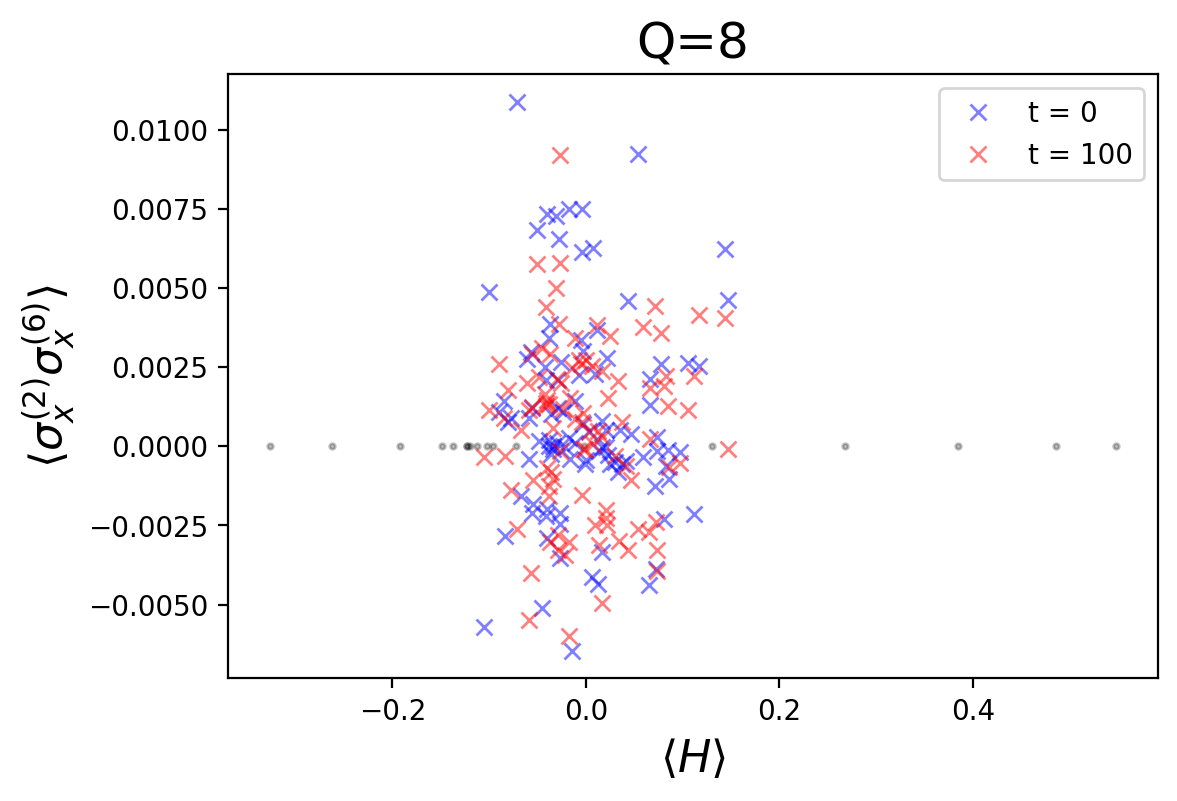}
	\end{subfigure}
	\caption{Each panel shows a comparison between the expectation value of a non-local operator product of Pauli operators $\langle \sigma^{(2)}_x \sigma^{(6)}_x \rangle$ placed halfway down the lattice and energies of eigenstates (marked by the black dots) within a fixed charge sector $Q$ and the time evolution of initially unentangled states with fixed $\langle Q\rangle$ (blue dots $t=0$, red dots $t=100$).}
	\label{fig: charge pauli non loc x}
\end{figure}

In figure \ref{fig: charge pauli non loc x} we study a nonlocal, two-site operator $\langle\sigma_x^{(2)}\sigma_x^{(6)}\rangle$. The same operator was studied in \ref{fig: microcanonical pauli non loc x window} for random microcanonical states.  We focus our discussion on the largest sectors ($Q=2,3,4$) for clarity.  We once again we find that these states settle to values with much smaller fluctuations than the microcanonical states of the previous section.  However the equilibrium value itself is visibly offset from the microcanonical prediction (which can be visualized inferred as threading the vertical center of the black dots).  Both of these effects should be considered in light of the discussion in section \ref{subsec: generic ETH in theory}: they arise due to the broadened support across both charge sectors and energy ranges which distinguishes these states from microcanonical states.  The fluctuations decrease simply due to averaging over a larger number of states.  The offset from the microcanonical values we interpret as arising from the concavity of the smooth function approximating this particular operator.  Across the dominant, central energy range, this function is approximately flat in the direction of varying charge, whereas it has noticeable positive concavity in the direction of varying energy.  In contrast to a generalized microcanonical ensemble, the generalized Gibbs ensemble\footnote{With only a single conserved charge present, this generalized Gibbs ensemble could also be called a grand canonical ensemble.} that is relevant to these states probes over a large range of both variables and is sensitive to this concavity, pulling the thermal values upward.

\subsubsection{State-averaged thermalization}
\label{subsubsec: state-averaged thermalization}

In this subsection we present one final study of the generic, fixed-$\langle Q \rangle$ states from the previous subsection.  Randomly generated unentangled states, without imposing any additional constraints, tend to have $\langle H \rangle \sim 0$ simply due to state space considerations.  This continues to be true for the unentangled states at fixed $\langle Q \rangle$, however different $\langle Q\rangle$ will admit different energy ranges due to the differing amount of support placed in the qubit subsystem versus the locally charged state (the ``third slot''). Additionally, for states constructed using randomly oriented qubits, the energy distributions of each particular state will differ slightly.  This results in states which thermalize along different trajectories, and are prone to fluctuations of varying sizes.  To further study the general behavior of the states used in the previous section, we average the system dynamics over 500 random initial states at fixed $\langle Q\rangle$, making sure to use the same 500 embedded qubits when comparing the dynamics of different values of $Q$. Averaging the behavior suppresses fluctuations from individual states and allows for a clearer view of the hallmark features of thermalization in the dynamics. The typical states have energies $\langle H \rangle \sim 0$. In principle we can also select for states at different $\langle H\rangle$, but we will focus on $\langle H \rangle \sim 0$ for concreteness.

\begin{figure}[H]
	\centering
	\includegraphics[width=\linewidth]{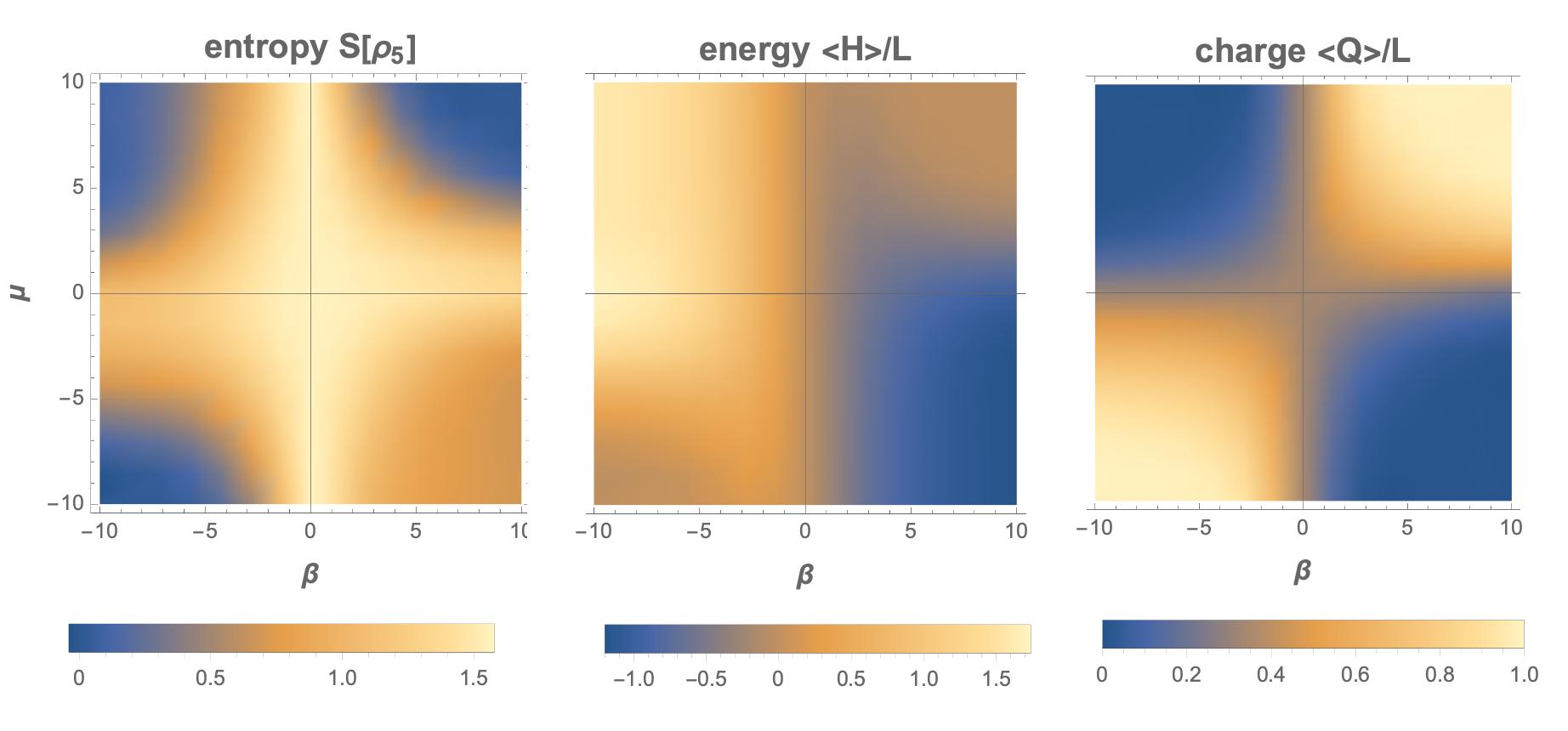}
	\caption{Density profiles are shown for several quantities associated with the thermal density matrix $\rho_{th} = e^{-\beta(H - \mu Q)}/\Tr\left(e^{-\beta(H - \mu Q)}\right)$ in a system of size $L=8$ qutrits.}
	\label{fig: qutrit thermal profiles}
\end{figure}

These states are characterized by both the expectation values of $\langle H\rangle$ and $\langle Q \rangle$. The corresponding thermal predictions of simple observables and entropy for these states are then computed from the thermal density matrix $\rho(\beta,\mu) = e^{-\beta(H - \mu Q)}/\Tr\left(e^{-\beta(H - \mu Q)}\right)$, which is specified by the inverse temperature $\beta$ and chemical potential $\mu$. In principle, the values of $\langle H\rangle$ and $\langle Q \rangle$ can be related to $\beta$ and $\mu$ through a Legendre transform to compute the predicted thermal values of our generated states. This is not possible analytically for our Hamiltonian (\ref{eq: qutrit hamiltonian}) due to the charge spreading terms required for chaotic dynamics. Instead we must turn to numerical methods to compute thermal predictions. For this purpose, we first generated values of $\langle Q \rangle$, $\langle H \rangle$, and $S(\rho_i)$ in the mixed state $\rho(\beta,\mu)$ each across a grid of $\beta$ and $\mu$ values (see figure \ref{fig: qutrit thermal profiles}), and then used a multidimensional interpolation to associate a prediction of the single qutrit entanglement entropy $S(\rho_i)$ for states with given values of $\langle H \rangle$ and $\langle Q\rangle$. However such an approach is too computationally intensive for system sizes larger than $L=9$, so when we come to these systems we will instead show equilibration rather than thermalization (comparing only to late time averages rather than predictions of a density matrix).

\begin{figure}[h]
	\centering
	\includegraphics[width=\linewidth]{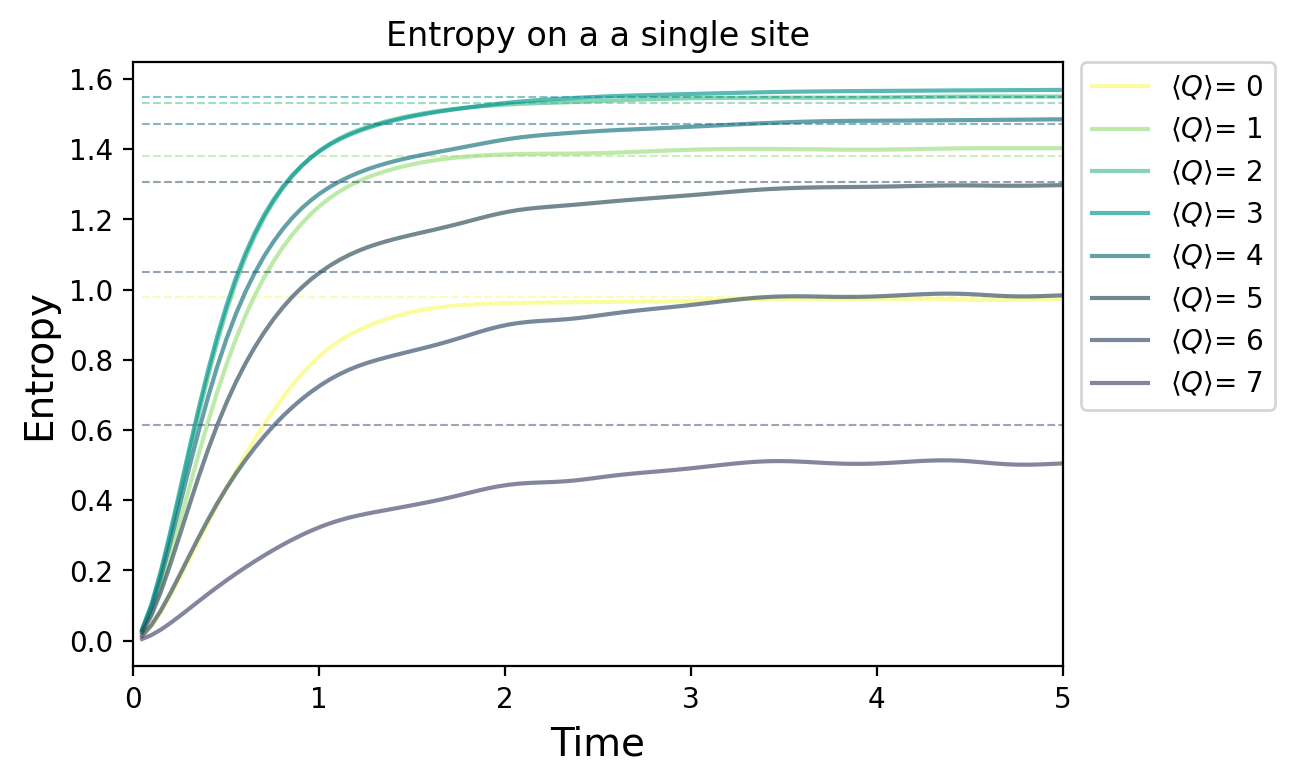}
	\caption{Growth curves of entanglement entropy on a single lattice site halfway down a spin chain of size $L=9$. Each dashed line shows the thermal expectation value of entropy computed from the density matrix $\rho_{th} = e^{-\beta(H - \mu Q)}/\Tr\left(e^{-\beta(H - \mu Q)}\right)$. The total charge for the spin chain, $\langle Q\rangle$, is distributed evenly along each site on the lattice. The curves shown are averages of the entropy growth of $100$ random initially unentangled states. }
	\label{fig:ent th}
\end{figure}

For a system size of $L=8$, the entropy growth and saturation to values predicted by $\rho(\beta,\mu)$ can be seen in figure \ref{fig:ent th}.  We see that the entropy saturation values increase as a function of $Q$, until $1/3$ of the maximal value of $Q$ is reached, when the entropy saturation values begin to decrease. The sector where $Q=1/3Q_{max}$ admits the largest charge sector, which correspondingly has the largest entropy. The dashed lines in the figure show the predicted values of entropy for each $\langle Q \rangle$ that were computed numerically from $\rho(\beta,\mu)$ for states with $\langle H\rangle=0$ and variable $\langle Q \rangle$. We see that the generated entropy curves match the predicted values well for low values of $\langle Q \rangle$, but the match becomes worse as $\langle Q \rangle$ approaches $L$.  We expect a breakdown for these states, which are mostly supported on at the edges of the charge spectrum (sectors $Q=6$, $Q=7$), just as we observed a breakdown of ETH in these sectors in the previous two subsections.

For a larger spinchain of $L=12$, we are unable to compute the exact thermal predictions for these states from $\rho(\beta,\mu))$ due to numerical constraints.  Instead, we show the equilibration of curves akin to thermalization. The dashed lines in the plots below correspond to a late-time equilibrated value of each entropy curve evolved to $t=50$ (thus the late time ``fit'' of these curves is guaranteed in this case). We find that the behavior of curves in the large system follow the same trends as for the smaller system, admitting a maximal entropy at $\langle Q \rangle =4$, corresponding to $1/3$ of the maximal charge of 12. We cannot however determine if the values match the predicted ones from the generalized Gibbs ensemble, which we expect to occur for states across a larger range of $\langle Q \rangle$ as the system size increases. 

\begin{figure}[h]
	\centering
		\centering
		\includegraphics[width=\linewidth]{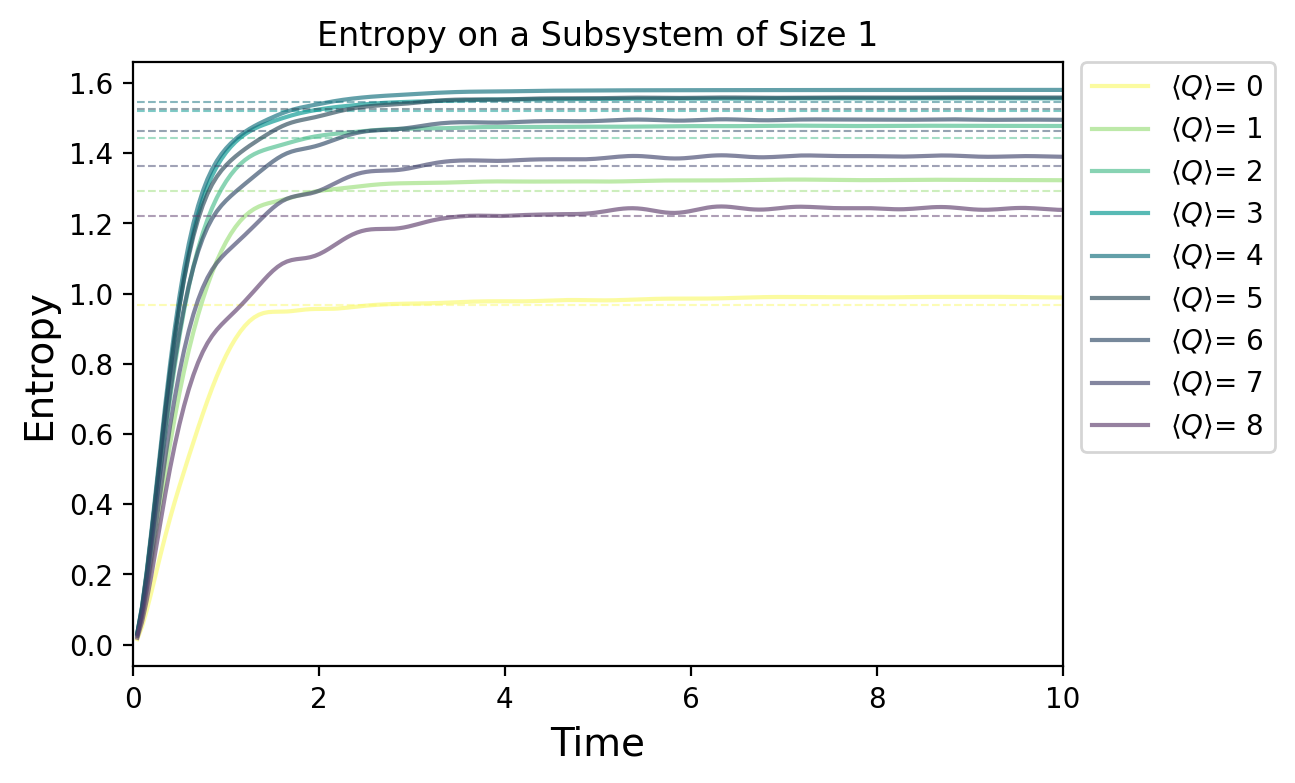}
		\label{fig:ent eq}
	\caption{Growth curves of entanglement entropy on a single lattice site halfway down a spin chain of size $L=12$. Each dashed line shows the equilibrium value of each curve for $t=50$. The total charge for the spin chain, $Q$, is distributed evenly along each site on the lattice. The curves shown are averages of the entropy growth of 100 random initially unentangled states.}
	\label{fig:entropygrowthequilibriation}
\end{figure}

\section{Conclusions}
\label{sec: conclusions}

In this work we have investigated the connection between several tightly linked phenomena in some simple quantum lattice systems: chaos, thermalization, and the ETH form of simple operators.  We have reviewed and illustrated standard results regarding the ETH mechanism and its implications for thermalization in a chaotic qubit spin chain model.  We then designed a related qutrit Hamiltonian with a quasi-local conserved charge.  This system, through its variable parameters, provides an arena to test the onset of generalized eigenstate thermalization and its implications.  We identify a regime where chaos and ETH can be clearly seen to emerge within individual charge sectors, with eigenstates far from the edges of the spectrum acting as approximate thermal states for local observables.  Eigenstate expectation values of simple operators across nearby charge sectors likewise show the emergence of smooth functions of both charge and energy, at least in the regimes with high density of eigenstates, which illustrates the onset of generalized eigenstate thermalization.  We anticipate that future studies on larger systems will only show these behaviors more clearly.  

Throughout this work, we have emphasized that the (generalized) ETH form of simple operators, where it occurs, is more predictive than just its most common application to (generalized) microcanonical states.  For a large class of states with non-negligible fluctuations in conserved quantities, it clarifies that deviations between local observables' equilibrium behavior and the predictions of thermal distributions are best captured simply by modifying the latter to mimic higher central moments of the states' energy and charge distributions.  For observables with approximately linear dependence on microcanonical thermal parameters, thermalization occurs with exceptional insensitivity to the details of the initial state, which can be broadly supported well outside microcanonical windows.  We dub this behavior `generic ETH,' and illustrate with a few simple, but natural sets of states in our lattice systems.  

With our focus on exact diagonalization and direct demonstration of eigenstate thermalization in these systems, many interesting phenomenological aspects of the designed qutrit system have been left unexplored.  It would be interesting to see in more detail how the time scales involved in thermalization are affected by the presence of the quasilocal charge.  An analysis in the manner of \cite{Volya_2020} concerning the detailed time evolution of thermalizing observables would likewise be of interest in this system. The way in which a chaotic qubit system is embedded into an enlarged lattice with an overlaid local charge dynamics may lead to interesting interplay of multiple timescales, associated with the chaotic qubit system dynamics and the charge diffusion separately.   Since the size of the state space that is ``active'' on the chaotic qubit subsystem depends directly on the amount of charge on a lattice site, there may be a rich phenomenology not just of thermalization and entanglement growth, but also other measures of information scrambling and chaos not studied in this work.  It would also be interesting to allow for local ``frustrations'' of the charge spreading ability that partially segregate otherwise homogeneous chaotic dynamics.  All these possibilities we leave to future work.

\acknowledgments

The authors thank Josh Kirlin, Juan Maldacena, and Grant Remmen for helpful discussions.  The work of E.C. was supported by the National Science Foundation
under grants PHY-2112725 and PHY–2210562.  S.E. was supported by funding from the Okinawa Institute of Science and Technology Graduate University, and thanks Perimeter Institute and Syracuse University for hosting visits during which part of this work was completed.  J.P. was supported in initial stages of this work by the Simons Foundation through \emph{It from Qubit: Simons Collaboration on Quantum Fields, Gravity, and Information}. This work was completed while they were visiting the Simons Institute for the Theory of Computing, supported by DOE QSA grant \#FP00010905. S.R. was supported by the National Science Foundation under grant PHY–2210562.

\bibliographystyle{unsrt}
\bibliography{paper}

\end{document}